\documentclass[10pt,journal,compsoc]{IEEEtran}

\usepackage[T1]{fontenc}
\usepackage{mathptmx}                  %

\usepackage{graphicx}
\graphicspath{{figures/}{pictures/}{images/}{./}}

\usepackage{amsmath,amssymb}
\usepackage{booktabs}                  %
\usepackage{array}                     %
\newcolumntype{L}[1]{>{\raggedright\arraybackslash}p{#1}}
\usepackage[caption=false,font=footnotesize]{subfig} %
\usepackage{cite}                      %
\usepackage{url}
\usepackage{CJKutf8}

\usepackage[colorlinks,allcolors=black,bookmarks]{hyperref}
\usepackage[all]{hypcap}
\usepackage[nameinlink,capitalise]{cleveref}
\crefname{table}{Table}{Tables}
\Crefname{table}{Table}{Tables}
\crefname{section}{Section}{Sections}
\Crefname{section}{Section}{Sections}
\crefname{figure}{Fig.}{Figs.}
\Crefname{figure}{Fig.}{Figs.}

\begin{document}

\title{Beyond the TV: Investigating Continuous\\ Real-to-Virtual Transitions Along the\\ Reality-Virtuality Continuum}

\author{Toru~Kowada$^{\ast}$,
        Yuki~Uranishi$^{\ast}$,~\IEEEmembership{Member,~IEEE,}
        Koji~Momota,
        Daiki~Tomohiro,
        Masato~Kobayashi,~\IEEEmembership{Member,~IEEE,}
        Shizuka~Shirai,~\IEEEmembership{Member,~IEEE,}
        Naoya~Chiba,~\IEEEmembership{Member,~IEEE,}
        Photchara~Ratsamee,~\IEEEmembership{Member,~IEEE,}
        and~Kiyoshi~Kiyokawa,~\IEEEmembership{Member,~IEEE}
\IEEEcompsocitemizethanks{%
\IEEEcompsocthanksitem This work has been submitted to the IEEE for possible publication. Copyright may be transferred without notice, after which this version may no longer be accessible.
\IEEEcompsocthanksitem $^{\ast}$T. Kowada and Y. Uranishi contributed equally to this work. Corresponding author: Y. Uranishi.
\IEEEcompsocthanksitem T. Kowada, Y. Uranishi, K. Momota, D. Tomohiro, M. Kobayashi, S. Shirai, and N. Chiba are with The University of Osaka, Osaka, Japan.\protect\\
E-mail: yuki.uranishi.cmc@osaka-u.ac.jp
\IEEEcompsocthanksitem P. Ratsamee is with Kansai University, Osaka, Japan.
\IEEEcompsocthanksitem K. Kiyokawa is with the Nara Institute of Science and Technology, Nara, Japan.
}}

\markboth{Preprint submitted to IEEE Transactions on Visualization and Computer Graphics}%
{Kowada \MakeLowercase{\textit{et al.}}: Beyond the TV: Investigating Continuous Real-to-Virtual Transitions}

\IEEEtitleabstractindextext{%
\begin{abstract}
We designed a transition into virtual reality (VR) around a physical landmark object. Users first interact with the object in augmented reality (AR), receiving visual, auditory, and tactile feedback. In \textit{Beyond the TV}, users tap a television prop to change channels and reach through its screen to grow a virtual flower, feeling mist or warmth. During the continuous transition, the room gradually disappears and the virtual television draws the user into its world. In a two-by-two study with 60 participants, we varied the number of AR activity repetitions and compared this transition with an abrupt cut. After adjustment for ratings from a common practice session, the estimated overall score on our 11-item version of the Igroup Presence Questionnaire was 0.38 points higher for the continuous transition on a seven-point scale (95\% confidence interval from 0.00 to 0.75). The difference was smaller when all items were weighted equally. Repeating the activities three times instead of once produced no clear difference in presence. An exploratory comparison with 15 separately recruited participants who watched the activities yielded no clear evidence of a difference in presence or engagement.
\end{abstract}

}

\maketitle

\IEEEdisplaynontitleabstractindextext

\IEEEpeerreviewmaketitle

\begin{figure*}[!t]
  \centering
  \includegraphics[width=0.8\linewidth]{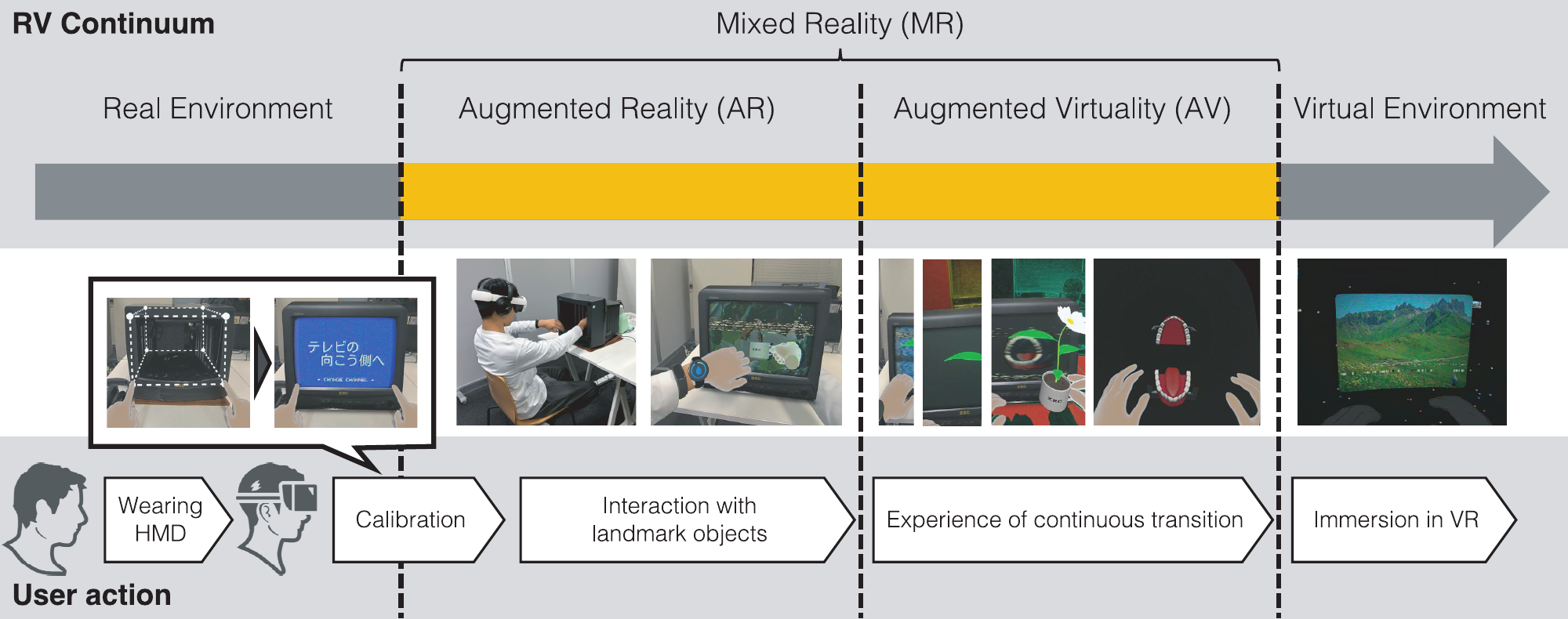}
  \caption{Overview of the proposed method. The user moves from the real room into a virtual environment through a continuous transition along Milgram's Reality-Virtuality Continuum~\cite{milgram1995augmented}; in the experience shown, the user is gradually drawn into a television.}
  \label{fig:teaser}
\end{figure*}

\IEEEraisesectionheading{\section{Introduction}\label{sec:introduction}}

Virtual reality (VR) systems can replace what users see and hear with a computer-generated environment, and augmented reality (AR) and mixed reality (MR) systems combine real and virtual content along the Reality-Virtuality Continuum~\cite{milgram1995augmented,slater1997framework,milgram1994taxonomy}. These technologies are applied in education~\cite{dalgarno2010learning,fowler2015virtual}, tourism~\cite{yung2019new,jung2017}, healthcare~\cite{aziz2018}, communication and simulation~\cite{rubio2017immersive}, and interactive storytelling~\cite{li2022interactive}, and also feature in research on digital tools for artistic practice~\cite{legino2024unveiling} and on students' acceptance of AR in education~\cite{matin2025aracceptance}. This study concerns how users enter such an experience.

In most current VR applications, the virtual environment is shown as soon as the user puts on a head-mounted display (HMD). Presence depends on the level of immersion~\cite{10.1007/s00779-019-01352-8} and on how users are transported into a virtual environment~\cite{10.1162/pres.1994.3.2.130}, and entering through a virtual replica of the real room has increased presence~\cite{steinicke2009} and improved distance estimation~\cite{10.1145/1620993.1620998}. Methods for a continuous transition from the physical to the virtual environment have therefore been studied, based on the Reality-Virtuality Continuum (RV Continuum) of Milgram et al.~\cite{milgram1995augmented}. The continuum places AR and augmented virtuality (AV) between a fully real and a fully virtual environment, and MR covers all mixtures of real and virtual elements between the two ends. Throughout this paper, following Slater and Wilbur~\cite{slater1997framework}, we use \textit{presence} to denote the subjective sense of being in the virtual environment and \textit{immersion} to denote the objective, technology-dependent properties of the system (\cref{sec:definitions}).

Schmidt and Yigitbas~\cite{schmidt2024development} proposed a conceptual framework for developing and evaluating transitional cross-reality interfaces and found high usability for an interface implemented with it. Pointecker et al.~\cite{pointecker2024real} inserted a replica environment, a virtual copy of the real room, into the transition; in their qualitative study, most participants stated that the replica supported spatial orientation. A replica preserves the room's layout but requires a model of each physical space. A physical object can also mark the entrance to VR, as in the door that users open and walk through in Fujisawa et al.~\cite{fujisawa2024door}. We asked how users might engage with such an object while still in the room. Our design gives them a task to perform in AR with the physical \textit{landmark object}. Changes they see and hear at the object are accompanied by tactile feedback from it. The object remains recognizable as they enter VR (\cref{fig:teaser}).

For the first two questions, we compared presence ratings after the main experience with those after a common practice session:
\begin{quote}
\begin{itemize}
\item[\textbf{RQ1}] How does this difference in presence ratings vary when the AR phase is extended by repeating the activities?
\item[\textbf{RQ2}] How does this difference vary between continuous and abrupt AR-to-VR transitions?
\item[\textbf{RQ3}] How do participants who perform the AR activities compare with those who watch them, in both the change in presence ratings from practice and their engagement with the main experience?
\end{itemize}
\end{quote}

Our contributions are:
\begin{itemize}
\item A transition design in which the entrance to VR first serves as the site of a multisensory AR activity. In \textit{Beyond the TV}, users grow a virtual flower at a television and feel mist and warmth at its screen before entering the world inside.
\item A $2 \times 2$ between-subjects comparison ($N=60$) of AR activity repetition and transition technique with the same prop. After adjustment for post-practice ratings, the estimated overall Igroup Presence Questionnaire (IPQ) score was 0.38 points higher for the designed transition than for an abrupt cut (95\% CI [0.00, 0.75] on a 7-point scale).
\item An exploratory comparison of presence and engagement between participants who performed the activities and a separately recruited group who watched them ($n=15$ per group).
\end{itemize}

The supplement (S-prefixed sections, figures, and tables) details the implementation, condition timelines, questionnaire statistics, coding of open-ended answers, and heart-rate analysis.

\section{Related Work}
\label{sec:related_work}

\subsection{RV Continuum, Immersion, Presence, and Engagement}
\label{sec:RVcontinuum}
\label{sec:definitions}
The RV Continuum of Milgram et al.~\cite{milgram1995augmented} arranges displays between a real and a virtual environment.
Ordered by the share of virtual content, it distinguishes the \textit{Real Environment}, \textit{Augmented Reality} (virtual objects added to the real world), \textit{Augmented Virtuality} (real elements added to a virtual world), and the \textit{Virtual Environment}. We use these categories to name the phases that a user passes through during a real-to-virtual transition.

The terms \textit{immersion} and \textit{presence} are often used interchangeably in everyday language, but the VR literature distinguishes them. Following Slater and Wilbur~\cite{slater1997framework}, we use \textit{immersion} to denote the objective, technology-dependent extent to which a system delivers an inclusive, extensive, surrounding, and vivid illusion of reality, and \textit{presence} to denote the user's subjective sense of ``being there'' in the mediated environment. Slater further characterizes presence as \textit{place illusion}---the illusion of being located in the displayed space even though one knows one is not there~\cite{slater2009place}. Presence is commonly assessed with questionnaires such as the Igroup Presence Questionnaire (IPQ)~\cite{schubert2001experience}, which includes a general presence item and three subscales for spatial presence, involvement, and experienced realism.

\textit{Engagement} describes the user's involvement with the content of an experience. A user may feel present in a virtual space yet find its content unrewarding. We assessed engagement with the User Engagement Scale--Short Form (UES-SF), which covers focused attention, perceived usability, aesthetic appeal, and reward~\cite{o2018practical}. We measured workload with the NASA Task Load Index (NASA-TLX)~\cite{nasa_tlx}.

\subsection{Transition Techniques in VR}
\label{sec:VRtransition}
Feld et al.~\cite{feld2023keep} compared six techniques for moving between virtual environments (Cut, Dissolve, Fade, Morph, Portal, and Orb) in terms of usability, preference, workload, presence, and task performance. In a memory task, Cut was preferred most often and rated higher than several other techniques in usability and presence, with no difference in error rate; the authors concluded that users valued efficiency over interactivity. Moghadam and Ragan~\cite{Moghadam2017} compared teleportation, animation, and pulse transitions in a nine-participant pilot study. Animated interpolation yielded the smallest orientation errors and the lowest reported sickness. In the pulsed condition, participants with more gaming and VR experience preferred fewer intermediate viewpoints.

\subsection{Continuous Transitions Between Real and Virtual Environments}
\label{sec:REVEtransition}
An early example is the \emph{MagicBook} of Billinghurst et al.~\cite{magicbook2001}, a ``transitional interface'' in which users moved from a real book to an AR view with virtual objects on its pages and then into a VR scene. In \emph{One Reality}, users moved through successive levels of augmentation while interacting with physical objects~\cite{solroo2017one}. Jung et al.~\cite{jung2018limbo} found that participants rated spatial presence and virtual body ownership higher after a gradual transition from the real body and room to their virtual counterparts than after an instantaneous transition. Pointecker et al.~\cite{pointecker2022bridging} proposed four AR--VR transition techniques (Fade, SimpleCut, TeleportBeam, and Portal). In SimpleCut, a clipping plane moves across the scene to reveal the next environment. Participants described Fade as suitable for frequent transitions and for environments with shared visual landmarks, and Portal for environments without them. Other work has emphasized continuity across phases. Replica-based intermediate environments can improve spatial orientation and perceived presence~\cite{pointecker2024real,steinicke2009}, and related approaches support gradual adaptation through smooth transformation of the environment~\cite{valkov2017}. Sra et al.~\cite{9790021} proposed ``situated VR,'' in which the virtual environment is designed to match the real context so that mismatches between the two do not disturb the experience. Feld et al.\ demonstrated transitions between VR and AR with a spatial memory game~\cite{feld2024navigating_realities} and discussed how the simplicity and the efficiency of a transition should be weighed against the requirements of the task~\cite{feld2024simple_efficient}. Regarding the return direction, Horst et al.~\cite{horst2024back} compared eight transitions from VR back to reality, and participants preferred non-interactive ones except one that allowed communication with the presenter; Soret et al.~\cite{soret2021door} reported that a door-based transition can facilitate returning to reality without increasing cybersickness, suggesting that structured transitional metaphors may improve comfort.

Fujisawa et al.~\cite{fujisawa2024door} let users enter a virtual forest by turning the knob of a physical door, opening it, and walking through. Users could feel the knob turn and hear the door creak. Both the door and a virtual portal yielded higher overall presence scores than a fade.

Schott et al.~\cite{schott2025onboarding} studied how collocated groups entered VR after an interactive tutorial in video-see-through MR. They compared entering individually, in sequence, and together.

\subsection{Interaction and Multimodal Feedback}
\label{sec:Interaction}
Clergeaud et al.~\cite{Damien2017} presented prototypes for asymmetric collaboration between a meeting room and a user in VR that address awareness of the remote location, remote interaction, and navigation between the locations, with first feedback from experts. Al-Sada et al.~\cite{al2020hapticsnakes} developed waist-worn robots that provide several forms of haptic feedback in VR. Participants favored their use.

\subsection{Summary of Related Work and Positioning of This Study}
Previous studies have used room replicas to preserve spatial orientation during entry into VR~\cite{pointecker2024real}, transitions that let users continue a task across environments~\cite{feld2024simple_efficient}, and interactive MR tutorials to prepare groups for entry~\cite{schott2025onboarding}. We build on the use of physical objects across stages in \emph{MagicBook} and \emph{One Reality}~\cite{magicbook2001,solroo2017one}. In the physical-door study~\cite{fujisawa2024door}, turning the knob and opening the door were part of entering the forest. We use the future entrance for a task with a goal of its own. Users grow a virtual flower at the television, feeling the rain and sunlight shown on its screen before they enter through it.

Within this design, we compare the number of activity repetitions (RQ1) and the transition technique (RQ2). A separate exploratory comparison examines performing versus watching the activities (RQ3).

\section{Transition Design}
\label{sec:proposed_method}
We describe the transition design and its three experimental factors. \Cref{BeyondTheTV} presents the implementation.

\subsection{Overview}
The design follows the RV Continuum from the real room through AR and AV into VR (\cref{fig:teaser}). A physical \textit{landmark object} has a spatially aligned virtual counterpart and remains recognizable as the surroundings change. While the room is still visible, users perform a task with the object. What they see and hear as they act is matched by what they feel at the object. We chose a cathode-ray-tube television because its casing gives the user a surface to tap and its screen can serve as an opening into the virtual scene. A tap changes the channel. Reaching in with a virtual flower lets the user feel mist or warmth corresponding to the rain or sunlight on screen. In the continuous transition, the virtual television then moves toward the seated user and swallows them (\cref{TV:nagare}).

Initially, the user puts on the HMD, which displays the real environment via video passthrough. Spatial calibration then aligns the virtual counterpart with the real landmark object. Once calibration is complete, the virtual object is superimposed onto the real landmark object while the remainder of the display continues to show the real world (AR phase), and the user performs activities that involve the landmark object. After these activities, the imagery of the real environment is gradually replaced until all elements except the landmark object are virtual (AV phase), and the VR experience begins. After the VR experience concludes, the system fades the display---except for the landmark object---to reveal the real environment. We chose this simple return following Horst et al.~\cite{horst2024back}, whose participants preferred non-interactive techniques for leaving VR.

\subsection{Design Factors}
\label{sec:design_factors}
The design has three factors that a designer must choose and that correspond to the research questions RQ1--RQ3: the duration of the activities in the AR phase, whether these activities are performed by the user, and the technique of the transition from the AR phase to the VR phase.

\textbf{Activities during the AR phase (RQ1 and RQ3).} In the AR phase, virtual content is overlaid on the real environment and attached to the landmark object, so that the virtual elements appear to belong to a physical object in the user's room. Activities with the landmark object are intended to let users become familiar with the virtual content while the real environment remains visible. Two properties of these activities can be chosen: their duration, that is, how long users engage with the landmark object before the transition (RQ1), and whether users perform the activities with their own hands or observe them being performed (RQ3). The expectation that interaction contributes to presence draws on work on haptic feedback, which is considered to enhance immersion and presence in VR~\cite{al2020hapticsnakes}. The bridging role of the landmark object and the familiarity built in the AR phase are design rationales rather than measured constructs; the user study tests only their predicted consequences for presence and engagement.

\textbf{Continuous transition (RQ2).} In the AV phase, the displayed imagery shifts progressively from the real to the virtual environment. Steinicke et al.~\cite{steinicke2009} found higher presence ratings when users entered VR through a replica room. Valkov and Flagge~\cite{valkov2017} evaluated continuous entry into VR by examining user awareness and judgments of what was real. In our design, the alternative to a continuous transition is an abrupt cut from the AR phase to the VR phase (RQ2).

\section{Implementation: Beyond the TV}
\label{BeyondTheTV}
In \textit{Beyond the TV}, a cathode-ray-tube television serves as the landmark object.

\subsection{Overview of the Experience}
\label{TV:gaiyou}
In \textit{Beyond the TV}, the screen of a television is the boundary between the real and the virtual environment (``screen-as-boundary''). The HMD presents the images and sound, and a physical prop built from a television gives tactile cues timed to the on-screen events, so that the boundary at the screen can be felt as well as seen.

The participant's task is to grow a flower using virtual ``channels'' shown on the television. After a short calibration, the AR phase begins: the participant taps the prop to change channels and, holding a cup with the flower, reaches into the screen so that the ``rain'' and ``sunlight'' of the channels make the flower grow. Each time the hand and the flower pass into the screen, the participant crosses the same boundary that the whole body later crosses in the transition to VR.

\subsection{Flow of the Experience}
\label{TV:nagare}
\begin{figure*}[t]
\centering
\includegraphics[width=0.87\textwidth]{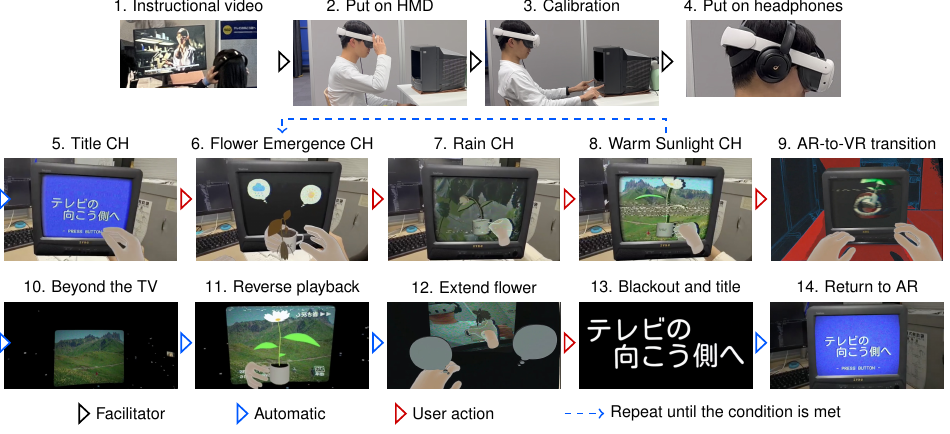}
\caption[Flow of the experience]{Flow of the experience (Proc.~1--14). The steps are described in \cref{TV:nagare}.}
\label{fig:experience_flow}
\end{figure*}

\Cref{fig:experience_flow} shows the flow of the experience. After an instructional video on the story, the operations, and the precautions (Proc.~1), the participant puts on the HMD, sees the room through passthrough video (Proc.~2), and, guided by the facilitator, aligns the virtual television with the prop (Proc.~3). From then on, the participant wears headphones that block ambient sound and follows only the audio prompts of the system (Proc.~4).

\begin{figure}[t]
\centering
\includegraphics[width=0.7\columnwidth]{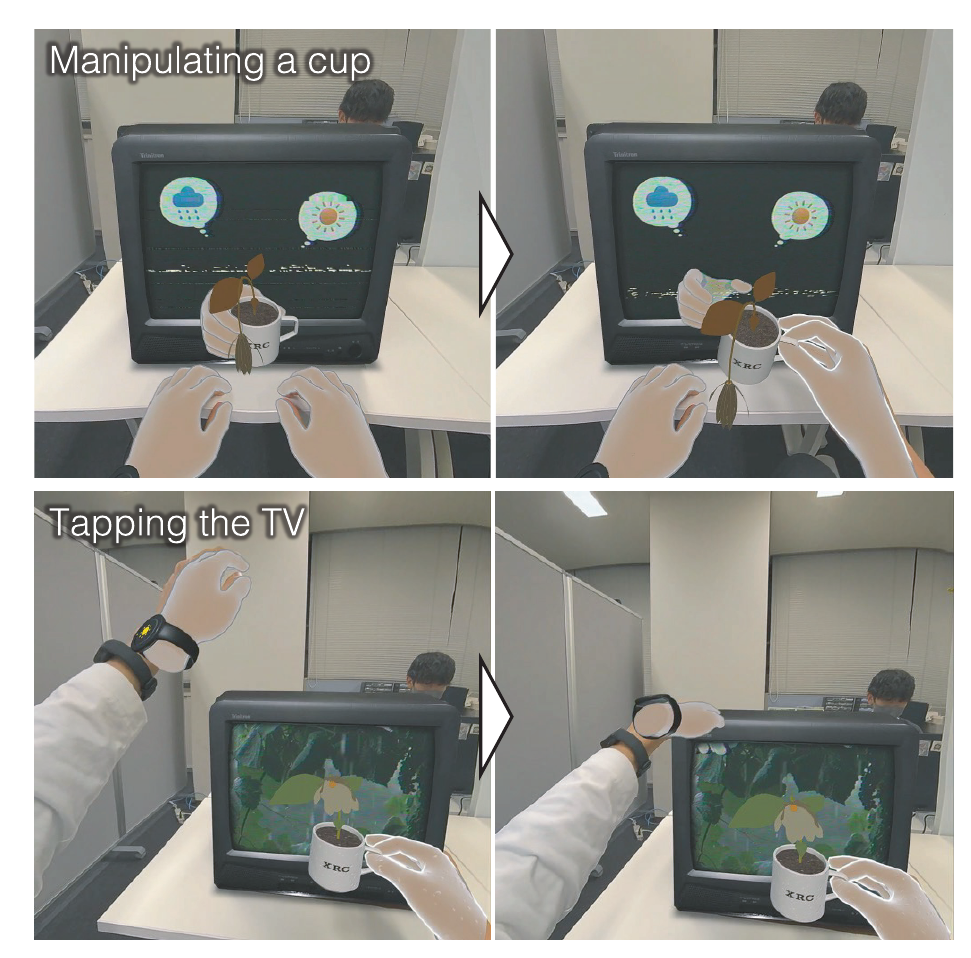}
\caption[Interactions during the AR phase]{Interactions during the AR phase. The participant holds the virtual cup in the right hand and taps the top of the television with the left hand to switch channels.}
\label{fig:operation}
\end{figure}

In the AR phase (Proc.~5--8), the participant switches channels by tapping the top of the television with the left hand and holds the virtual cup in the right hand (\cref{fig:operation}). In the ``Flower Emergence CH,'' a virtual hand reaches out of the screen and passes the participant a cup holding a flower (Proc.~6). In the ``Rain CH'' (Proc.~7) and then the ``Warm Sunlight CH'' (Proc.~8), holding the flower inside the screen fills a growth meter on a wristwatch, and a full meter advances the flower one stage (\cref{S-fig:flowerWatch}). The number of repetitions sets the AR-phase duration (\cref{sec:design_factors}).

\begin{figure*}[t]
\centering
\includegraphics[width=0.6\textwidth]{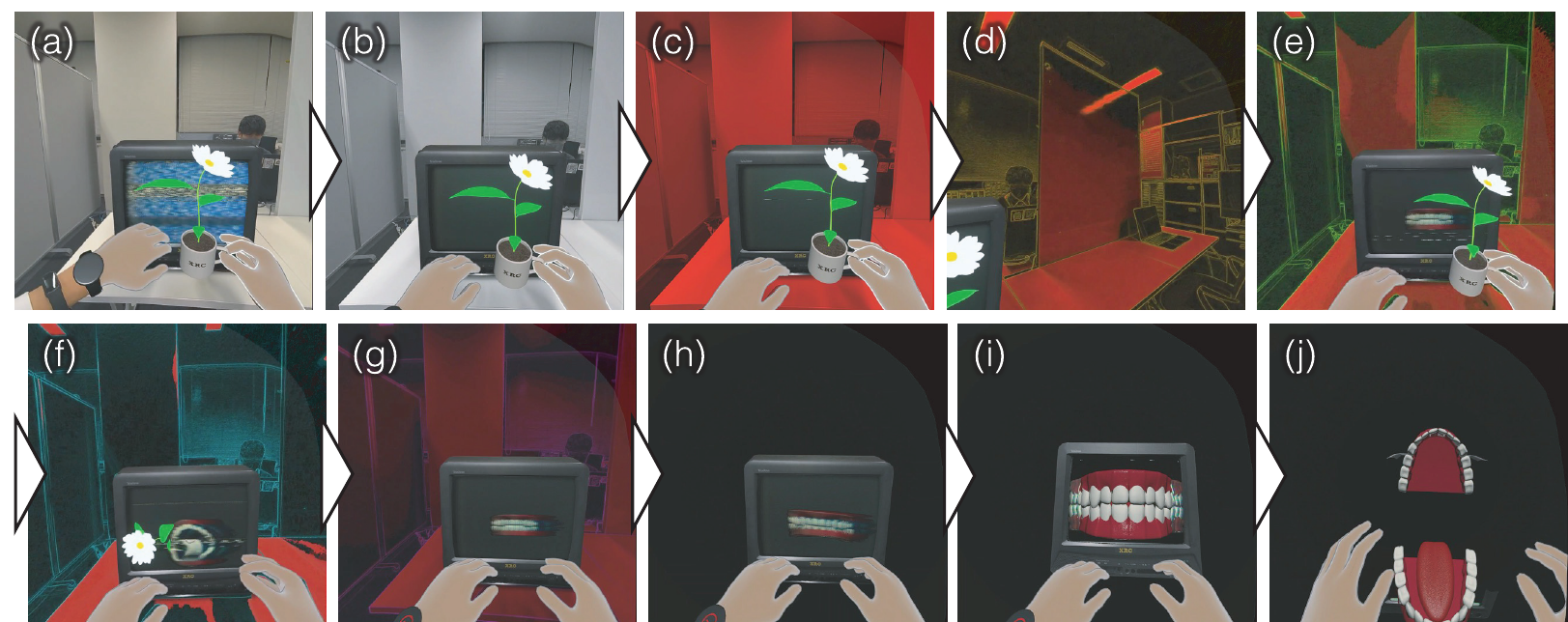}
\caption[Continuous transition during the AV phase]{Continuous transition. (a) AR phase. (b)--(g) AV phase, in which the passthrough image of the room is processed in stages and replaced by virtual objects. (h)--(j) Fully virtual imagery: the virtual television detaches from the prop and swallows the participant, after which the VR phase (Proc.~10) begins. The stages are described in the text.}
\label{fig:transitionFlow}
\end{figure*}

Once the flower is fully grown, a further tap turns the screen black and the continuous transition begins (Proc.~9; \cref{fig:transitionFlow}). The screen turns to static noise [\cref{fig:transitionFlow}(a)], and the passthrough image of the room is desaturated [\cref{fig:transitionFlow}(b)], shifted to a red-and-black gradient [\cref{fig:transitionFlow}(c)], and reduced to outlines whose hue cycles over time [\cref{fig:transitionFlow}(d)--(f)]; meanwhile, teeth appear on the screen and the flower is pulled into the television [\cref{fig:transitionFlow}(e) and (f)]. The image then darkens until only the virtual television remains in a dark virtual space [\cref{fig:transitionFlow}(g) and (h)], and the television, whose screen has become a mouth, moves toward the participant and swallows them [\cref{fig:transitionFlow}(i) and (j)]. Because the participant stays seated, it is the television that moves the participant from outside to inside the screen, so that the object that anchored the AR phase also carries out the boundary crossing. This sequence belongs to the CT condition only, and some participants described it as surprising or scary (\cref{sec:limitations}).

Inside the television (Proc.~10), screens extend in every direction. The screen ahead replays backward the channel scenes in which the participant handled the flower, mirrored because they are seen from inside (Proc.~11), and the replay reveals that the hand that passed the cup in Proc.~6 was the participant's own future hand. The participant then escapes by extending the flower through the screen, which now shows the webcam footage recorded in Proc.~5--6 (Proc.~12). The screen fades to black and shows the title (Proc.~13). The passthrough view of the room then fades back in (Proc.~14).

\subsection{System Configuration and Hardware}
\label{TV:system}
\label{TV:hard}
\begin{figure}[t]
\centering
\includegraphics[width=0.9\columnwidth]{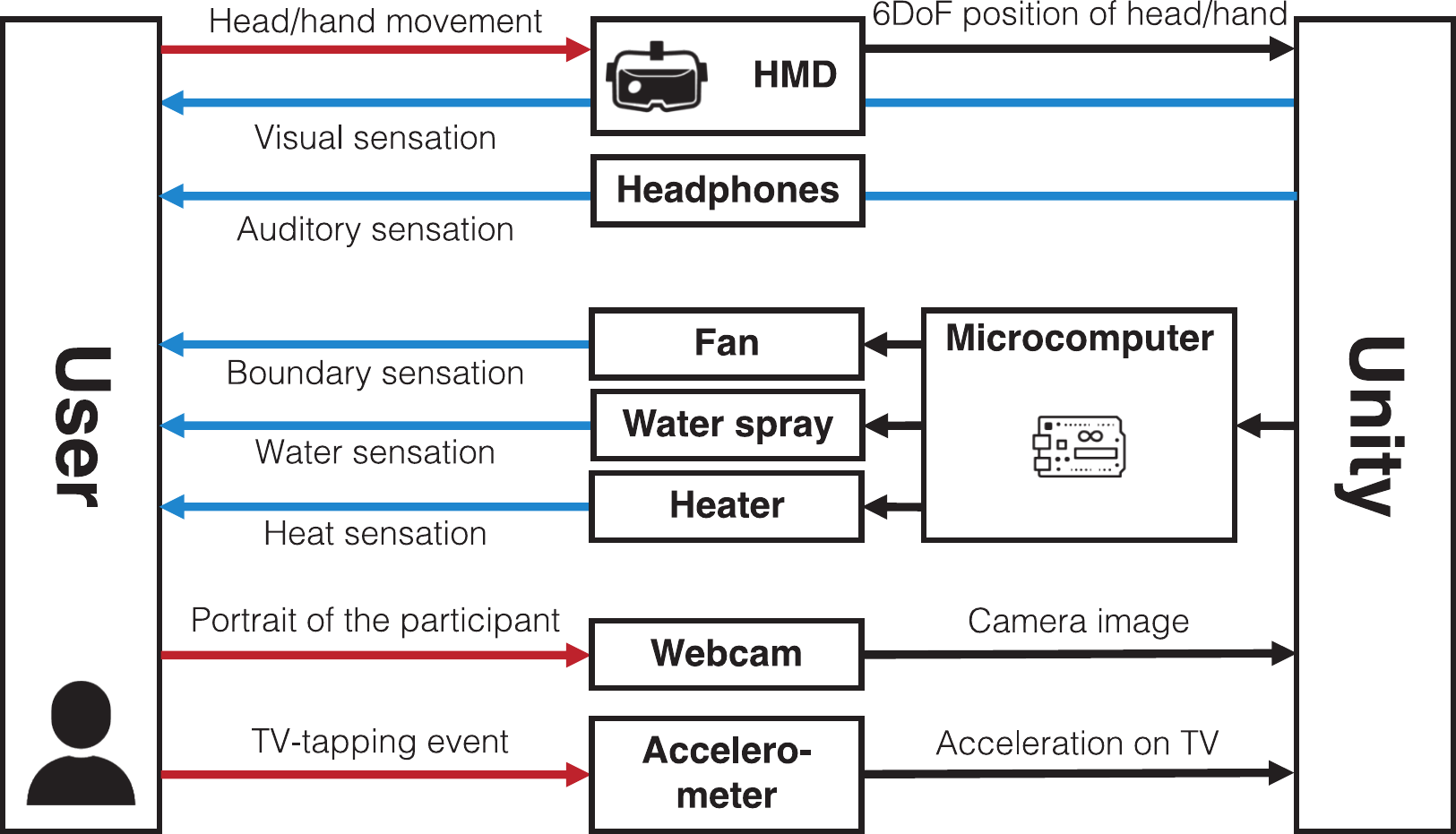}
\caption[System configuration]{System configuration. The Unity application on the PC drives the HMD, records the webcam footage replayed in the VR phase, and sends commands to the Arduino that controls the fan, sprayer, and heater in the prop. An accelerometer in the prop detects taps.}
\label{fig:system}
\end{figure}

\cref{fig:system} shows the system configuration: a PC running Unity connects to the HMD and to an Arduino microcontroller that drives the actuators embedded in a television-like prop, and a webcam in the prop records footage for later replay. The prop uses the frame of a cathode-ray-tube television.

Inside the prop, a mist sprayer and a heater, each switched by a servomotor, and a fan at the screen opening let the participant feel the rain of the ``Rain CH,'' the warmth of the ``Warm Sunlight CH,'' and the boundary at the screen on the hand; in CT, the fan is switched off when the flower is drawn into the television (Proc.~9). The accelerometer of a Nintendo Joy-Con~\cite{nintendo_joycon_support} detects taps on the prop, and a webcam records the participant in Proc.~5--6 for the replay in Proc.~12; the images are deleted when the experience ends. The application was implemented in Unity 2021.3 LTS with the Meta XR SDK and runs on a desktop PC (Intel Core i7-13700F, 32~GB RAM, NVIDIA GeForce RTX 3060 Ti) connected to a Meta Quest 3 via Quest Link. Photographs of the prop and the timing of each device are given in \cref{S-sec:supp_hw}.

\subsection{Software}
\label{TV:soft}

Hand interaction uses the pinch gesture recognized by the hand tracking of the Meta Quest 3, and the AR and AV phases are built on its color video passthrough. Objects inside the virtual screen are rendered as 2D footage by an orthographic camera, so that the channels look like television images, and a short calibration, in which the participant points at the four corners of the prop, aligns the virtual television with the real one. \Cref{S-sec:supp_sw} describes the rendering and the calibration.

\subsubsection{Implementation of the AR-to-VR Continuous Transition}
We transformed the passthrough image across the whole field of view, preserving the layout of the room around the virtual television (\cref{fig:transitionFlow}). We first desaturated the image, then applied a red-and-black gradient and displayed the room edges with a cycling hue. These colors were chosen to convey the unease of the story. Periodic reductions in brightness gradually obscured the room until only the television remained visible in the dark virtual scene [\cref{fig:transitionFlow}(h)]. We use \textit{continuous} to describe this change from passthrough to virtual imagery without a cut.

\section{User Study}
\label{sec:jikken}

\subsection{Hypotheses}
\label{sec:hypotheses}
For the three design factors in \cref{sec:design_factors}, we examined the change in presence ratings from practice to the main experience. For the comparison of performing and watching the activities, we also examined engagement with the main experience.

\begin{quote}
\begin{itemize}
\item[\textbf{H1}] The increase in presence ratings from practice is larger after three sets of AR activities than after one.
\item[\textbf{H2}] The increase in presence ratings from practice is larger with the designed continuous AR-to-VR transition than with an abrupt transition.
\item[\textbf{H3}] Participants who perform the AR activities show a larger increase in presence ratings from practice and report higher engagement with the main experience than those who watch.
\end{itemize}
\end{quote}

H1 and H3 follow from the expectation that activities with the landmark object let users become familiar with the virtual content before the real environment is removed (\cref{sec:design_factors}), and H2 draws on studies that found higher presence ratings after gradual entry into VR than after direct entry~\cite{steinicke2009,jung2018limbo}. Presence was operationalized as the change in IPQ scores from a practice session to the main experience, engagement as UES-SF scores, and workload as NASA-TLX scores; heart rate was recorded as a complementary physiological index (\cref{sec:measures}). \Cref{S-table:operationalization} lists the groups, measures, and tests for each hypothesis.

The study was approved by the Institutional Review Board of the Cybermedia Center, the University of Osaka (currently the D3 Center) (approval no.~2024-13), and all participants provided written informed consent prior to participation.

\subsection{Design and Conditions}
\label{sec:conditions}
Each participant took part in only one condition (between-subjects design). The three factors (AR-phase duration, transition technique, and interaction) are listed in Table~\ref{table:jikken_syuhou}.

\begin{table}[b]
 \caption[Five experience design groups]{Details of the Five Experience Design Groups}
 \label{table:jikken_syuhou}
 \footnotesize
 \centering
 \begin{tabular}{cccc}
  \toprule
  Group & \shortstack{AR Phase\\Duration} & \shortstack{Transition\\Technique} & \shortstack{User\\Interaction} \\
  \midrule
  Group A & LD & CT & I \\
  Group B & SD & CT & I \\
  Group C & LD & DT & I \\
  Group D & SD & DT & I \\
  Group E & LD & CT & NI \\
  \bottomrule
 \end{tabular}
\end{table}

We varied \emph{AR-phase duration} (RQ1) through the number of activity sets, with three sets for long duration (LD) and one for short duration (SD) (\cref{S-fig:Syurui_AR}).
One set consists of performing the sequence of actions---namely, (1) tapping the television to switch channels, (2) inserting the flower into the television, and (3) the flower growing---twice, once on each of the ``Rain CH'' and the ``Warm Sunlight CH.'' According to the system logs, the AR phase (from the appearance of the virtual television to the start of the transition) lasted 159~s on average in the LD groups ($SD = 22$) and 75~s in the SD groups ($SD = 16$).

For the \emph{AR-to-VR transition technique} (RQ2), two levels were defined (\cref{S-fig:Syurui_AV}):
\begin{itemize}
    \item Continuous Transition (CT)
    \item Discontinuous Transition (DT)
\end{itemize}
In the CT condition, the continuous transition described in \cref{BeyondTheTV} is applied. In the DT condition, no continuous transition is performed; instead, the system abruptly transitions to a black virtual environment at the moment the transition would begin, and the VR phase then starts in the same virtual scene as in the CT condition. The DT condition thus omits all stages shown in \cref{fig:transitionFlow}(b)--(j), namely the color transformation of the passthrough imagery, the teeth on the screen, the flower being drawn into the television, and the participant being swallowed by the virtual television. Because the fan is switched off when the flower is drawn into the television, which does not occur in DT, the fan remained on in DT at least until the VR phase began (system logs), whereas in CT it stopped about 18~s before the VR phase.

\begin{figure}[t]
\centering
\subfloat[Interactive (I)]{%
    \includegraphics[width=0.48\columnwidth]{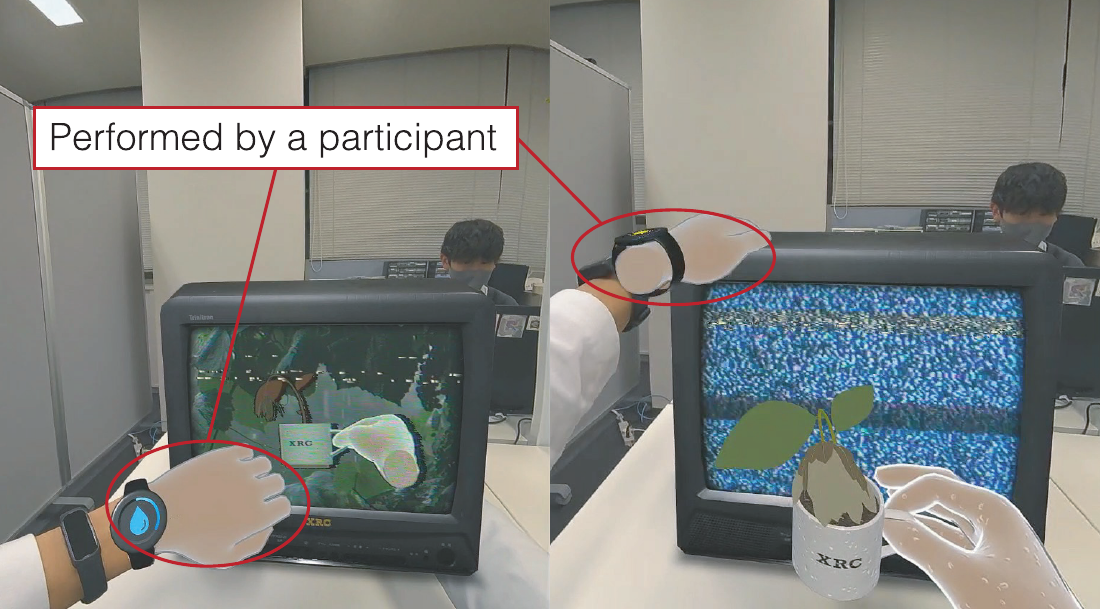}%
    \label{fig:Interaction_syurui_I}}
\hfil
\subfloat[Non-Interactive (NI)]{%
    \includegraphics[width=0.48\columnwidth]{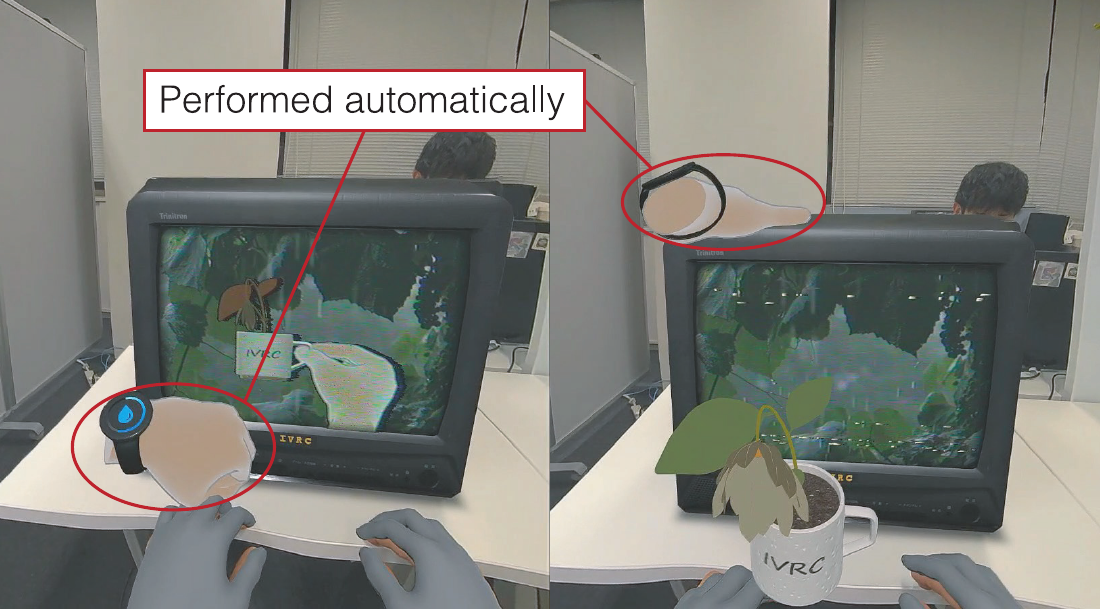}%
    \label{fig:Interaction_syurui_NI}}
\caption[User interaction conditions]{User interaction conditions during the AR phase: (a) Interactive (I), where the participant performs the actions; (b) Non-Interactive (NI), where the system performs the actions automatically.}
\label{fig:Interaction_syurui}
\end{figure}

\emph{User interaction} (RQ3) had two levels (\cref{fig:Interaction_syurui}):
\begin{itemize}
    \item Interactive (I)
    \item Non-Interactive (NI)
\end{itemize}
In the I condition, the participant taps the television and grows the flower as described in \cref{BeyondTheTV}. In the NI condition, a virtual hand automatically performs these operations while the participant watches, so that the displayed sequence of events is the same as in the I condition; the participant does not perform any operation. The automated sequence made the AR phase somewhat longer than in the corresponding interactive group (Group E: 175~s, $SD = 4$; Group A: 159~s, $SD = 21$). The escape operation in the VR phase (Proc.~12) was also triggered automatically in the NI condition, so that the VR phase had a constant length of 92~s, compared with 101~s ($SD = 17$) in Group A. Because participants in the NI condition did not insert their hands into the prop, they did not feel the airflow at the screen opening, the mist, or the warmth on their hands. The fan, sprayer, and heater were nevertheless driven by the same rules as in the I condition (system logs), so that the sounds of the fan and of the servomotors that drive the sprayer and heater, audible through the headphones, occurred at the corresponding points of the sequence.

For RQ1 and RQ2, Groups A--D crossed AR-phase duration (LD/SD) with transition technique (CT/DT) in a $2 \times 2$ between-subjects design. All four groups performed the activities. For RQ3, we compared Groups A and E under LD/CT: Group A performed the activities and Group E watched them. This exploratory comparison addresses interaction only for the long AR phase followed by the continuous transition. The groups also differed in AR- and VR-phase duration, how the escape was triggered, and the period and room of data collection (\cref{sec:participants}).

\Cref{table:constant_varied} lists which elements of the experience were identical across groups and which differed. Elements identical in all groups, such as the introduction of the story, the activities on each channel, and the VR-phase content, cannot by themselves produce group differences, although they may modify the effects of the manipulated factors. The elements that differ between CT and DT are bundled by design: CT presents the designed transition as a whole---the transformation of the passthrough image, the story events of the television drawing in the flower and swallowing the participant, an interval about 19~s longer between the end of the AR phase and the start of the VR phase, and the stopping of the fan---and DT removes it. \Cref{S-table:timeline} gives the time course of these events in each condition.

\begin{table}[t]
 \caption[Elements held constant and varied across groups]{Elements Held Constant and Varied Across Groups}
 \label{table:constant_varied}
 \footnotesize
 \centering
 \begin{tabular}{L{0.44\columnwidth}L{0.46\columnwidth}}
  \toprule
  Element & Across groups \\
  \midrule
  Instruction video, story introduction, and audio prompts (Proc.~1, 4) & Identical in all groups \\
  VR practice session and post-practice questionnaire & Identical in all groups \\
  Calibration and channel content (Proc.~3, 5--8) & Identical in all groups \\
  Commands sent to the fan, sprayer, and heater in the AR phase & Same rules in each set; in I, the number of sprayer switchings depends on the participant \\
  Fan after the AR phase & CT: off when the flower is drawn in (Proc.~9); DT: on at least until the VR phase begins \\
  Airflow, mist, and warmth felt on the hand & I: felt when the hand is inserted; NI: not felt \\
  Number of activity sets in the AR phase & LD: three sets; SD: one set \\
  Agent of tapping and flower insertion; escape in the VR phase & I: participant; NI: virtual hand; escape triggered automatically \\
  Transition stages in \cref{fig:transitionFlow}(b)--(j) (color transformation, teeth, swallowing) & CT only; DT cuts to black \\
  VR-phase content and webcam replay (Proc.~10--13) & Identical in all groups \\
  Return to the real environment (Proc.~14) & Identical in all groups \\
  \bottomrule
 \end{tabular}
\end{table}

\subsection{Participants}
\label{sec:participants}
Recruitment was limited to university students. Seventy-five participants (41 male [54.7\%] and 34 female [45.3\%]) aged between 18 and 30, who had not previously experienced \textit{Beyond the TV}, were recruited from our university. Pilot sessions run before recruitment are not included, and no participant of the main study was excluded from the analysis. Each group consisted of 15 participants (demographics per group in \cref{S-table:hikensya}). Data were collected in two periods. Groups A--D were run from September to October 2024. Group E was run in late November 2024 with a separate recruitment call that excluded anyone who had taken part in the first period. Group E used the same apparatus, the same HMD model and strap, the same instructions and questionnaires, and the same procedure, but the sessions took place in a different room of our facility.
Of the 75 participants, 37 had never used VR, 35 used it a few times a year, 1 a few times a month, 2 a few times a week, and none almost daily.
In the first period, participants were assigned to Groups A--D manually before each session, with VR experience as the only criterion, so that the proportion of participants with and without VR experience was similar across groups (7 or 8 of 15 without VR experience in each group; \cref{S-table:hikensya}); the assignment was not randomized. The first 56 sessions followed an A--B--C--D rotation. The last four ran in the order A, C, B, D. Group E was recruited separately after data collection for Groups A--D had ended (\cref{sec:limitations}).

To make explicit which effects this sample can detect, we conducted a sensitivity analysis ($\alpha = .05$, power $= .80$). For the main effects in the $2 \times 2$ design (Groups A--D, $N = 60$), the minimum detectable effect is Cohen's $f = 0.37$ (partial $\eta^2 = .12$). For the two-group comparison between Groups A and E ($n = 15$ per group), it is Cohen's $d = 1.06$ for an unpaired $t$-test, or approximately $d = 1.09$ for a Wilcoxon rank-sum test when its asymptotic relative efficiency of 0.955 is taken into account.

\subsection{Procedure}
Participants sat in front of the prop for the baseline heart-rate measurement, the practice session, and the main experience, and answered the questionnaires at an adjacent seat (\cref{S-fig:ex_environment}). After the procedure had been explained, the heart-rate sensor was put on and a 10-minute baseline was recorded. Next, participants received an explanation of the experience and practiced the operational procedures in a VR environment containing a virtual television, covering the operations required in the main experience (tapping to switch channels and reaching into the screen with the hand). The practice took 84~s on average ($SD = 16$) in Groups A--D, as recorded in the system logs; it took place entirely in VR and included neither the story nor the real-to-virtual transition of the main experience, and the fan, sprayer, and heater were not operated.
Participants then completed the post-practice questionnaire (demographics and the IPQ~\cite{schubert2001experience}), the main experience with continuous heart-rate recording, and the final questionnaire (IPQ, UES-SF~\cite{o2018practical}, NASA-TLX~\cite{nasa_tlx}, and the additional questions).

\subsection{Measures}
\label{sec:measures}
The overall IPQ change score was the primary measure for H1 and H2. H3 concerns both presence and engagement, so the overall IPQ change score and the overall UES-SF score were its primary measures. The factor scores, the UES-SF for H1 and H2, the NASA-TLX, and the additional questions were secondary measures, and heart rate was an exploratory physiological measure.

\subsubsection{Heart Rate}
Heart rate was recorded with a myBeat WHS-1 wearable sensor (Union Tool Co., Ltd.)~\cite{myBeat}. The sensor was worn on a chest strap for the whole session and sent its data wirelessly to a USB receiver on the recording PC.

We computed heart rate from the beat-to-beat (RR) intervals after removing artifacts, as the mean of each phase delimited by the event times recorded by the system, and, to examine the response to the transition itself, as the change around the end of the AR phase and the onset of the VR phase relative to the 20~s before the end of the AR phase. The practice session, which was identical in all groups, served as the within-participant reference.

The preprocessing and the number of participants in each analysis are given in \cref{S-sec:supp_hr}. The sensor software also computes the frequency-domain index LF/(LF+HF), but over a 300-s window that is longer than every phase of the experience; because the index cannot be attributed to individual phases, we omitted it.

\subsubsection{IPQ}
Presence was measured with the Igroup Presence Questionnaire (IPQ). We used the Japanese items from the IPQ database~\cite{IPQ_database}, with minor wording differences (\cref{S-table:ipq_items_ja}). The IPQ includes one general presence item (G) and three subscales: Spatial Presence (SP), Involvement (INV), and Experienced Realism (REAL). In this study, 11 of the 14 IPQ items were administered (G1; SP1--SP5; INV2--INV4; REAL1 and REAL3). INV1 asks how aware participants were of the real world while moving through the virtual world; it was omitted because participants do not move through the virtual environment in this experience. REAL2 and REAL4 were judged not to apply to the content of this experience and were omitted. In the Japanese version administered, REAL1 and REAL3 ask nearly the same question and differ in their response anchors (\cref{S-table:ipq_items_ja}); both were retained.
Each item was rated on a 7-point scale, once after the practice session and once after the main experience. We reverse-scored SP2, INV3, and REAL1 ($8-x$) and averaged the administered items within each subscale. For this study, we defined the overall score (ALL) as the mean of G and the three subscale scores. G thus contributes 25\% of ALL and each REAL item contributes 12.5\%. We also report the 11-item mean, the mean of G, SP, and INV, and G alone. We specified these alternatives after the original analysis and before calculating them (\cref{sec:results_robust}). Internal consistency after the main experience was $\alpha = .79$ for SP, $\alpha = .64$ for INV, and $\alpha = .84$ for all 11 items; the two REAL items correlated at $r = .05$ (after practice: SP $\alpha = .43$, INV $\alpha = .57$, REAL $r = .20$; $N = 75$). The items, response anchors, and scoring are listed in \cref{S-table:ipq_items}. The main outcome was the change score, the post-experience IPQ score minus the post-practice IPQ score. Each participant rated the practice session and the main experience as a whole. Their difference compares the two experiences and can account for a consistent tendency to rate both high or low.
Post-experience scores are additionally analyzed on their own in \cref{sec:results}. One participant in Group E wrote that they had misread the direction of the REAL1 scale after practice. The experimenter asked the participant for the intended response before departure, but it was not entered into the data, so we analyzed the recorded value of 6. Using 2 instead gave the same conclusion for the A--E comparison of overall IPQ change ($p = .41$ vs.\ $.51$).

\subsubsection{UES-SF}
Engagement was measured with the User Engagement Scale--Short Form (UES-SF). Its 12 items cover four factors with three items each: Focused Attention (FA), Perceived Usability (PU), Aesthetic Appeal (AE), and Reward Factor (RW)~\cite{o2018practical}. In the post-experience questionnaire, participants responded on a 5-point Likert scale (1 = Strongly Disagree to 5 = Strongly Agree), and factor scores were calculated after reverse-scoring the three negatively worded PU items ($6 - x$). The mean of the four factor scores served as the overall score (ALL).

\subsubsection{NASA-TLX}
We administered the NASA-TLX in Japanese to measure workload~\cite{nasa_tlx}. The NASA-TLX comprises six subscales: Mental Demand (MD), Physical Demand (PD), Temporal Demand (TD), Performance (PE), Effort (EF), and Frustration (FR). Participants weighted the subscales by pairwise comparison and rated each subscale from 0 to 100, and the overall score was the weighted mean of the ratings.

\subsubsection{Additional Questions}
The post-practice questionnaire asked for demographics and the frequency of VR use. The post-experience questionnaire also contained five additional questions, the same for all groups (\cref{S-table:After_question}): Q1 (the practice was sufficient), Q2 (more absorbed than in the practice session), and Q4 (the heart-rate sensor did not disturb the experience), rated on a 7-point scale (1 = Strongly Disagree, 7 = Strongly Agree), and two open-ended questions, Q3 (reason for the Q2 rating) and Q5 (other comments). Q1 and Q4 served to check that factors unrelated to the manipulations did not differ between groups. The questionnaire was administered in Japanese; ``immersed'' in Q2 translates the everyday Japanese word \textit{botsunyu} (absorption) and does not refer to immersion in the technical sense defined in \cref{sec:definitions}. The same applies to participants' open-ended comments quoted in \cref{sec:results_qual,sec:discussion}.

\subsection{Statistical Analysis}
\label{sec:analysis}
\Cref{S-table:operationalization} summarizes the analyses for each hypothesis. \Cref{S-sec:supp_tests} tabulates ART and covariate-adjusted questionnaire results for Groups A--D and rank-sum results for Groups A and E. Unless otherwise noted, we treat $p < .05$ as statistically significant and $.05 \le p < .10$ as trend-level evidence; trend-level findings are interpreted as suggestive rather than conclusive.

For the $2 \times 2$ design (Groups A--D), questionnaire measures were analyzed with the Aligned Rank Transform (ART)~\cite{wobbrock2011aligned}, implemented in Python and checked against the original ARTool results, for which we report $F$ values with partial $\eta^2$ ($\eta_p^2$). Where a main effect was significant, we compared the raw scores of all pairs of the four groups with Steel--Dwass tests, a one-way rank-based procedure that is not tied to the ART model. For the comparison between Groups A and E, we used Wilcoxon rank-sum tests and report $U$, $Z$ (positive values indicate higher scores in Group A), two-sided $p$ values from the tie-corrected normal approximation, and the effect size $r=|Z|/\sqrt{N}$. For each effect, we applied the Holm adjustment separately to the four IPQ component scores and the four UES-SF factors; for IPQ covariate analyses, we applied the correction separately to parametric and rank-based tests. We report adjusted values as $p_\mathrm{Holm}$ where they change the interpretation. The NASA-TLX was tested on the overall score only, and its six subscales are shown descriptively. No adjustment was made across questionnaires or hypotheses, and results other than the primary measures are treated as exploratory. For the CT--DT and LD--SD comparisons and the A--E comparison of the primary measures, we also report mean differences with 95\% confidence intervals.

Because the groups could differ in their post-practice IPQ ratings before the manipulation, we also analyzed the post-experience IPQ scores with an analysis of covariance (ANCOVA) that included the post-practice score as a covariate, which compares the conditions at a common post-practice level~\cite{vickers2001analysing}. We checked the homogeneity of regression slopes and the normality of residuals, and repeated the adjustment with a rank-based ANCOVA (Quade's method) where the residuals deviated from normality. Heart rate was analyzed with two-way ANOVAs for Groups A--D and with Welch's $t$-tests for Groups A and E.

The open-ended answers to Q3 and Q5 were analyzed by thematic coding. An initial set of themes and code assignments was drafted from the Japanese answers with the assistance of an AI tool (see Acknowledgments). One of the authors then checked the theme definitions and every assignment against the original answers. Each answer could be assigned to one or more themes (definitions and examples in \cref{S-table:codes}), and we counted the participants per theme and group; a participant was counted at most once per theme and question. Because the coding is descriptive and the numbers per group are small, we report counts without statistical tests.

\section{Results}
\label{sec:results}
We report RQ1 and RQ2 for Groups A--D, RQ3 for Groups A and E, and open-ended responses for all groups (analysis in \cref{sec:analysis}).

\subsection{RQ1 and RQ2: AR-Phase Duration and Transition Technique (Groups A--D)}
We analyzed Groups A--D in a $2 \times 2$ between-subjects design with AR-phase duration (LD/SD) and transition technique (CT/DT).

\subsubsection{IPQ Change Scores}
\begin{figure}[t]
    \centering
    \includegraphics[width=0.65\columnwidth]{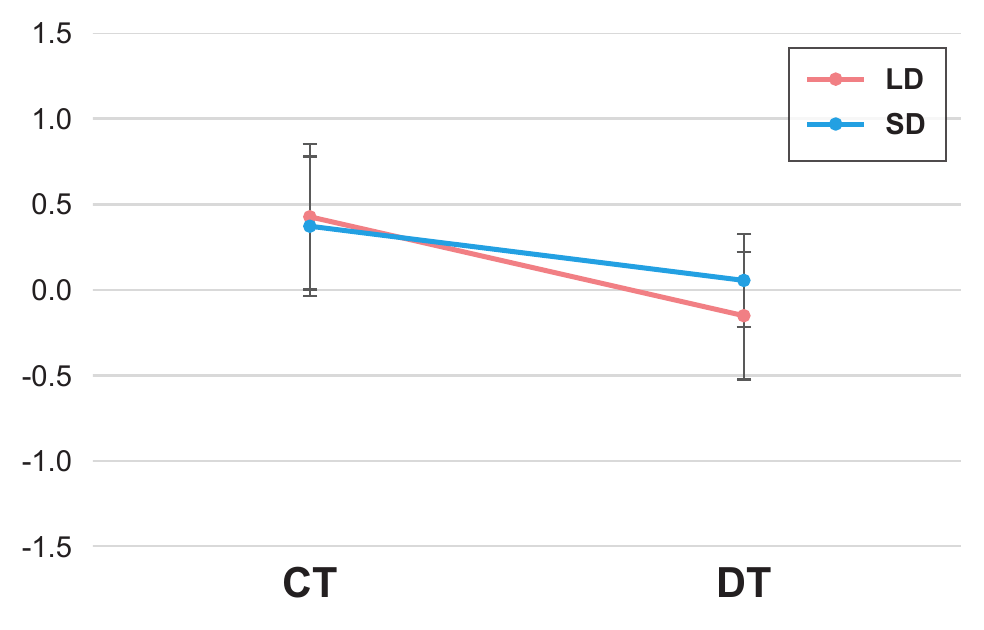}
    \caption[Overall IPQ scores for Groups A--D]{Overall IPQ change scores (post-experience minus post-practice) for Groups A--D under the $2 \times 2$ design (AR-phase duration: LD/SD; transition technique: CT/DT). Mean; error bars, 95\% confidence intervals. In ART, only the main effect of transition technique was significant. Post-hoc Steel--Dwass tests showed a significant difference between Groups A and C.}
    \label{fig:result_ex1_IPQ-ALL}
\end{figure}

\cref{fig:result_ex1_IPQ-ALL} shows the overall IPQ change scores (post-experience minus post-practice). ART revealed a significant main effect of transition technique ($F(1, 56) = 8.46$, $p = .005$, $\eta_p^2 = .131$), with larger gains in the CT condition (mean change $0.40$ vs.\ $-0.05$). Neither the main effect of AR-phase duration ($F(1, 56) = 0.35$, $p = .555$, $\eta_p^2 = .006$) nor the interaction ($F(1, 56) = 1.26$, $p = .266$, $\eta_p^2 = .022$) was significant. Post-hoc Steel--Dwass tests on the change scores of the four groups showed a significant difference between Groups A and C ($p = .047$); the other pairs did not differ (A--D: $p = .233$; B--C: $p = .233$; B--D: $p = .651$; C--D: $p = .417$; A--B: $p > .99$).

\begin{figure}[t]
\centering
\subfloat[G]{\includegraphics[width=0.48\columnwidth]{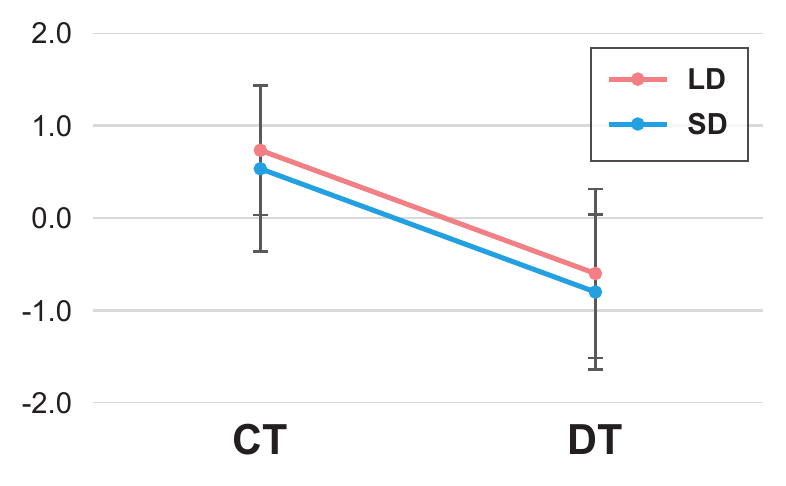}}
\hfil
\subfloat[SP]{\includegraphics[width=0.48\columnwidth]{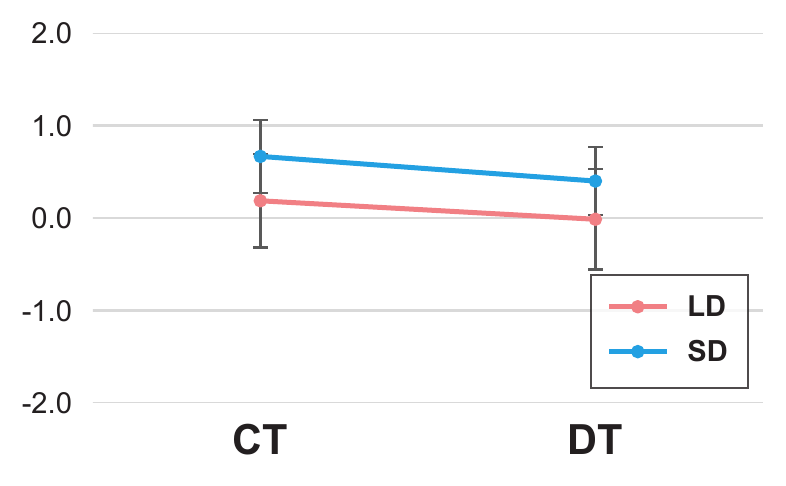}}

\subfloat[INV]{\includegraphics[width=0.48\columnwidth]{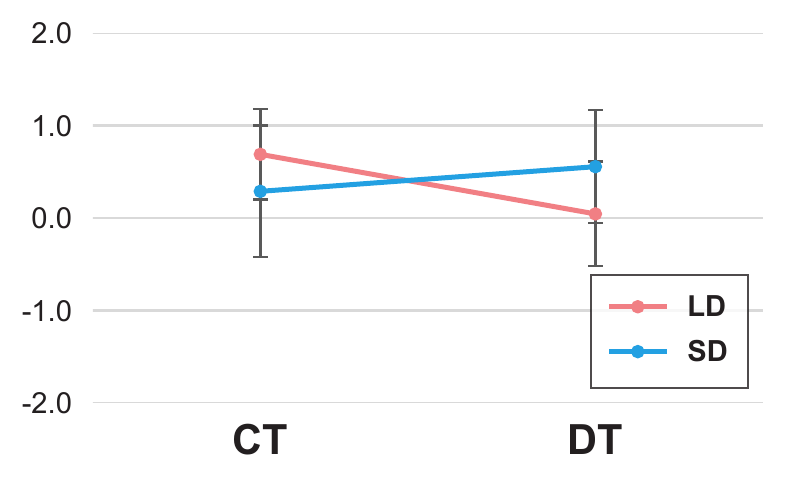}}
\hfil
\subfloat[REAL]{\includegraphics[width=0.48\columnwidth]{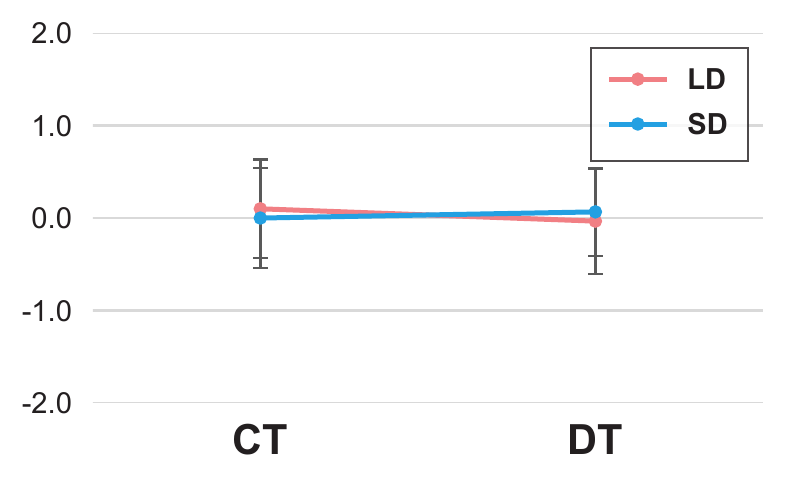}}
\caption[IPQ component scores for Groups A--D]{IPQ component change scores for Groups A--D: (a) General Presence (G), (b) Spatial Presence (SP), (c) Involvement (INV), and (d) Realism (REAL) (mean; error bars, 95\% confidence intervals). A significant effect of transition technique was observed for G, the INV interaction had unadjusted $p=.076$ ($p_\mathrm{Holm}=.305$), and no significant differences were found for SP or REAL.}
\label{fig:result_ex1_IPQ-others}
\end{figure}

The largest CT--DT difference in change scores was on G ($F(1,56)=12.16$, $p<.001$, $\eta_p^2=.178$, Holm-adjusted $p=.004$), partly because the CT groups rated G lower after practice (\cref{sec:ipq_pre_post}). Steel--Dwass comparisons gave $p=.043$ for A vs.\ D and $.065$ for A vs.\ C. The INV interaction had unadjusted $p=.076$ and Holm-adjusted $p=.305$. SP and REAL showed no significant effects (\cref{fig:result_ex1_IPQ-others,S-table:tests_art}).

\subsubsection{Post-Practice and Post-Experience IPQ Scores}
\label{sec:ipq_pre_post}
\Cref{S-table:ipq_pre_post} lists the post-practice and post-experience IPQ scores per group. We analyzed each administration on its own and the post-experience scores with the post-practice scores as a covariate.

Post-experience scores showed no significant effects on the overall score or its components (\cref{S-table:tests_art}). The smallest $p$ was .089 for duration on REAL, with higher ratings in SD.

Before the manipulation, CT participants rated the common practice session lower on G than DT participants (4.77 vs.\ 5.67, $p=.010$). No other post-practice comparison was significant. After the main experience, G was higher in CT (5.40 vs.\ 4.97, $p=.275$). Both differences contributed to the larger change in CT (\cref{S-table:tests_art}).

To account for this imbalance, we fitted an analysis of covariance (ANCOVA) to the post-experience scores with duration, transition technique, their interaction, and the post-practice score as predictors. Allowing slopes to vary across Groups A--D did not improve model fit for the overall score ($p=.665$) or G ($p=.962$). For the overall IPQ score, the main effect of transition technique reached $p = .0496$ ($F(1, 55) = 4.03$, $\eta_p^2 = .068$; adjusted means 5.14 for CT and 4.77 for DT), with no significant effect of duration ($F(1, 55) = 0.35$, $p = .556$) or interaction ($F(1, 55) = 0.54$, $p = .467$). For G, the adjusted means were 5.52 (CT) and 4.85 (DT). No component effect reached significance after Holm correction ($p_\mathrm{Holm}=.333$ for transition on G and $.258$ for duration on SP; \cref{S-table:tests_ancova}). Because the G residuals deviated slightly from normality (Shapiro--Wilk $p = .049$), we also used rank-based ANCOVA. For the transition effect, Quade's test gave $p=.140$ for the overall score.

\subsubsection{Size and Robustness of the Transition Difference}
\label{sec:results_robust}
\Cref{table:robust} gives the CT--DT difference as an estimate with its 95\% confidence interval (CI), for the overall score used throughout and for the three alternative composites defined in \cref{sec:measures}: the mean of the 11 administered items, which weights every item equally; the mean of G, SP, and INV, which omits the two weakly correlated Realism items (\cref{sec:measures}); and G alone. For the overall score, the unadjusted CT--DT difference was $+0.19$ [$-0.27$, 0.65] after the experience and $-0.25$ [$-0.65$, 0.14] after practice. With adjustment for the post-practice score (common slope 0.72), the difference was 0.38 points on the 7-point scale for the overall score (95\% CI [0.00, 0.75]; $d = 0.42$ relative to the pooled within-cell SD) and between 0.26 and 0.67 points for the alternatives, whose CIs included 0. With the 11-item mean, the difference was not significant in either the change score or the ANCOVA ($p = .13$ and $p = .17$); this composite gives the General Presence item a weight of 1/11 instead of 1/4. For comparison, the adjusted difference between LD and SD was $-0.11$ [$-0.48$, 0.26] for the overall score and ranged from $-0.22$ to $+0.27$ for the alternatives.

\begin{table}[t]
 \caption[CT--DT difference across composites]{CT--DT Difference in IPQ Scores (7-Point Scale) for the Overall Score and Three Alternative Composites, Groups A--D ($n = 30$ vs.\ 30), with 95\% Confidence Intervals}
 \label{table:robust}
 \scriptsize
 \centering
 \setlength{\tabcolsep}{2.5pt}
 \begin{tabular}{lcccc}
  \toprule
  Composite & Change score & ANCOVA-adjusted & $p$ & Quade $p$ \\
  \midrule
  Overall & $+0.45$ [0.07, 0.83] & $+0.38$ [0.00, 0.75] & .0496 & .14 \\
  11-item mean & $+0.28$ [$-0.09$, 0.66] & $+0.26$ [$-0.11$, 0.63] & .17 & .25 \\
  G, SP, INV & $+0.59$ [0.11, 1.06] & $+0.46$ [$-0.01$, 0.94] & .055 & .077 \\
  G (1 item) & $+1.33$ [0.47, 2.19] & $+0.67$ [$-0.09$, 1.43] & .083 & .060 \\
  \bottomrule
  \multicolumn{5}{@{}p{0.97\columnwidth}@{}}{\rule{0pt}{2.6ex}Differences [95\% CI] are averaged over AR-phase duration in a $2 \times 2$ linear model; the ANCOVA adds the post-practice score of the same composite as a covariate; $p$ is that of the ANCOVA. Quade: rank-based ANCOVA.} \\
 \end{tabular}
\end{table}

\subsubsection{Direct Comparison with the Practice Session (Q2)}
Participants rated absorption relative to practice higher after CT (6.43 vs.\ 5.87, $p=.061$). Neither duration nor its interaction with transition technique was significant (\cref{S-table:tests_art,S-table:ipq_pre_post}).

\subsubsection{UES-SF}
\Cref{S-fig:result_ex1_UES} shows the overall and factor-level UES-SF scores for Groups A--D. ART found no significant effects for the overall score (duration: $F(1, 56) = 0.71$, $p = .404$; transition: $F(1, 56) = 0.29$, $p = .595$; interaction: $F(1, 56) = 0.35$, $p = .556$). At the factor level, Aesthetic Appeal (AE) showed a trend for duration ($F(1, 56) = 3.28$, $p = .075$, $\eta_p^2 = .055$; $p_\mathrm{Holm} = .302$), with higher scores in the LD condition (4.28 vs.\ 4.01); no other effect on FA, PU, AE, or RW approached significance (all $p \geq .25$).

\subsubsection{NASA-TLX}
NASA-TLX overall scores showed no significant effects (\cref{S-table:tests_art}). The CT--DT mean difference was $-4.7$ points on the 0--100 scale (95\% CI [$-13.1$, 3.7]). Subscale ratings are shown descriptively in \cref{S-fig:result_ex1_NASATLX}.

\subsubsection{Heart Rate (Exploratory)}
Mean heart rate in the VR phase, and its change from the practice session, did not differ by AR-phase duration, transition technique, or their interaction (all $p \geq .14$). Mean heart rate fell in all groups just after the AR phase ended. Relative to the final 20~s of AR, mean heart rate in the first 10~s of VR was higher after CT than after DT ($F(1, 54) = 8.26$, $p = .006$, $\eta_p^2 = .133$). This window began about 27~s after AR ended for CT and 8--9~s after AR ended for DT (\cref{S-table:timeline}). The CT--DT main effect was not significant in the later VR windows, and heart-rate changes did not correlate with the IPQ or UES-SF scores (all $|\rho| \leq .21$). Phase means and test statistics appear in \cref{S-sec:supp_hr}.

\subsubsection{Control Questions}
Practice sufficiency (Q1) was rated slightly higher after CT (6.33 vs.\ 5.90, $p=.061$). Q4 ratings were high across groups, with no significant effects (\cref{S-table:ipq_pre_post,S-table:tests_art}).

\subsection{RQ3: Effects of Interaction (Groups A and E)}
Group A (interactive) was compared with Group E (non-interactive); this comparison is exploratory (\cref{sec:conditions}).

\subsubsection{IPQ}
Overall IPQ change averaged 0.19 points higher in Group A than in Group E (95\% CI [$-0.46$, 0.84], \cref{S-fig:result_ex2_IPQ}). Wilcoxon rank-sum tests found no significant differences in the change scores, either for the overall score ($U = 128.5$, $Z = 0.66$, $p = .507$) or for any component (G: $Z = 1.08$, $p = .280$; SP: $Z = 1.06$, $p = .288$; INV: $Z = -0.33$, $p = .738$; REAL: $Z = -0.19$, $p = .850$). The post-experience scores alone did not differ either (overall: $Z = 0.44$, $p = .663$; G: $p = .428$; SP: $Z = 1.46$, $p = .146$; INV: $p = .722$; REAL: $p = .932$), nor did the post-experience scores adjusted for the post-practice scores by ANCOVA (all $p \geq .21$).

\subsubsection{UES-SF}
\begin{figure}[t]
\centering
\includegraphics[width=0.7\columnwidth]{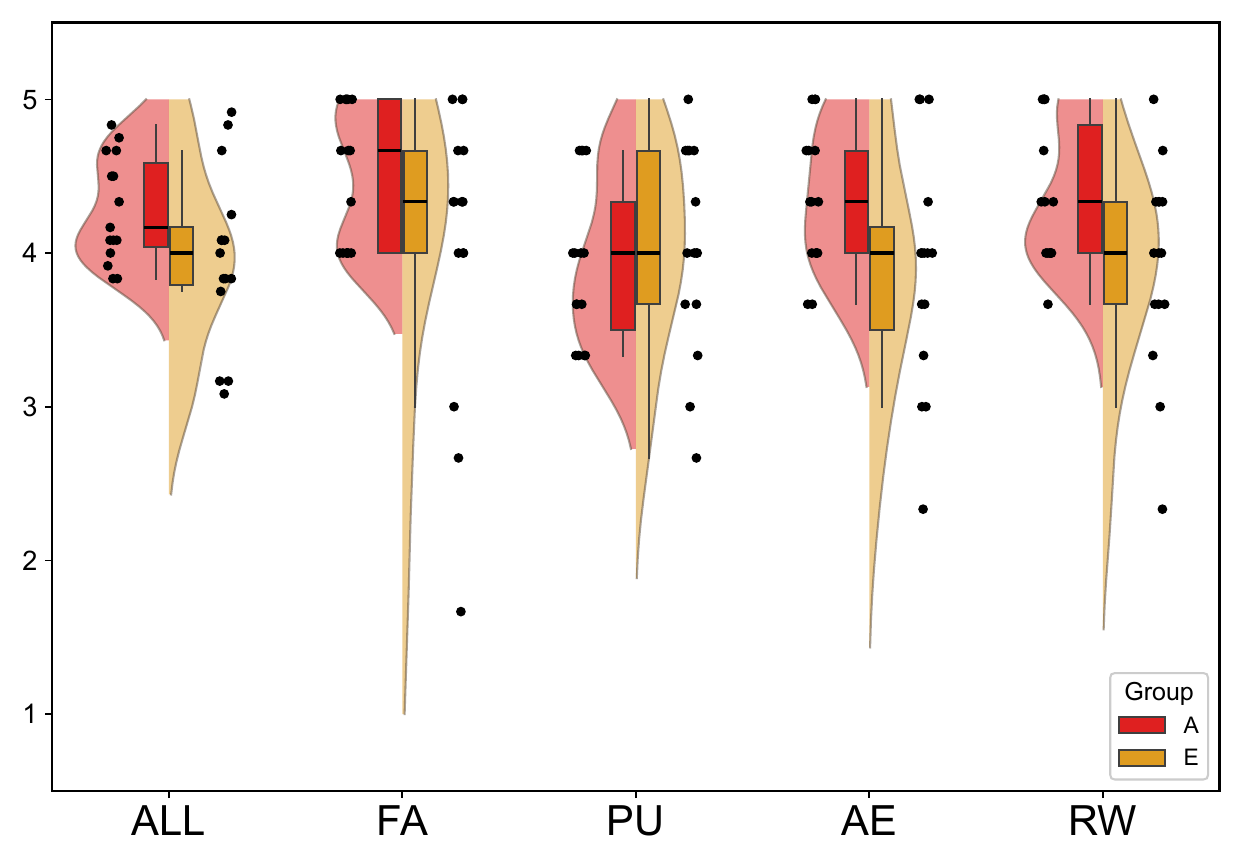}
\caption[UES-SF results for Groups A and E]{UES-SF results for Groups A and E. Half violins show the distribution of each group (A, left; E, right), boxes show the median and interquartile range, and dots show individual participants. The differences favoring Group A for the overall score, AE, and RW were not significant ($p = .07$--$.10$ uncorrected).}
\label{fig:result_ex2_UES}
\end{figure}

\cref{fig:result_ex2_UES} shows the UES-SF comparison between Groups A and E. Scores were higher in Group A than in Group E for the overall score (4.28 vs.\ 3.97), Aesthetic Appeal (AE; 4.36 vs.\ 3.89), and Reward Factor (RW; 4.36 vs.\ 3.91), but none of these differences was significant in Wilcoxon rank-sum tests (overall: $U = 152.5$, $Z = 1.67$, $p = .095$, $r = .305$; AE: $U = 155.0$, $Z = 1.80$, $p = .072$, $r = .328$; RW: $U = 154.0$, $Z = 1.76$, $p = .078$, $r = .322$; $p_\mathrm{Holm} = .29$ for AE and RW). The mean differences on the 5-point scale were 0.31 (95\% CI [$-0.04$, 0.66]) for the overall score, 0.47 [$-0.01$, 0.95] for AE, and 0.44 [0.01, 0.88] for RW (Welch intervals). Focused Attention (FA; $U = 135.5$, $Z = 0.98$, $p = .327$, $r = .179$) and Perceived Usability (PU; $U = 98.0$, $Z = -0.62$, $p = .538$, $r = .112$) showed no significant differences.

\subsubsection{NASA-TLX}
The A--E mean difference in overall workload was 4.6 points (95\% CI [$-9.7$, 18.8]; Wilcoxon $p=.590$; \cref{S-fig:result_ex2_NASATLX,S-table:tests_ae}).

\subsubsection{Heart Rate (Exploratory)}
Groups A and E did not differ in mean heart rate in any phase or in the response at VR onset. From practice to VR, mean heart rate fell by 2.3~bpm in A and rose by 1.9~bpm in E (Welch $p=.068$, \cref{S-sec:supp_hr}). Differences in session dates, rooms, and VR-phase duration prevent us from attributing this contrast to performing the activities.

\subsubsection{Additional Questions}
Groups A and E showed no significant differences on Q1, Q2, or Q4 (\cref{S-table:ipq_pre_post,S-table:tests_ae}).

\subsection{Open-Ended Responses (All Groups)}
\label{sec:results_qual}
All 75 participants answered Q3, which asked for the reason for their Q2 rating of absorption relative to the practice session. Of the 58 participants who answered Q5, 57 provided a substantive comment. \Cref{table:qual} lists the themes and the number of participants per group who mentioned each theme; the quotes below were translated from Japanese by the authors, and the letter after each quote gives the group.

\begin{table}[t]
 \caption[Themes of the open-ended responses]{Themes of the Open-Ended Responses: Number of Participants per Group (15 per Group) Who Mentioned Each Theme}
 \label{table:qual}
 \footnotesize
 \centering
 \setlength{\tabcolsep}{3.5pt}
 \begin{tabular}{L{0.50\columnwidth}ccccc}
  \toprule
  Theme & A & B & C & D & E \\
  \midrule
  \textit{Q3: reason for being more (or less) absorbed than in practice} & & & & & \\
  Physical effects felt on the body (water, mist, wind, warmth) & 7 & 5 & 8 & 7 & 0 \\
  Story or content (plot, unexpected development, scary content) & 5 & 6 & 2 & 6 & 5 \\
  Surroundings replaced by the virtual space; real room no longer visible & 3 & 2 & 3 & 5 & 4 \\
  Sound (headphones blocking ambient sound, audio) & 0 & 1 & 3 & 1 & 3 \\
  Operations (knowing, getting used to, or having more of them) & 1 & 1 & 2 & 3 & 0 \\
  Absence of own operations (lower absorption) & 0 & 0 & 0 & 0 & 4 \\
  Other detracting factors (lower absorption) & 1 & 0 & 2 & 3 & 0 \\
  \midrule
  \textit{Q5: other comments} & & & & & \\
  Participants who answered Q5 & 12 & 7 & 12 & 14 & 13 \\
  Enjoyment or positive overall impression & 11 & 3 & 9 & 9 & 11 \\
  Positive remark on physical effects & 1 & 1 & 4 & 3 & 0 \\
  Device or ergonomic problems (headset weight, posture, audio volume) & 2 & 0 & 0 & 4 & 0 \\
  Operational difficulty (grasping the cup, hand passing through objects) & 0 & 1 & 2 & 2 & 0 \\
  Difficulty understanding the setting, instructions, or plot & 1 & 0 & 1 & 1 & 0 \\
  \bottomrule
 \end{tabular}
\end{table}

In Groups A--D, which performed the activities themselves, the most frequent reason given in Q3 was the physical effects felt on the body (27 of 60 participants), for example: ``In the main session, the feeling of my own hands getting wet matched the visual information in the VR space, so I could become more immersed'' (A). No participant in Group E, who did not insert their hands into the prop, mentioned these effects; instead, 4 of 15 attributed lower absorption to the absence of operations (``There was nothing to do by myself; I just watched the VR space'' (E)). The story was named in all groups (19 of 60 in Groups A--D and 5 of 15 in Group E). Three participants reported a vibration that the system did not present; two of them also mentioned water and were counted under physical effects for that reason.

Mentions of the surroundings being replaced by the virtual space were not more frequent after the continuous transition (5 of 30 in the CT groups) than after the abrupt cut (8 of 30 in the DT groups); participants in both conditions described the end state, in which nothing of the real room remained visible (``Everything I could see had been repainted by the virtual space, so I completely lost the feeling of being in the real environment'' (A); ``It became pitch dark, and nothing from the real world was shown anymore'' (D)). Factors that lowered absorption were named by 6 participants in Groups A--D, 5 of whom were in the DT groups; only one of them concerned the cut itself (``when effects such as the VR space going dark came in, I conversely found myself wondering what was happening in reality'' (D)). Unexpected or sudden plot developments were mentioned by 3 participants in the LD groups and 3 in the SD groups. In Q5, most comments expressed enjoyment; the problems reported concerned the headset and posture, the volume of the narration, and grasping the virtual cup, and difficulties in understanding the setting, the plot, or the final instruction were reported by one participant each in Groups A, C, and D.

\section{Discussion}
\label{sec:discussion}
\Cref{table:hypotheses} summarizes the findings for the three hypotheses.

\begin{table}[t]
 \caption[Hypothesis outcomes]{Summary of Hypothesis Outcomes}
 \label{table:hypotheses}
 \small
 \centering
 \begin{tabular}{cp{0.27\columnwidth}p{0.53\columnwidth}}
  \toprule
  & Hypothesis & Outcome \\
  \midrule
  H1 & Longer AR phase $\rightarrow$ higher presence & Not supported: adjusted LD--SD difference $-0.11$ [$-0.48$, 0.26]. \\
  H2 & Continuous transition $\rightarrow$ higher presence & Partly supported. All estimates favored CT, but the strength of evidence depended on scoring and analysis (adjusted overall difference 0.38 [0.00, 0.75]). \\
  H3 & Interaction $\rightarrow$ higher presence and engagement & Not supported: no clear IPQ difference; higher but non-significant UES-SF scores in Group A (exploratory). \\
  \bottomrule
 \end{tabular}
\end{table}

\subsection{RQ2: Transition Technique and Presence}
Participants who experienced the designed continuous transition (CT) rated presence higher, relative to the practice session, than those who experienced the abrupt cut (DT). All composites favored CT, with the largest difference in change scores on the General Presence item (\cref{table:robust}). Part of the difference in change scores reflects the lower practice-session ratings of the CT groups; compared at a common practice-session rating, the difference in the overall score was 0.38 points on the 7-point scale (95\% CI [0.00, 0.75], $d = 0.42$), and with the 11-item mean it was smaller and not significant. Q2, which asked directly for a comparison with the practice session, pointed in the same direction. We read this as a small-to-moderate presence advantage of the transition design whose lower bound is close to zero.

Adding about 85~s of repeated AR activities produced no clear presence difference. The additional 19~s in CT contained different events, so the contribution of transition duration remains uncertain.

Five of 30 CT participants and eight of 30 DT participants mentioned feeling surrounded by the virtual space or losing sight of the room. These comments describe an experience shared across conditions and do not explain the higher CT ratings.

Spatial Presence and Realism did not differ between conditions. The manipulations changed the temporal structure of the transition but not the spatial layout of the VR scene, and the two Realism items were nearly uncorrelated (\cref{sec:measures}). Mean heart rate over the VR phase was similar across groups. In the first 10~s of VR, it was near or above the pre-transition level in the CT groups and below it in the DT groups. VR began later after CT. We did not detect a difference in workload between CT and DT ($-4.7$ points on the 0--100 scale, [$-13.1$, 3.7]).

\subsection{RQ1: AR-Phase Duration}
Repeating the AR-phase activities three times instead of once (about 160 vs.\ 75~s) produced no clear difference in presence ($-0.11$ [$-0.48$, 0.26] for the adjusted overall score), engagement, or heart rate, so H1 was not supported, although the interval allows LD to be up to about half a point lower than SD. Even the SD condition contained every type of interaction with the landmark object once, so LD added repetitions rather than new content, and the IPQ, which rates the experience as a whole, may be dominated by the transition and the VR phase. Some participants named the weight of the headset and the downward viewing posture as obstacles (``the weight of the goggles considerably hindered my absorption'' (A)), which a longer AR phase would make more relevant.

\subsection{RQ3: Interaction (Exploratory)}
H3 was not supported. Overall UES-SF, AE, and RW scores were higher in Group A without reaching significance (0.31--0.47 points on the 5-point scale). The comments point in the same direction: physical effects felt on the hands were the most frequent reason for absorption in Group A (7 of 15) and were not mentioned in Group E, whose participants did not insert their hands, and 4 of 15 Group E participants attributed lower absorption to the absence of operations. One observer wrote, ``I could not operate anything, but when everything around me became virtual space I felt strong immersion'' (E). A concurrent study could test whether performing the activities increases engagement.

\subsection{Relation to Prior Work}
Prior work has studied continuous real-to-virtual transitions and shared landmarks in other situations. Feld et al.\ compared transitions between virtual environments~\cite{feld2023keep} and transitions that users make repeatedly while continuing a task in cross-reality experiences; in both, short and simple transitions were preferred~\cite{feld2024simple_efficient}. Our study concerns a single, narrative entry into VR, in which a cut also removes the physical room and the object the user has just touched. Pointecker et al.\ reported that participants linked their AR--VR transition preferences to shared landmarks~\cite{pointecker2022bridging}. In a later study, the authors used a replica of the room as an intermediate stage~\cite{pointecker2024real}; MagicBook moved users from a real book into VR~\cite{magicbook2001}. The higher overall IPQ score for CT than for DT is consistent with earlier findings for entry through a replica room versus direct entry~\cite{steinicke2009} and for a physical door or virtual portal versus a fade~\cite{fujisawa2024door}. In a group setting, participants preferred onboarding through MR to direct entry into VR~\cite{schott2025onboarding}. Our design adds a task at the object that serves as the entrance to VR. All groups performed or observed that task, so its contribution to presence remains open.

\subsection{Limitations and Future Work}
\label{sec:limitations}
\label{sec:future_work}
\textbf{Composite transition.} CT and DT differed in how the room disappeared, whether the television remained visible as it did so, the story events, the transition duration, and the timing of the fan (\cref{S-table:timeline}). Some participants found the swallowing sequence surprising or scary, which may also have affected arousal. The difference therefore belongs to this transition design as a whole. Isolating visual continuity would require matching story events, transition duration, and fan timing.

\textbf{Role of the landmark.} All conditions included the television during AR. Only CT kept it visible as the room disappeared. Whether a landmark is needed, how the design compares with replica-based transitions, and whether crossing from the real room adds presence beyond the same introduction presented entirely in VR remain to be tested.

\textbf{Single content and implementation.} We tested one narrative experience, one landmark object with purpose-built hardware, and one transformation of the passthrough image. Other objects (e.g., a desk, a door, or a book~\cite{magicbook2001}), non-narrative content, other forms of gradual replacement, and measures of VR sickness and spatial orientation~\cite{Moghadam2017} remain to be studied.

\textbf{Precision.} The main effects of the $2 \times 2$ design compare 30 with 30 participants and had a power of .80 for effects of partial $\eta^2 \geq .12$; the A--E comparison, with 15 per group, only for $d \geq 1.06$. Detecting the engagement differences observed for RQ3 ($d \approx 0.7$) with a power of .80 would require about 30--40 participants per group.

\textbf{Assignment and comparison groups.} Groups A--D were assigned manually rather than randomly, and the CT groups rated the practice session lower. The covariate adjustment addresses this measured difference but not group differences that were not measured. Group E was recruited about six weeks later and run in a different room.

\textbf{Measurement.} Eleven of the 14 IPQ items were used, so the overall score is not directly comparable with full-scale IPQ scores, and the CT--DT difference depended on how the items were combined. The open-ended answers were coded by one author. Participants were aged 18--30, and about half had never used VR.

\section{Conclusion}
\label{sec:conclusion}
We designed a real-to-virtual transition in which users first carry out a multisensory AR task with the object through which they will enter VR. In \textit{Beyond the TV}, users tap a television to change channels and reach through the screen to tend a virtual flower, feeling mist or warmth on their hand. Within this experience, we compared AR activity repetition and transition technique in a $2 \times 2$ study ($N=60$). A separately recruited group ($N=15$) watched the activities.

After adjustment for post-practice ratings, the estimated overall IPQ score was 0.38 points higher for the designed transition than for an abrupt cut on the 7-point scale (95\% CI [0.00, 0.75]). The adjusted difference was smaller when all 11 items were weighted equally. Repeating the AR-phase activities three times instead of once produced no clear difference, and the exploratory comparison of self-performed and observed activities yielded no clear evidence of a difference in presence or engagement.

Future work should separate the visual, narrative, and tactile transition components, test the need for a landmark, compare performing and watching activities in a concurrent randomized study, and examine other objects and content.

\section*{Acknowledgments}
Anthropic Claude assisted with manuscript revision, statistical reanalysis scripts, and initial thematic coding of open-ended answers (\cref{sec:analysis}). OpenAI Codex assisted with checking citations and revising the text. The authors checked all analyses against the raw data and all code assignments against the original Japanese answers, reviewed, edited, and verified the final wording, and are solely responsible for the content of the manuscript.

\bibliographystyle{IEEEtran}
\bibliography{kowada_reference}

\begin{thebibliography}{10}
\providecommand{\url}[1]{#1}
\csname url@samestyle\endcsname
\providecommand{\newblock}{\relax}
\providecommand{\bibinfo}[2]{#2}
\providecommand{\BIBentrySTDinterwordspacing}{\spaceskip=0pt\relax}
\providecommand{\BIBentryALTinterwordstretchfactor}{4}
\providecommand{\BIBentryALTinterwordspacing}{\spaceskip=\fontdimen2\font plus
\BIBentryALTinterwordstretchfactor\fontdimen3\font minus
  \fontdimen4\font\relax}
\providecommand{\BIBforeignlanguage}[2]{{%
\expandafter\ifx\csname l@#1\endcsname\relax
\typeout{** WARNING: IEEEtran.bst: No hyphenation pattern has been}%
\typeout{** loaded for the language `#1'. Using the pattern for}%
\typeout{** the default language instead.}%
\else
\language=\csname l@#1\endcsname
\fi
#2}}
\providecommand{\BIBdecl}{\relax}
\BIBdecl

\bibitem{milgram1995augmented}
P.~Milgram, H.~Takemura, A.~Utsumi, and F.~Kishino, ``Augmented reality: A
  class of displays on the reality-virtuality continuum,'' in
  \emph{Telemanipulator and telepresence technologies}, vol. 2351.\hskip 1em
  plus 0.5em minus 0.4em\relax {SPIE}, 1995, pp. 282--292.

\bibitem{slater1997framework}
M.~Slater and S.~Wilbur, ``A framework for immersive virtual environments
  ({FIVE}): Speculations on the role of presence in virtual environments,''
  \emph{Presence: Teleoperators \& Virtual Environments}, vol.~6, no.~6, pp.
  603--616, 1997.

\bibitem{milgram1994taxonomy}
P.~Milgram and F.~Kishino, ``A taxonomy of mixed reality visual displays,''
  \emph{IEICE Transactions on Information and Systems}, vol. E77-D, no.~12, pp.
  1321--1329, 1994.

\bibitem{dalgarno2010learning}
B.~Dalgarno and M.~J. Lee, ``What are the learning affordances of
  {{\mbox{3-D}}} virtual environments?'' \emph{British Journal of Educational
  Technology}, vol.~41, no.~1, pp. 10--32, 2010.

\bibitem{fowler2015virtual}
C.~Fowler, ``Virtual reality and learning: Where is the pedagogy?''
  \emph{British Journal of Educational Technology}, vol.~46, no.~2, pp.
  412--422, 2015.

\bibitem{yung2019new}
R.~Yung and C.~Khoo-Lattimore, ``New realities: a systematic literature review
  on virtual reality and augmented reality in tourism research,'' \emph{Current
  Issues in Tourism}, vol.~22, no.~17, pp. 2056--2081, 2019.

\bibitem{jung2017}
T.~Jung, M.~C. tom Dieck, N.~Moorhouse, and D.~tom Dieck, ``Tourists'
  experience of virtual reality applications,'' in \emph{2017 IEEE
  International Conference on Consumer Electronics (ICCE)}, 2017, pp. 208--210.

\bibitem{aziz2018}
H.~A. Aziz, ``Virtual reality programs applications in healthcare,''
  \emph{Journal of Health \& Medical Informatics}, vol.~9, no.~1, p. 305, 2018.

\bibitem{rubio2017immersive}
J.~L. Rubio-Tamayo, M.~Gertrudix~Barrio, and F.~Garc{\'\i}a~Garc{\'\i}a,
  ``Immersive environments and virtual reality: Systematic review and advances
  in communication, interaction and simulation,'' \emph{Multimodal Technologies
  and Interaction}, vol.~1, no.~4, p.~21, 2017.

\bibitem{li2022interactive}
C.~Li, W.~Li, H.~Huang, and L.-F. Yu, ``Interactive augmented reality
  storytelling guided by scene semantics,'' \emph{ACM Transactions on
  Graphics}, vol.~41, no.~4, pp. 1--15, 2022.

\bibitem{legino2024unveiling}
R.~Legino, M.~S. Romat, and M.~K. Shamsudin, ``Unveiling the advantages of
  digitalization in fine art and artistic studio research,'' in
  \emph{International Conference on Innovation \& Entrepreneurship in
  Computing, Engineering \& Science Education (InvENT 2024)}.\hskip 1em plus
  0.5em minus 0.4em\relax Atlantis Press, 2024, pp. 15--21.

\bibitem{matin2025aracceptance}
F.~Matin and E.~Mangina, ``Exploring students' acceptance of augmented reality
  technologies in education: An extended technology acceptance model
  approach,'' in \emph{Virtual Reality and Mixed Reality: Proceedings of EuroXR
  2024}, ser. Lecture Notes in Computer Science.\hskip 1em plus 0.5em minus
  0.4em\relax Springer, 2025, pp. 243--258.

\bibitem{10.1007/s00779-019-01352-8}
\BIBentryALTinterwordspacing
I.~Ghani, A.~Rafi, and P.~Woods, ``The effect of immersion towards place
  presence in virtual heritage environments,'' \emph{Personal and Ubiquitous
  Computing}, vol.~24, no.~6, pp. 861--872, Dec. 2020. [Online]. Available:
  \url{https://doi.org/10.1007/s00779-019-01352-8}
\BIBentrySTDinterwordspacing

\bibitem{10.1162/pres.1994.3.2.130}
\BIBentryALTinterwordspacing
M.~Slater, M.~Usoh, and A.~Steed, ``Depth of presence in virtual
  environments,'' \emph{Presence: Teleoperators and Virtual Environments},
  vol.~3, no.~2, pp. 130--144, Jan. 1994. [Online]. Available:
  \url{https://doi.org/10.1162/pres.1994.3.2.130}
\BIBentrySTDinterwordspacing

\bibitem{steinicke2009}
F.~Steinicke, G.~Bruder, K.~Hinrichs, A.~Steed, and A.~Gerlach, ``Does a
  gradual transition to the virtual world increase presence?'' in
  \emph{Proceedings of IEEE Virtual Reality (VR)}, Mar. 2009, pp. 203--210.

\bibitem{10.1145/1620993.1620998}
\BIBentryALTinterwordspacing
F.~Steinicke, G.~Bruder, K.~Hinrichs, M.~Lappe, B.~Ries, and V.~Interrante,
  ``Transitional environments enhance distance perception in immersive virtual
  reality systems,'' in \emph{Proceedings of the 6th Symposium on Applied
  Perception in Graphics and Visualization}, ser. APGV '09.\hskip 1em plus
  0.5em minus 0.4em\relax New York, NY, USA: Association for Computing
  Machinery, 2009, pp. 19--26. [Online]. Available:
  \url{https://doi.org/10.1145/1620993.1620998}
\BIBentrySTDinterwordspacing

\bibitem{schmidt2024development}
\BIBentryALTinterwordspacing
L.~Schmidt and E.~Yigitbas, ``Development and usability evaluation of
  transitional cross-reality interfaces,'' \emph{Proceedings of the ACM on
  Human-Computer Interaction}, vol.~8, no. EICS, pp. 1--32, 2024. [Online].
  Available: \url{https://doi.org/10.1145/3664637}
\BIBentrySTDinterwordspacing

\bibitem{pointecker2024real}
F.~Pointecker, J.~Friedl-Knirsch, H.-C. Jetter, and C.~Anthes, ``From real to
  virtual: Exploring replica-enhanced environment transitions along the
  reality-virtuality continuum,'' in \emph{Proceedings of the CHI Conference on
  Human Factors in Computing Systems}, 2024, pp. 1--13.

\bibitem{fujisawa2024door}
\BIBentryALTinterwordspacing
T.~Fujisawa, D.~Hagimori, M.~Perusqu{\'i}a-Hern{\'a}ndez, N.~Isoyama,
  H.~Uchiyama, and K.~Kiyokawa, ``Seamless multi-modal transitions between real
  and virtual environments using a physical door enhances presence and user
  engagement,'' in \emph{International Conference on Artificial Reality and
  Telexistence and Eurographics Symposium on Virtual Environments
  (ICAT-EGVE)}.\hskip 1em plus 0.5em minus 0.4em\relax The Eurographics
  Association, 2024. [Online]. Available:
  \url{https://doi.org/10.2312/egve.20241373}
\BIBentrySTDinterwordspacing

\bibitem{slater2009place}
M.~Slater, ``Place illusion and plausibility can lead to realistic behaviour in
  immersive virtual environments,'' \emph{Philosophical Transactions of the
  Royal Society B: Biological Sciences}, vol. 364, no. 1535, pp. 3549--3557,
  2009.

\bibitem{schubert2001experience}
T.~Schubert, F.~Friedmann, and H.~Regenbrecht, ``The experience of presence:
  Factor analytic insights,'' \emph{Presence: Teleoperators \& Virtual
  Environments}, vol.~10, no.~3, pp. 266--281, 2001.

\bibitem{o2018practical}
H.~L. O’Brien, P.~Cairns, and M.~Hall, ``A practical approach to measuring
  user engagement with the refined user engagement scale ({UES}) and new {UES}
  short form,'' \emph{International Journal of Human-Computer Studies}, vol.
  112, pp. 28--39, 2018.

\bibitem{nasa_tlx}
{NASA Human Systems Integration Division}, ``{NASA} task load index ({TLX}),''
  \url{https://www.nasa.gov/human-systems-integration-division/nasa-task-load-index-tlx/},
  (Accessed on November 18, 2026).

\bibitem{feld2023keep}
N.~Feld, P.~Bimberg, B.~Weyers, and D.~Zielasko, ``Keep it simple? evaluation
  of transitions in virtual reality,'' in \emph{Extended Abstracts of the 2023
  CHI Conference on Human Factors in Computing Systems}, 2023, pp. 1--7.

\bibitem{Moghadam2017}
K.~R. Moghadam and E.~D. Ragan, ``Towards understanding scene transition
  techniques in immersive 360 movies and cinematic experiences,'' in \emph{2017
  IEEE Virtual Reality (VR)}, 2017, pp. 375--376.

\bibitem{magicbook2001}
M.~Billinghurst, H.~Kato, and I.~Poupyrev, ``The {MagicBook}: A transitional
  {AR} interface,'' \emph{Computers \& Graphics}, vol.~25, no.~5, pp. 745--753,
  2001.

\bibitem{solroo2017one}
J.~S. Roo and M.~Hachet, ``One reality: Augmenting how the physical world is
  experienced by combining multiple mixed reality modalities,'' in
  \emph{Proceedings of the 30th Annual ACM Symposium on User Interface Software
  and Technology}, 2017, pp. 787--795.

\bibitem{jung2018limbo}
S.~Jung, P.~J. Wisniewski, and C.~E. Hughes, ``In limbo: The effect of gradual
  visual transition between real and virtual on virtual body ownership illusion
  and presence,'' in \emph{2018 IEEE Conference on Virtual Reality and 3D User
  Interfaces (VR)}, 2018, pp. 267--272.

\bibitem{pointecker2022bridging}
F.~Pointecker, J.~Friedl, D.~Schwajda, H.-C. Jetter, and C.~Anthes, ``Bridging
  the gap across realities: Visual transitions between virtual and augmented
  reality,'' in \emph{2022 IEEE International Symposium on Mixed and Augmented
  Reality (ISMAR)}.\hskip 1em plus 0.5em minus 0.4em\relax IEEE, 2022, pp.
  827--836.

\bibitem{valkov2017}
\BIBentryALTinterwordspacing
D.~Valkov and S.~Flagge, ``Smooth immersion: the benefits of making the
  transition to virtual environments a continuous process,'' in
  \emph{Proceedings of the 5th Symposium on Spatial User Interaction}, ser. SUI
  '17, 2017, pp. 12--19. [Online]. Available:
  \url{https://doi.org/10.1145/3131277.3132183}
\BIBentrySTDinterwordspacing

\bibitem{9790021}
M.~Sra, V.~Danry, and P.~Maes, ``Situated {VR}: Toward a congruent hybrid
  reality without experiential artifacts,'' \emph{IEEE Computer Graphics and
  Applications}, vol.~42, no.~3, pp. 7--18, May 2022.

\bibitem{feld2024navigating_realities}
N.~Feld, P.~Bimberg, B.~Weyers, and D.~Zielasko, ``Navigating realities:
  Assessing cross-reality transitions through a spatial memory game in {VR} and
  {AR} environments,'' in \emph{2024 IEEE Conference on Virtual Reality and 3D
  User Interfaces Abstracts and Workshops (VRW)}, 2024, pp. 1194--1195.

\bibitem{feld2024simple_efficient}
------, ``Simple and efficient? evaluation of transitions for task-driven
  cross-reality experiences,'' \emph{IEEE Transactions on Visualization and
  Computer Graphics}, vol.~30, no.~12, pp. 7601--7618, 2024.

\bibitem{horst2024back}
R.~Horst, R.~Naraghi-Taghi-Off, L.~Rau, and R.~D{\"o}rner, ``Back to reality:
  Transition techniques from short {HMD}-based virtual experiences to the
  physical world,'' \emph{Multimedia Tools and Applications}, vol.~83, no.~15,
  pp. 46\,683--46\,706, 2024.

\bibitem{soret2021door}
R.~Soret, A.-M. Montes-Solano, C.~Manzini, V.~Peysakhovich, and E.~F. Fabre,
  ``Pushing open the door to reality: On facilitating the transitions from
  virtual to real environments,'' \emph{Applied Ergonomics}, vol.~97, p.
  103535, 2021.

\bibitem{schott2025onboarding}
E.~Schott, T.~J. Zoeppig, P.~Viriyathomrongul, A.~B. Lammert, and B.~Froehlich,
  ``{VR} onboarding procedures for multiple collocated users: See-through
  tutorials and group transitions,'' in \emph{2025 IEEE International Symposium
  on Mixed and Augmented Reality (ISMAR)}, 2025, pp. 315--325.

\bibitem{Damien2017}
\BIBentryALTinterwordspacing
D.~Clergeaud, J.~S. Roo, M.~Hachet, and P.~Guitton, ``Towards seamless
  interaction between physical and virtual locations for asymmetric
  collaboration,'' in \emph{Proceedings of the 23rd ACM Symposium on Virtual
  Reality Software and Technology (VRST)}, ser. VRST '17, 2017, pp. 1--4.
  [Online]. Available: \url{https://doi.org/10.1145/3139131.3139165}
\BIBentrySTDinterwordspacing

\bibitem{al2020hapticsnakes}
M.~Al-Sada, K.~Jiang, S.~Ranade, M.~Kalkattawi, and T.~Nakajima,
  ``{HapticSnakes}: multi-haptic feedback wearable robots for immersive virtual
  reality,'' \emph{Virtual Reality}, vol.~24, no.~2, pp. 191--209, 2020.

\bibitem{nintendo_joycon_support}
{Nintendo}, ``{Joy-Con - Nintendo Switch Support},''
  \url{https://www.nintendo.com/my/support/switch/controller/joycon.html},
  (Accessed on March 13, 2026).

\bibitem{myBeat}
{Union Tool Co.}, ``my{B}eat product information,''
  \url{https://www.uniontool-mybeat.com/}, (Accessed on March 13, 2026).

\bibitem{IPQ_database}
igroup.org, ``igroup presence questionnaire ({IPQ}) item download,''
  \url{https://www.igroup.org/pq/ipq/download.php}, (Accessed on March 13,
  2026).

\bibitem{wobbrock2011aligned}
J.~O. Wobbrock, L.~Findlater, D.~Gergle, and J.~J. Higgins, ``The aligned rank
  transform for nonparametric factorial analyses using only {ANOVA}
  procedures,'' in \emph{Proceedings of the SIGCHI Conference on Human Factors
  in Computing Systems (CHI '11)}, 2011, pp. 143--146.

\bibitem{vickers2001analysing}
A.~J. Vickers and D.~G. Altman, ``Analysing controlled trials with baseline and
  follow up measurements,'' \emph{BMJ}, vol. 323, no. 7321, pp. 1123--1124,
  2001.

\end{thebibliography}
\clearpage

\begin{IEEEbiography}[{\includegraphics[width=1in,height=1.25in,clip,keepaspectratio]{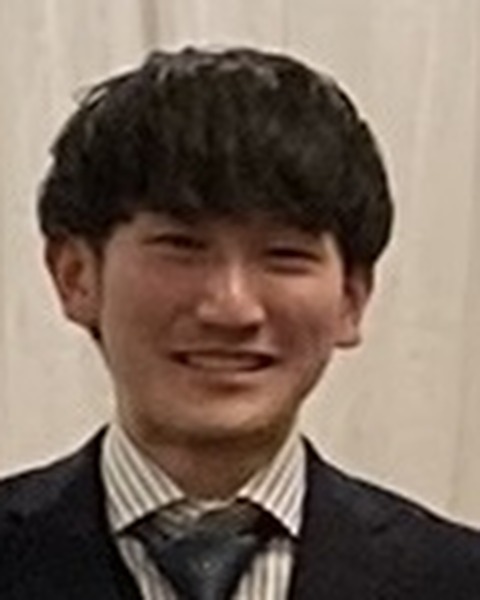}}]{Toru Kowada}
received the B.E. degree and the Master of Information Science degree from the University of Osaka, Japan, in 2023 and 2025, respectively. His research interest is in virtual reality and immersive experiences.
\end{IEEEbiography}

\begin{IEEEbiography}[{\includegraphics[width=1in,height=1.25in,clip,keepaspectratio]{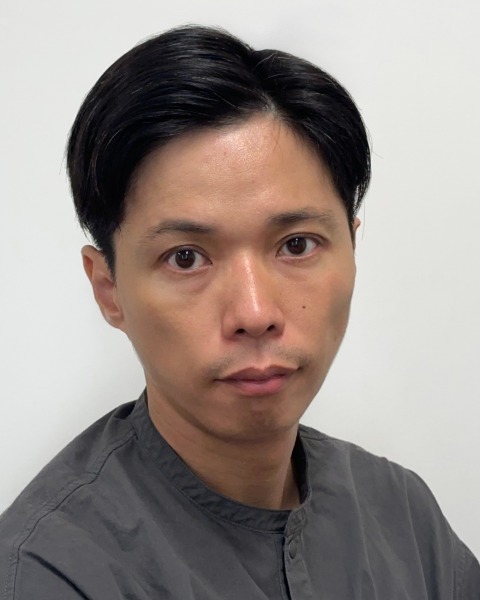}}]{Yuki Uranishi}
received the Ph.D. degree from the Graduate School of Information Science, Nara Institute of Science and Technology, Japan, in 2008. He is a professor at D3 Center, the University of Osaka, Japan. Before joining the University of Osaka, he was a research fellow of Japan Society for the Promotion of Science and an assistant professor at Nara Institute of Science and Technology from 2009 to 2012, at the University of Osaka from 2012 to 2014, and at Kyoto University Hospital from 2014 to 2016. He was an associate professor of Cybermedia Center, the University of Osaka, from 2016 to 2024. His research interests include computer vision, augmented reality, virtual reality, and human-computer interaction.
\end{IEEEbiography}

\begin{IEEEbiography}[{\includegraphics[width=1in,height=1.25in,clip,keepaspectratio]{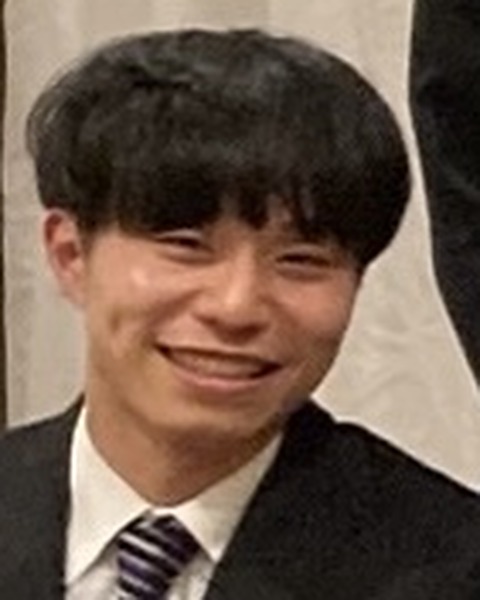}}]{Koji Momota}
received the B.E. degree and the Master of Information Science degree from the University of Osaka, Japan, in 2023 and 2025, respectively. His research interest is in avatar-based virtual reality environments for hybrid meetings.
\end{IEEEbiography}

\begin{IEEEbiography}[{\includegraphics[width=1in,height=1.25in,clip,keepaspectratio]{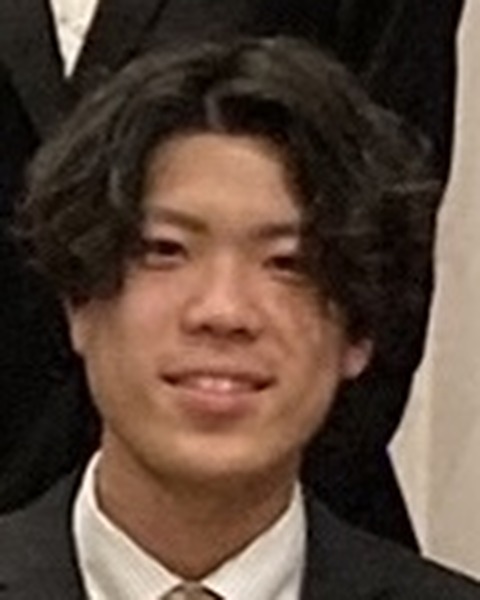}}]{Daiki Tomohiro}
received the B.E. degree from Kyoto Institute of Technology, Japan, in 2023, and the Master of Information Science degree from the University of Osaka, Japan, in 2025. His research interest is in thermal perception in virtual reality.
\end{IEEEbiography}

\begin{IEEEbiography}[{\includegraphics[width=1in,height=1.25in,clip,keepaspectratio]{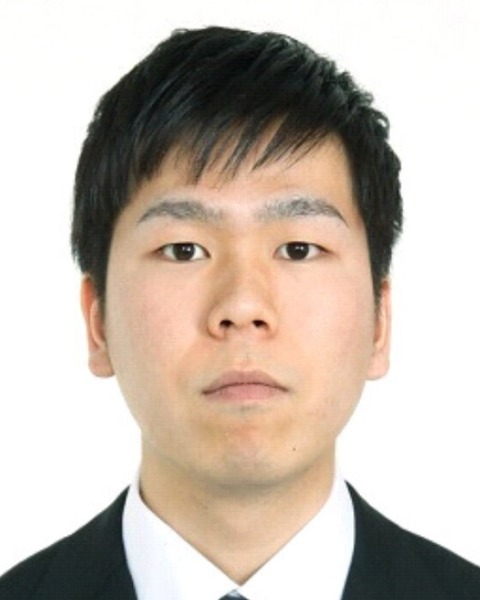}}]{Masato Kobayashi}
received the Bachelor of Maritime Sciences, Master of Maritime Sciences, and Doctor of Engineering degrees from Kobe University, Japan, in 2017, 2019, and 2022. From 2019 to 2021, he was an engineer and researcher with the Technology Development Division, Seiko Epson Corporation, Japan. From 2021 to 2022, he worked as a research intern at OMRON SINIC X Corporation, Japan. Since 2022, he has been Academic Researcher at Kobe University, Japan. He has worked with the Matsuo Lab at The University of Tokyo, researching and developing Robotics and AI. Since 2023, he has been with the University of Osaka, Japan, where he is currently Assistant Professor. His current research interests include robot learning, motion control, XR, HCI, and AI.
\end{IEEEbiography}

\begin{IEEEbiography}[{\includegraphics[width=1in,height=1.25in,clip,keepaspectratio]{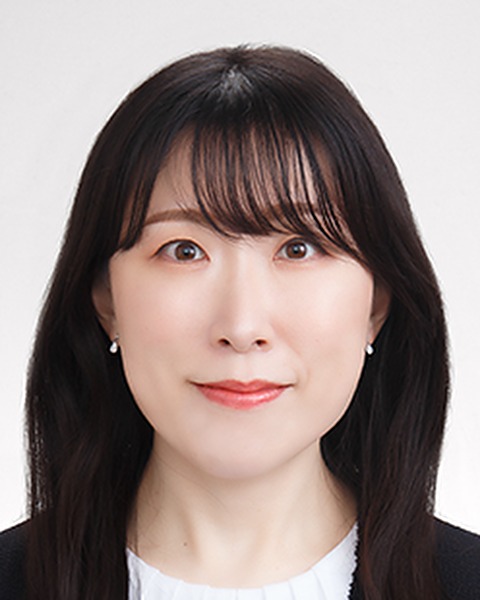}}]{Shizuka Shirai}
received the B.S., M.S., and Ph.D. degrees in Informatics and Mediology from Mukogawa Women's University, Hyogo, Japan, in 2007, 2012, and 2015, respectively. She was an Assistant Professor at the School of Human Environmental Sciences, Mukogawa Women's University from 2015 to 2018. She was an Associate Professor at the Cybermedia Center, the University of Osaka from 2018 to 2024. She is currently an Associate Professor at the D3 Center, the University of Osaka. Her research interests include human-computer-interaction, educational technology, learning analytics, and computer science education.
\end{IEEEbiography}

\begin{IEEEbiography}[{\includegraphics[width=1in,height=1.25in,clip,keepaspectratio]{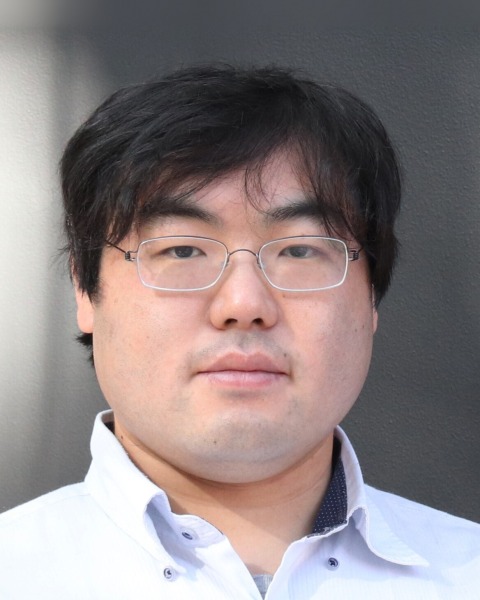}}]{Naoya Chiba}
received the B.S. degree in engineering, and the M.S. and Ph.D. degrees in information science from Tohoku University, Japan, in 2016, 2018, and 2020, respectively. He was a researcher at Waseda University and OMRON SINIC X Corporation from 2020 to 2023, and an Assistant Professor at the Graduate School of Information Sciences, Tohoku University from 2023 to 2024. He is currently an Associate Professor at the D3 Center, the University of Osaka. His research interests include robot vision, 3D measurement, and 3D data processing, with a focus on point cloud processing and 3D deep learning, as well as their applications.
\end{IEEEbiography}

\begin{IEEEbiography}[{\includegraphics[width=1in,height=1.25in,clip,keepaspectratio]{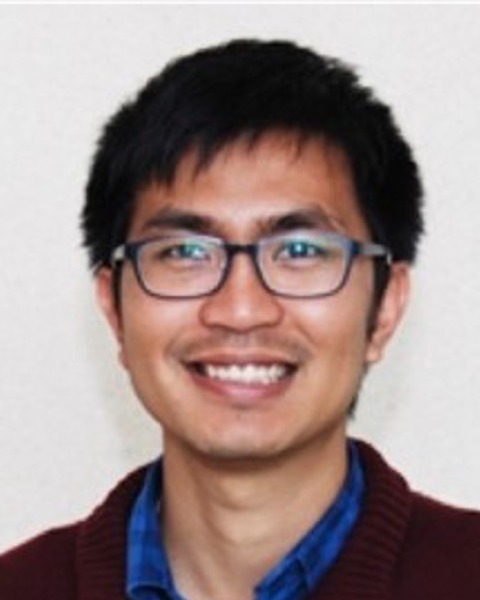}}]{Photchara Ratsamee}
received the M.E. and Ph.D. degrees from the Graduate School of Engineering Science, the University of Osaka, Japan, in 2012 and 2015, respectively. From 2015 to 2022, he served as an Assistant Professor at the Cybermedia Center, the University of Osaka. He then joined the Faculty of Robotics and Design, Osaka Institute of Technology, as an Associate Professor (Lecturer) from 2022 to 2024. He is currently an Associate Professor at the Faculty of Engineering Science, Department of Mechanical Engineering, Kansai University. His research interests include robot vision, rescue robotics, human--robot interaction, and haptics in mixed reality. He is a member of the Robotics Society of Japan (RSJ) and the IEEE Robotics and Automation Society (RAS).
\end{IEEEbiography}

\begin{IEEEbiography}[{\includegraphics[width=1in,height=1.25in,clip,keepaspectratio]{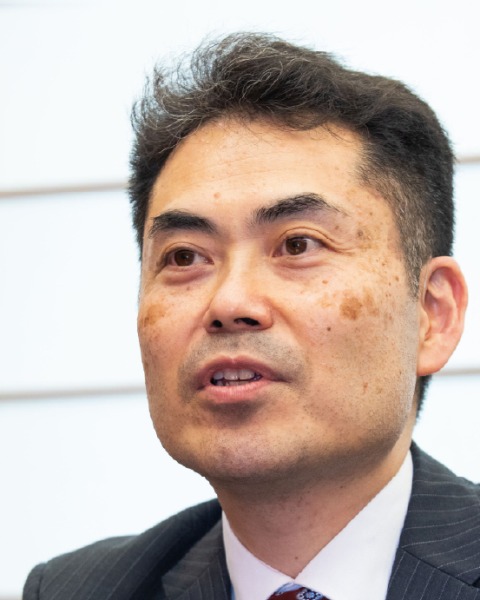}}]{Kiyoshi Kiyokawa}
received his Ph.D. degree in information systems from Nara Institute of Science and Technology (NAIST) in 1998. After working at the National Institute of Information and Communications Technology (NICT) and the University of Osaka, he is currently a Professor at NAIST, leading the Cybernetics and Reality Engineering Laboratory. His research interests include virtual reality, augmented reality, human augmentation, 3D user interfaces, CSCW, and context awareness. He is also a Fellow of the Virtual Reality Society of Japan. He is an associate editor-in-chief of IEEE TVCG and an inductee of the IEEE VGTC Virtual Reality Academy (Inaugural Class).
\end{IEEEbiography}

\clearpage
\setcounter{section}{0}\setcounter{figure}{0}\setcounter{table}{0}\setcounter{equation}{0}
\renewcommand{\thesection}{S\arabic{section}}\renewcommand{\theHsection}{S\arabic{section}}
\renewcommand{\thefigure}{S\arabic{figure}}\renewcommand{\theHfigure}{S\arabic{figure}}
\renewcommand{\thetable}{S\arabic{table}}\renewcommand{\theHtable}{S\arabic{table}}
\renewcommand{\theequation}{S\arabic{equation}}\renewcommand{\theHequation}{S\arabic{equation}}
\twocolumn[{\centering\LARGE\sffamily Supplementary Material\par\vspace{1.5em}}]
This supplementary material accompanies the paper above. Unprefixed section, figure, and table numbers refer to the main paper. Bibliographic citations use the reference list above.

\section{Implementation Details}
\label{S-sec:supp_impl}

\subsection{Prop and Hardware}
\label{S-sec:supp_hw}
\begin{figure}[t]
\centering
\includegraphics[width=\columnwidth]{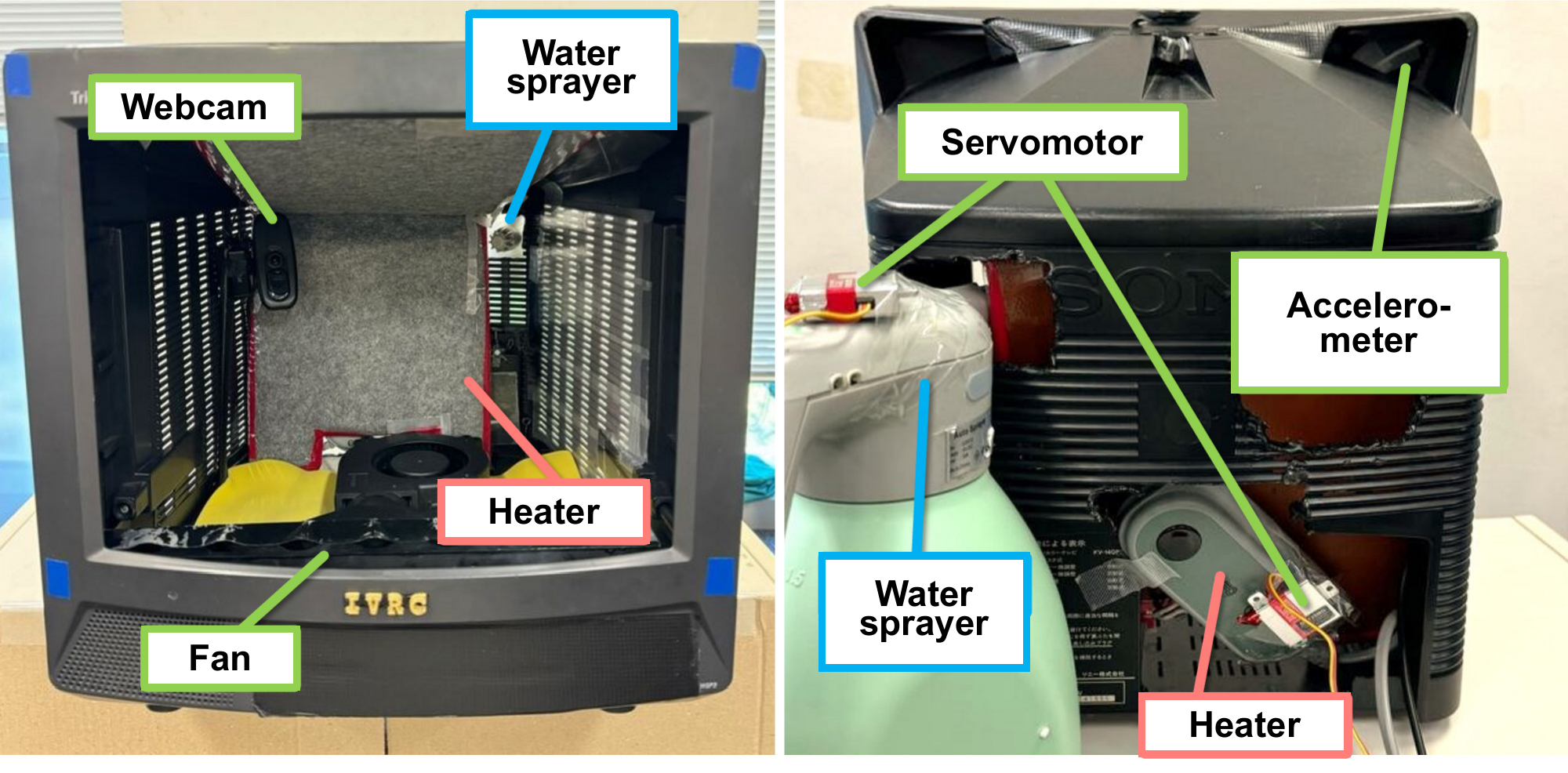}
\caption[Television-like prop]{The television-like prop. Left: front view with the webcam, the nozzle of the water sprayer, the heater, and the fan inside the prop. Right: rear view with the water sprayer, the heater, the servomotors that press their switches, and the accelerometer that detects tapping. Label colors: blue, water; red, heat; green, other components.}
\label{S-fig:tv_device}
\end{figure}

\Cref{S-fig:tv_device} shows the television prop, and \cref{S-table:hardware} lists its components and their roles.

\begin{table}[t]
 \caption[Hardware components]{Hardware Components of \textit{Beyond the TV}}
 \label{S-table:hardware}
 \footnotesize
 \centering
 \begin{tabular}{L{0.16\columnwidth}L{0.31\columnwidth}L{0.38\columnwidth}}
  \toprule
  Component & Device and control & Role and timing \\
  \midrule
  Tap sensor & Accelerometer of a Nintendo Joy-Con~\cite{nintendo_joycon_support}, mounted beneath the tapping position & Detects taps on the prop; triggers channel switching \\
  Water sprayer & Mist sprayer; power button pressed by a servomotor (Arduino) & Sprays the hand inside the prop during the ``Rain CH''; toggled periodically, off when the channel changes \\
  Heater & Heater switched by a servomotor (Arduino) & Warms the interior of the prop during the ``Warm Sunlight CH'' \\
  Fan & Fan near the screen opening (Arduino) & Blows air outward at the screen boundary; on from the start; in CT, switches off when the flower is drawn into the screen (Proc.~9) \\
  Webcam & Camera inside the prop & Records the participant in Proc.~5--6; footage replayed in the VR phase \\
  HMD & Meta Quest 3 with Elite strap, tethered via Quest Link & Presents passthrough and rendered imagery; application runs on the PC \\
  \bottomrule
 \end{tabular}
\end{table}

The sprayer, heater, and fan render the content of the channels on the participant's hand. The sprayer is toggled on and off during the ``Rain CH'' to approximate the intermittent feel of rain, and a waterproof tray and absorbent pads beneath it collect the water. The fan is intended to make the border between the two worlds perceptible: when participants move a hand from outside to inside the prop, they feel the airflow at the screen opening. In CT, the fan is switched off when the flower is drawn into the television during the transition, about 18~s before the participant is swallowed and the VR phase begins, so that the end of the airflow marks the closing of the border between the two worlds. The webcam footage is displayed on the in-VR screen in Proc.~12, allowing participants to see themselves from inside the television; the captured images are deleted automatically when the experience ends.

\subsection{Software}
\label{S-sec:supp_sw}
\subsubsection{Hand Tracking}
Interactions are based on the hand-tracking capability of the Meta Quest 3, using its pinch-gesture recognition: contact between the index finger and thumb is detected as a ``pinch'' and used as the input for object manipulation. To make grasping feel natural under pinch-based interaction, the graspable virtual object (the cup holding the flower) was designed as a mug-shaped cup with a handle rather than a simple cylinder, so that the pinch pose corresponds to a plausible grip.

\subsubsection{Video Passthrough}
The Meta Quest 3 displays the room via real-time color passthrough and overlays virtual elements with low latency. The AR phase superimposes the virtual television content on this passthrough imagery, and the AV phase progressively transforms it as described below.

\subsubsection{2D Projection inside the TV Screen}
\begin{figure}[t]
\centering
\includegraphics[width=0.9\columnwidth]{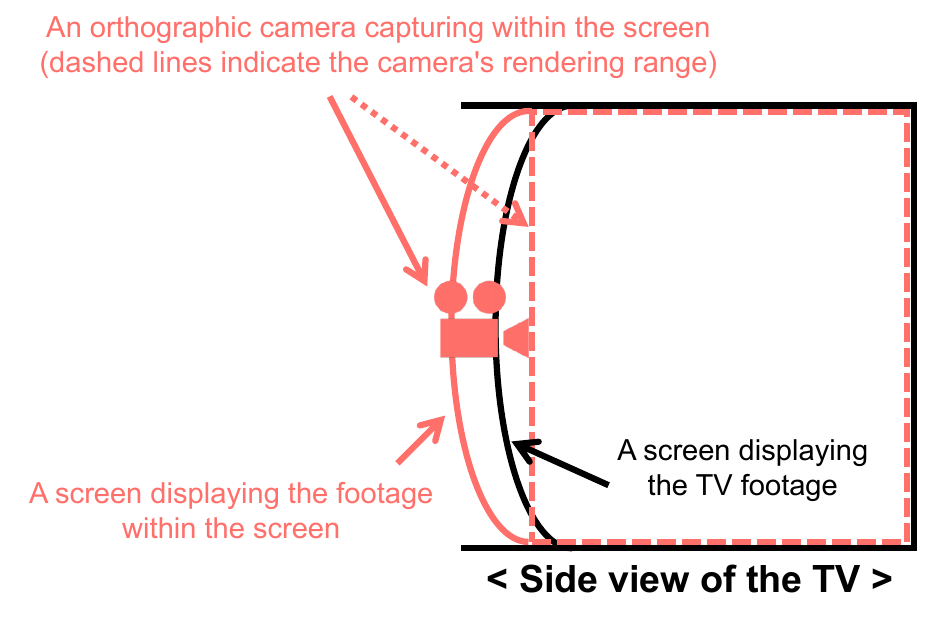}
\caption[Rendering method inside the TV screen]{Rendering method inside the TV screen (side view). An orthographic camera captures the objects inside the screen, and the captured footage is displayed on the screen plane as 2D TV footage.}
\label{S-fig:projection}
\end{figure}

To make the content shown on the virtual television read as ``TV footage,'' we render virtual objects differently depending on whether they are outside or inside the screen (\cref{S-fig:projection}). Outside the screen, objects are rendered in 3D with a perspective camera; once an object enters the screen, it is captured by an orthographic camera and rendered as 2D footage projected onto the screen plane. Specific objects such as the hand model and the flower are chroma-key composited in front of the playing channel footage so that their spatial relationships and motion appear natural within the 2D image. In addition, a VHS-style retro filter is applied to both the screen and its footage, which unifies the look of the imagery and makes the switch between the 3D exterior and the 2D interior feel smooth and coherent.

\subsubsection{Calibration}
\begin{figure}[t]
\centering
\includegraphics[width=\columnwidth]{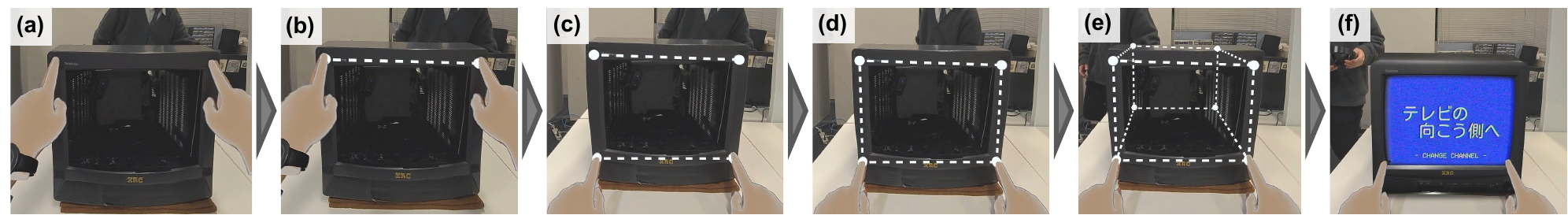}
\caption[Calibration procedure]{Calibration procedure. (a)--(d) The participant points at the four corners of the real television. (e) The four rear corners are placed a distance $d$ behind the front corners along the screen normal. (f) The virtual television is aligned with the real prop.}
\label{S-fig:calibration}
\end{figure}

A short calibration aligns the virtual television with the real prop: the participant points at the four corners of the real television, and the recorded coordinates are used to reproduce the television's position and orientation in the virtual environment (\cref{S-fig:calibration}). The computation proceeds as follows.

Let $\mathbf{P_0}$ (top-left), $\mathbf{P_1}$ (top-right), $\mathbf{P_2}$ (bottom-left), and $\mathbf{P_3}$ (bottom-right) be the four pointed coordinates [\cref{S-fig:calibration}(a)--(d)]. The normal vector $\mathbf{n}$ of the screen plane is computed from $\mathbf{v_1} = \mathbf{P_1} - \mathbf{P_0}$ and $\mathbf{v_2} = \mathbf{P_2} - \mathbf{P_0}$ as
\begin{equation}
\mathbf{n} = \frac{\mathbf{v_1} \times \mathbf{v_2}}{\| \mathbf{v_1} \times \mathbf{v_2} \|}.
\end{equation}
The center position of the television is obtained as $\mathbf{C} = (\mathbf{P_0}+\mathbf{P_1}+\mathbf{P_2}+\mathbf{P_3})/4$, with offsets added as needed to adjust height and depth in the virtual space. The orientation is determined as a quaternion $\mathbf{Q}$ from the normal vector with the up vector $\mathbf{u} = (0,1,0)$, using Unity's \texttt{Quaternion.LookRotation}$(\mathbf{n}, \mathbf{u})$. To reproduce the depth of the television, four additional points are computed as $\mathbf{P_i'} = \mathbf{P_i} + d \cdot \mathbf{n}$ $(i=0,\ldots,3)$ with a depth offset $d$ [\cref{S-fig:calibration}(e)]. Finally, the computed center $\mathbf{C}$, rotation $\mathbf{Q}$, and an appropriate scale are applied to the virtual television model so that it appears aligned with the real prop [\cref{S-fig:calibration}(f)].

\subsection{Flower Growth}
\label{S-sec:supp_flower}
\begin{figure}[t]
\centering
\includegraphics[width=0.9\columnwidth]{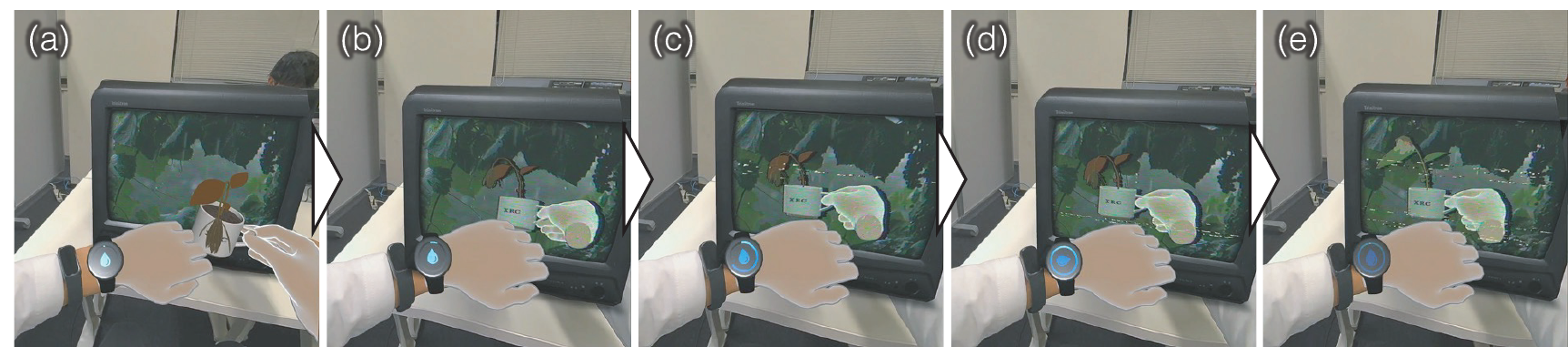}
\caption[Flower growth process]{Flower growth process. (a) The display is switched to the \textit{Rain CH}. (b) The participant inserts the flower into the television. (c) The growth meter on the wristwatch progresses. (d) When the meter becomes full, (e) the flower grows to the next stage.}
\label{S-fig:flowerWatch}
\end{figure}

\Cref{S-fig:flowerWatch} shows how the flower grows in the ``Rain CH'' and the ``Warm Sunlight CH'' (Proc.~7--8).

\section{User Study Details}
\label{S-sec:supp_study}

\subsection{Conditions}
\begin{figure}[t]
\centering
\includegraphics[width=0.9\columnwidth]{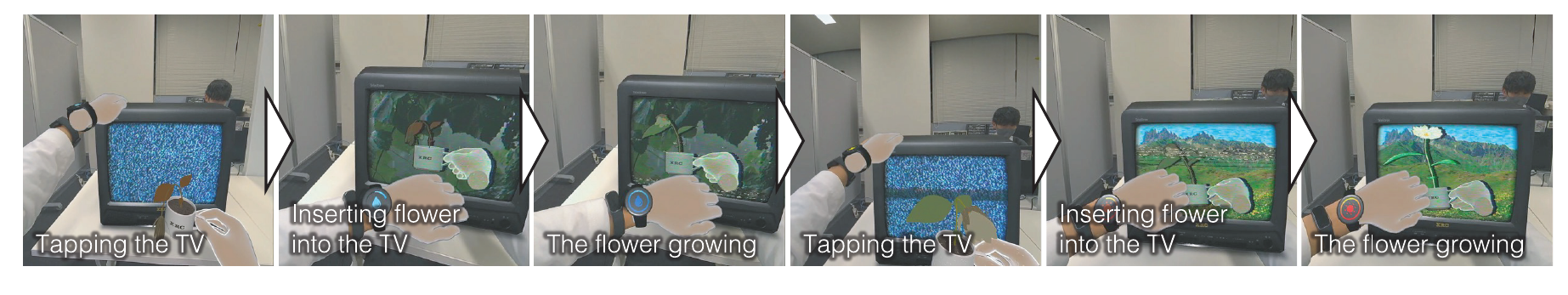}
\caption[AR-phase duration conditions]{AR-phase duration conditions used in the experiment: one set consists of the action sequence performed once on each channel; the set is repeated three times in the long-duration (LD) condition, whereas in the short-duration (SD) condition, the set is performed only once.}
\label{S-fig:Syurui_AR}
\end{figure}

\begin{figure}[t]
\centering
\includegraphics[width=0.9\columnwidth]{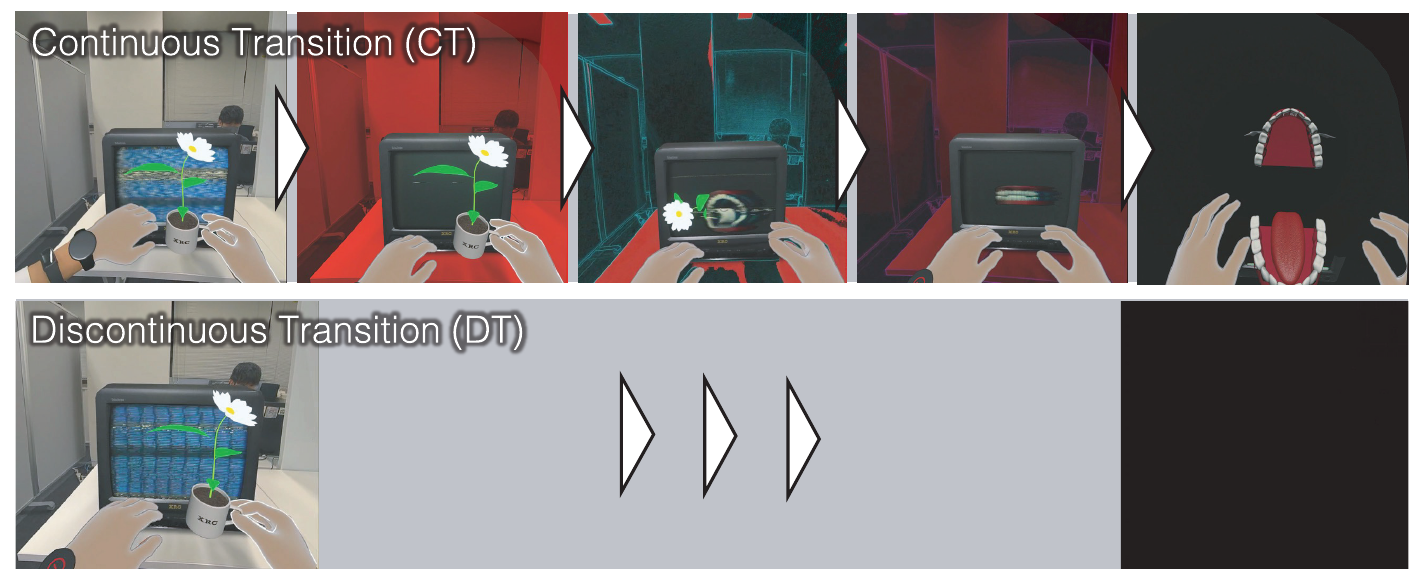}
\caption[Types of AR-to-VR transition technique]{AR-to-VR transition technique conditions used in the experiment: continuous transition (CT) continuously carries over the AR-phase scene into VR (see Section~\ref{BeyondTheTV}), whereas discontinuous transition (DT) does not apply the continuous transition and instead switches abruptly to a black virtual environment.}
\label{S-fig:Syurui_AV}
\end{figure}

\Cref{S-fig:Syurui_AR,S-fig:Syurui_AV} illustrate the levels of AR-phase duration and transition technique (\cref{sec:conditions}). \Cref{S-table:timeline} gives the logged times of events that differ by condition.

\begin{table*}[t]
 \caption[Time course of the conditions]{Time Course of the Events That Differ Between Conditions (System Logs; Times in Seconds, Mean (SD) or Range)}
 \label{S-table:timeline}
 \footnotesize
 \centering
 \begin{tabular}{L{0.30\textwidth}L{0.20\textwidth}L{0.20\textwidth}L{0.20\textwidth}}
  \toprule
  Event & CT (Groups A, B) & DT (Groups C, D) & NI (Group E; CT) \\
  \midrule
  AR phase (from the appearance of the virtual television) & LD 159 (21), SD 72 (17) & LD 159 (24), SD 77 (15) & 175 (4); operations by a virtual hand \\
  End of the AR phase ($t = 0$) & Tap turns the screen black; continuous transition starts & Tap; cut to a black virtual environment & As CT \\
  Flower drawn into the television; fan switched off & $t \approx 9$ & Does not occur; no fan-off command sent & As CT \\
  Teeth, darkening, swallowing (\cref{fig:transitionFlow}(e)--(j)) & $t \approx 9$--27 & Does not occur; black screen & As CT \\
  Start of the VR phase (Proc.~10) & $t = 27$ & $t = 8$--9 & $t = 27$ \\
  Fan at VR onset & Off & On & Off \\
  VR phase (Proc.~10--13) & 101 (17) in A, 92 (10) in B; escape by the participant & 100 (18) in C, 100 (20) in D; escape by the participant & 92 (constant); escape triggered automatically \\
  Data collection & Sept.--Oct.\ 2024 & Sept.--Oct.\ 2024 & Late Nov.\ 2024, different room \\
  \bottomrule
 \end{tabular}
\end{table*}

\subsection{Participants and Environment}
\begin{figure}[t]
\centering
\includegraphics[width=0.65\columnwidth]{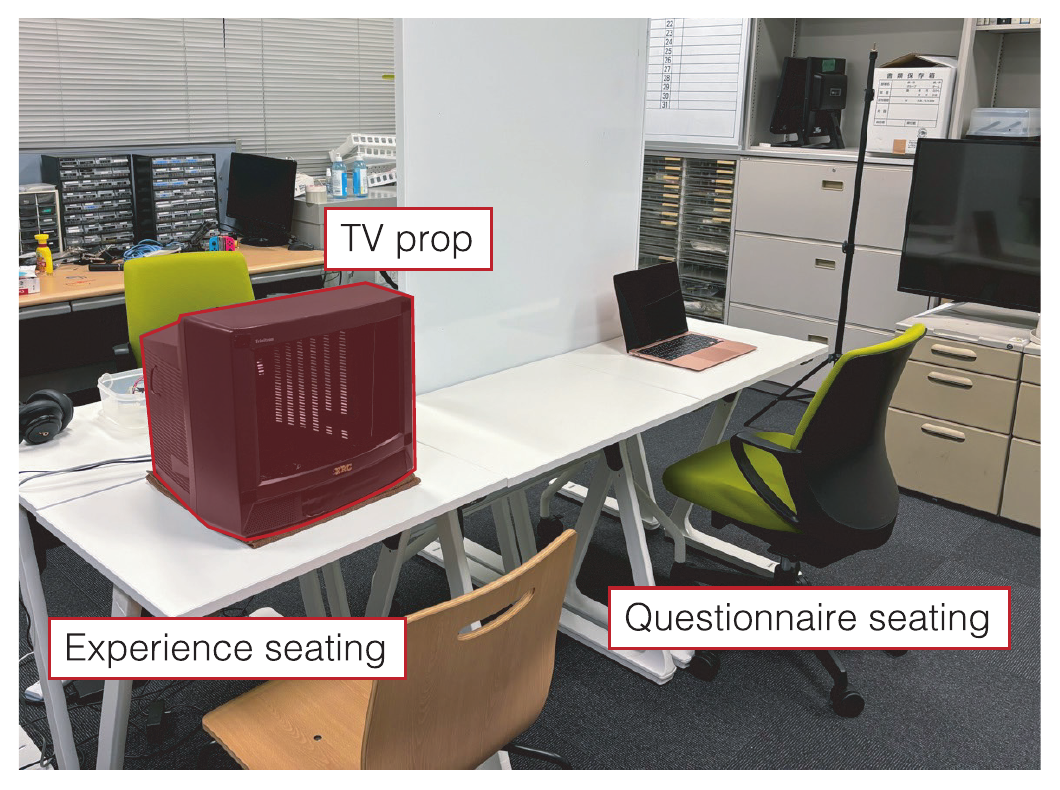}
\caption[Experimental environment]{Experimental environment. The experience seat faces the television prop; the questionnaires were answered at the seat next to it.}
\label{S-fig:ex_environment}
\end{figure}

\Cref{S-table:hikensya} gives the demographics per group, and \cref{S-fig:ex_environment} shows the experimental environment.

\begin{table}[t]
 \caption[Participants in each group]{Participants in Each Group. VR Use: Never / A Few Times a Year / A Few Times a Month or More}
 \label{S-table:hikensya}
 \footnotesize
 \centering
 \setlength{\tabcolsep}{3pt}
 \begin{tabular}{cccccc}
  \toprule
  & N & Males & Females & Age (Range, Mean, SD) & VR use \\
  \midrule
  Group A  & 15 & 6  & 9  & 19--24 (21.3, $SD$ = 1.63) & 8 / 7 / 0 \\
  Group B  & 15 & 9  & 6  & 18--24 (21.2, $SD$ = 2.01) & 7 / 6 / 2 \\
  Group C  & 15 & 10 & 5  & 18--30 (21.8, $SD$ = 3.49) & 8 / 7 / 0 \\
  Group D  & 15 & 7  & 8  & 18--26 (20.9, $SD$ = 2.26) & 7 / 8 / 0 \\
  Group E  & 15 & 9  & 6  & 18--23 (20.3, $SD$ = 1.94) & 7 / 7 / 1 \\
  \midrule
  Total       & 75 & 41 & 34 & 18--30 (21.1, $SD$ = 2.35) & 37 / 35 / 3 \\
  \bottomrule
 \end{tabular}
\end{table}

\subsection{Questionnaire Items and Analyses}
\Cref{S-table:ipq_items,S-table:ipq_items_ja} list the IPQ items that were administered, in English and in the Japanese wording used, \cref{S-table:After_question} lists the additional post-experience questions, and \cref{S-table:operationalization} lists the groups, measures, and tests for each hypothesis.

\begin{table*}[t]
 \caption[IPQ items administered]{IPQ Items Administered (11 of 14). The items were administered in Japanese (\cref{S-table:ipq_items_ja}); the published English wording and anchors are shown here}
 \label{S-table:ipq_items}
 \footnotesize
 \centering
 \begin{tabular}{lL{0.45\textwidth}L{0.30\textwidth}c}
  \toprule
  Item & Question (published English wording) & Anchors (1 / 7) & Reversed \\
  \midrule
  G1 & In the computer generated world I had a sense of ``being there.'' & not at all / very much & \\
  SP1 & Somehow I felt that the virtual world surrounded me. & fully disagree / fully agree & \\
  SP2 & I felt like I was just perceiving pictures. & fully disagree / fully agree & yes \\
  SP3 & I did not feel present in the virtual space. & did not feel / felt present & \\
  SP4 & I had a sense of acting in the virtual space, rather than operating something from outside. & fully disagree / fully agree & \\
  SP5 & I felt present in the virtual space. & fully disagree / fully agree & \\
  INV2 & I was not aware of my real environment. & fully disagree / fully agree & \\
  INV3 & I still paid attention to the real environment. & fully disagree / fully agree & yes \\
  INV4 & I was completely captivated by the virtual world. & fully disagree / fully agree & \\
  REAL1 & How real did the virtual world seem to you? & completely real / not real at all & yes \\
  REAL3 & How real did the virtual world seem to you? & about as real as an imagined world / indistinguishable from the real world & \\
  \midrule
  \multicolumn{4}{@{}p{0.95\textwidth}@{}}{Item sources as listed by the IPQ database~\cite{IPQ_database}: G1, Slater and Usoh (1994); REAL1, Hendrix (1994); REAL3, Carlin, Hoffman, and Weghorst (1997). The database credits Thomas Schubert, Holger Regenbrecht, and Frank Friedmann for SP1, SP2, SP4, SP5, and INV2--INV4. It does not identify a source for SP3.} \\
  \multicolumn{4}{@{}p{0.95\textwidth}@{}}{Not administered: INV1 (awareness of the real world while navigating; participants did not move through the virtual world), REAL2 and REAL4 (judged not to apply to the content). Reversed items were scored as $8 - x$. For this study, we averaged G and the three subscale means to obtain the overall score. In Japanese, REAL1 and REAL3 differ in one particle of the question and in their anchors (\cref{S-table:ipq_items_ja}).} \\
  \bottomrule
 \end{tabular}
\end{table*}

\begin{table*}[t]
 \caption[IPQ items in Japanese]{IPQ Items as Administered in Japanese (Post-Practice and Post-Experience Questionnaires; 7-Point Scale)}
 \label{S-table:ipq_items_ja}
 \footnotesize
 \centering
 \begin{CJK}{UTF8}{ipxm}
 \begin{tabular}{lL{0.50\textwidth}L{0.37\textwidth}}
  \toprule
  Item & Question & Anchors (1 / 7) \\
  \midrule
  G1 & コンピュータで作られた世界の中で、私はそこにいる感じがした。 & 全くなし / 非常に強い \\
  SP1 & 私は仮想世界に取り囲まれている気がした。 & 全くその通りでない / 全くその通り \\
  SP2 & 私はただ単に映像を見ているような気がした。 & 全くその通りでない / 全くその通り \\
  SP3 & 私は仮想空間にいる気がしなかった。 & 気がしなかった / 気がした \\
  SP4 & 私は何かを外部から操作しているのではなく、仮想空間の中でふるまっているような気がした。 & 全くその通りでない / 全くその通り \\
  SP5 & 私は仮想空間の中に居合わせているように感じた。 & 全くその通りでない / 全くその通り \\
  INV2 & 私は現実世界をもはや意識しなかった。 & 全くその通りでない / 全くその通り \\
  INV3 & 私はいまだ現実環境を注意していた。 & 全くその通りでない / 全くその通り \\
  INV4 & 私は仮想世界によって完全に魅了されていた。 & 全くその通りでない / 全くその通り \\
  REAL1 & あなたには仮想世界がどのくらい現実のように見えましたか？ & 全くの現実のよう / 全くの現実のようでない \\
  REAL3 & あなたの仮想世界がどのくらい現実のように見えましたか？ & 想像された世界のよう / 現実環境と区別ができない \\
  \midrule
  \multicolumn{3}{@{}p{0.95\textwidth}@{}}{Item sources are given in \cref{S-table:ipq_items}. The questions are reproduced from the response records of the online questionnaires. The response records do not contain the anchors, which are taken from the description of the questionnaire in the first author's master's thesis. Compared with the Japanese translation distributed with the IPQ~\cite{IPQ_database}, INV2 reads 現実世界 instead of 現実環境, REAL3 reads あなたの instead of あなたには and has the anchor 現実環境 instead of 現実世界 (in that translation, the questions of REAL1 and REAL3 are identical), and SP4 and INV3 differ only in the kana spelling of one word.} \\
  \bottomrule
 \end{tabular}
 \end{CJK}
\end{table*}

\begin{table*}[t]
 \caption[Post-experience questionnaire items]{Post-Experience Questionnaire Items}
 \label{S-table:After_question}
 \centering
 \begin{tabular}{cp{0.72\textwidth}c}
  \toprule
  Item & Question & Response Format \\
  \midrule
  Q1
  & The practice session for the XR immersion experience was sufficiently conducted so that I did not struggle with the operational procedures during the main experiment.
  & 7-point scale \\
  Q2
  & Compared to the practice session, I felt more immersed in the VR space.
  & 7-point scale \\
  Q3
  & Please explain the reason for your response to the previous question.
  & Open-ended \\
  Q4
  & The heart-rate monitor did not interfere with the XR immersion experience.
  & 7-point scale \\
  Q5
  & Please provide any additional comments or impressions.
  & Open-ended \\
  \bottomrule
 \end{tabular}
\end{table*}

\begin{table*}[t]
 \caption[Operationalization of the hypotheses]{Operationalization of the Hypotheses}
 \label{S-table:operationalization}
 \footnotesize
 \centering
 \begin{tabular}{L{0.08\textwidth}L{0.20\textwidth}L{0.16\textwidth}L{0.18\textwidth}L{0.24\textwidth}}
  \toprule
  Hypothesis & Comparison & Primary measure & Secondary measures & Analysis \\
  \midrule
  H1 & LD vs.\ SD: Groups A, C vs.\ B, D (main effect of duration) & Overall IPQ change score & IPQ components, UES-SF, NASA-TLX, heart rate, Q2 & ART on the $2 \times 2$ design; ANCOVA of post-experience IPQ with post-practice IPQ as covariate; two-way ANOVA for heart rate \\
  H2 & CT vs.\ DT: Groups A, B vs.\ C, D (main effect of transition technique) & Overall IPQ change score & IPQ components, UES-SF, NASA-TLX, heart rate, Q2 & ART on the $2 \times 2$ design with Steel--Dwass post-hoc tests; ANCOVA as for H1; two-way ANOVA for heart rate \\
  H3 & I vs.\ NI: Group A vs.\ Group E (exploratory; \cref{sec:conditions}) & Overall IPQ change score and overall UES-SF score & IPQ components, UES-SF factors, NASA-TLX, heart rate, Q2 & Wilcoxon rank-sum tests; ANCOVA as for H1; Welch's $t$-test for heart rate \\
  \bottomrule
 \end{tabular}
\end{table*}

\subsection{Coding of the Open-Ended Answers}
\Cref{S-table:codes} gives the definition of each theme used in \cref{table:qual} and an example answer, translated from Japanese.

\begin{table*}[t]
 \caption[Themes of the open-ended answers]{Themes of the Open-Ended Answers: Definitions and Examples (Participant Group in Parentheses)}
 \label{S-table:codes}
 \scriptsize
 \centering
 \begin{tabular}{L{0.22\textwidth}L{0.38\textwidth}L{0.32\textwidth}}
  \toprule
  Theme & Definition & Example \\
  \midrule
  \multicolumn{3}{@{}l}{\textit{Q3: reason for the Q2 rating (absorption compared with the practice session)}} \\
  Physical effects felt on the body & Water, mist, rain, wind, or warmth felt on the body. & ``Because I felt the mist and the wind.'' (A) \\
  Story or content & The plot or its development, scary content, or scenes that did not occur in the practice. & ``Because there was a story.'' (E) \\
  \quad Unexpected development (sub-theme) & A development that was not announced beforehand or came suddenly. & ``Something unexpected happened at the end; I was confused and forgot the real world for a moment.'' (A) \\
  Surroundings replaced by the virtual space & Being surrounded or enclosed by the virtual space, or no longer seeing the real room. & ``I could no longer see the surroundings, so I concentrated on the VR imagery.'' (B) \\
  Sound & Headphones blocking ambient sound, music, or narration. & ``Wearing the headphones cut off information from outside.'' (C) \\
  Operations & Knowing how to operate, getting used to the operations, or having more or longer operations. & ``Because I knew exactly how to operate.'' (A) \\
  Absence of own operations & Lower absorption because the participant did not perform the operations. & ``Because I was only watching.'' (E) \\
  Other detracting factors & Lower absorption for other reasons: anticipating events, mismatch between sensations and images, low resolution, thinking about the story or the questionnaire, or operations that did not go smoothly. & ``Operations sometimes did not go smoothly.'' (A) \\
  No difference from practice (not in \cref{table:qual}) & No difference from the practice session, or reasons that fit no other theme. & ``I was already immersed enough in the practice.'' (B) \\
  \midrule
  \multicolumn{3}{@{}l}{\textit{Q5: other comments}} \\
  Enjoyment or positive overall impression & The experience was enjoyable, interesting, or impressive. & ``It was a lot of fun.'' (B) \\
  Positive remark on physical effects & A positive remark on the water, mist, wind, or warmth. & ``The wind while growing the flower felt pleasant.'' (D) \\
  Device or ergonomic problems & Headset weight, posture, or audio volume. & ``The sound from the headphones was a little quiet.'' (D) \\
  Operational difficulty & Grasping the cup, the meter not being shown, the hand passing through objects, or both hands being occupied. & ``The cup at the end was hard to hold.'' (C) \\
  Difficulty understanding & Difficulty understanding the setting, the instructions, or the plot. & ``The setting was a little complicated.'' (C) \\
  Anticipation (not in \cref{table:qual}) & Anticipating the plot from the briefing or the instruction video. & ``From the instruction video, I could tell that the VR experience would try to surprise participants.'' (A) \\
  Insufficient realism (not in \cref{table:qual}) & The content did not look real enough. & ``The virtual space did not seem real; it felt like playing a game with 3D glasses.'' (E) \\
  Other (not in \cref{table:qual}) & Suggestions and remarks on the questionnaires or the study setting. & ``It was my first time in VR and I expected to get sick, but I did not, perhaps because I did not move.'' (E) \\
  \bottomrule
  \multicolumn{3}{@{}p{0.95\textwidth}@{}}{\rule{0pt}{2.6ex}An answer could receive several themes; a participant was counted at most once per theme and question. Reports of a vibration, which the system did not present, were counted under physical effects only when the answer also mentioned water.} \\
 \end{tabular}
\end{table*}

\section{Heart Rate}
\label{S-sec:supp_hr}

\subsection{Preprocessing}
We analyzed heart rate computed from the beat-to-beat (RR) intervals recorded by the sensor. We discarded RR intervals of 300~ms or less and 2000~ms or more. We then excluded intervals that differed by more than 20\% from the centered five-interval rolling median (at least three observations required). Together these steps removed 7.8\% of intervals on average. For each participant, the practice session and the AR, AV, and VR phases were delimited using the event times recorded by the system, and the mean heart rate of each phase was computed as $60000$ divided by the mean RR interval in milliseconds. A phase was treated as missing when the retained intervals covered less than half of its duration. Because the baseline measurement was not time-stamped for Group E, we used the practice session, which took place in VR immediately before the main experience and was identical for all groups, as the reference for within-participant changes. We aligned each participant's heart rate, interpolated at 1~Hz, separately to the end of the AR phase and to VR onset. VR began about 27~s after AR ended in CT and 8--9~s after AR ended in DT. For both alignments, we expressed heart rate relative to the mean over the final 20~s of AR, when the content was the same across conditions.

Mean heart rate per phase was available for all participants except for the practice session of two participants (one each in Groups C and D), the AV phase of two (B, D), and the VR phase of two (B, C). The two-way ANOVA of the VR phase therefore included 58 participants of Groups A--D, and the change from the practice session 57. For the event-locked analysis, the heart rate of each retained beat was interpolated linearly to 1-s bins. Bins lying in a gap of more than 2~s between retained beats were left out, and, as for the phases, a window was treated as missing when fewer than half of its bins remained. This excluded two to four participants of Groups A--D per window (one of them had no retained beat in the 20-s reference period), leaving 57 participants for the first 8~s after the end of the AR phase and 58, 57, and 56 participants for 0--10, 10--30, and 30--60~s after the onset of the VR phase; no participant of Group E was excluded.

The AV phase in these analyses is the interval from the end of the AR phase to the start of Proc.~10. In CT it lasts 27~s and includes the stages of \cref{fig:transitionFlow}(h)--(j), in which the imagery is already fully virtual; in DT it is the black screen after the cut (8--9~s).

The sensor software also computes the index LF/(LF+HF), which was analyzed in the thesis on which this study is based, over a sliding 300-s window. Because this window is longer than every phase of the experience, the index cannot be attributed to individual phases; we therefore omitted it from analysis.

\subsection{Results}
\Cref{S-table:hr} summarizes heart rate per phase and around the transition. Across participants with paired measurements, mean heart rate was lower in AV and VR than in AR (AV minus AR: $-1.6$~bpm, $t(72) = -3.68$, $p < .001$; VR minus AR: $-1.4$~bpm, $t(72) = -2.62$, $p = .011$). Two-way ANOVAs found no significant effect of AR-phase duration, transition technique, or their interaction on mean heart rate in the VR phase (duration: $F(1, 54) = 0.85$, $p = .361$; transition: $F(1, 54) = 0.07$, $p = .786$; interaction: $F(1, 54) = 0.12$, $p = .734$), on its change from the practice session (all $p \geq .14$), or on the VR-phase heart rate adjusted for the practice session by ANCOVA (all $p \geq .13$).

Aligned to the events of the transition, heart rate fell in all groups during the first 8~s after the end of the AR phase, most strongly in Group D (transition: $p = .130$; interaction: $p = .060$). By the onset of the VR phase it had returned to or above the pre-transition level in the CT groups, whereas in the DT groups, after the cut to black, it remained lower. During the first 10~s of the VR phase, heart rate relative to the pre-transition period was higher after CT than after DT ($F(1, 54) = 8.26$, $p = .006$, $\eta_p^2 = .133$; ART: $p < .001$), with no significant effect of duration ($p = .531$) and an interaction reflecting the low values in Group D ($p = .050$). The main effect of transition technique was not significant in the later VR windows (10--30~s: $p = .221$; 30--60~s: $p = .602$), whereas heart rate in Group D remained below the pre-transition level (interaction: $p = .042$ and $p = .047$). The heart-rate changes (AV and VR phases minus AR phase, VR phase minus practice session, and the two event-locked values) did not correlate with the IPQ change score, the post-experience IPQ score, or the UES-SF overall score, either across Groups A--E (Spearman, 72--73 participants per correlation, all $|\rho| \leq .19$, all $p \geq .10$) or within Groups A--D (57--58 participants per correlation, all $|\rho| \leq .21$, all $p \geq .11$).

\begin{table}[t]
 \caption[Heart rate]{Heart Rate (bpm) per Phase, Mean (SD), and Event-Locked Change (bpm, Mean)}
 \label{S-table:hr}
 \scriptsize
 \centering
 \setlength{\tabcolsep}{2pt}
 \begin{tabular}{lccccc}
  \toprule
  & A & B & C & D & E \\
  \midrule
  Practice & 76.4 (10.6) & 78.5 (14.3) & 75.0 (12.4) & 81.4 (8.3) & 75.6 (14.6) \\
  AR phase & 76.1 (10.2) & 78.1 (14.9) & 74.5 (12.7) & 82.3 (9.4) & 76.8 (13.8) \\
  AV phase & 75.1 (11.3) & 76.2 (14.7) & 72.5 (13.6) & 76.8 (9.8) & 77.5 (17.4) \\
  VR phase & 74.1 (9.6) & 76.0 (16.4) & 73.8 (14.8) & 78.0 (8.2) & 77.5 (16.8) \\
  \midrule
  End of AR, 0--8~s$^{a}$ & $-2.2$ & $-0.5$ & $-1.6$ & $-6.0$ & $-2.6$ \\
  VR onset, 0--10~s$^{a}$ & $+0.5$ & $+3.5$ & $-1.4$ & $-7.2$ & $+2.7$ \\
  \bottomrule
  \multicolumn{6}{@{}l}{\rule{0pt}{2.6ex}$^{a}$Relative to the mean of the 20~s before the end of the AR phase.} \\
 \end{tabular}
\end{table}

\begin{table}[t]
 \caption[Post-practice and post-experience IPQ scores]{Post-Practice (Pre) and Post-Experience (Post) IPQ Scores and Additional Post-Experience Measures, Mean (SD)}
 \label{S-table:ipq_pre_post}
 \footnotesize
 \centering
 \setlength{\tabcolsep}{2.4pt}
 \begin{tabular}{llccccc}
  \toprule
  & & A & B & C & D & E \\
  \midrule
  ALL & Pre  & 4.60 (0.94) & 4.70 (0.52) & 4.84 (0.85) & 4.97 (0.64) & 4.57 (0.59) \\
      & Post & 5.03 (0.89) & 5.08 (0.87) & 4.69 (0.97) & 5.03 (0.83) & 4.80 (0.92) \\
  G   & Pre  & 4.80 (1.32) & 4.73 (1.44) & 5.73 (1.33) & 5.60 (0.99) & 4.87 (0.99) \\
      & Post & 5.53 (1.19) & 5.27 (1.49) & 5.13 (1.41) & 4.80 (1.52) & 5.07 (1.33) \\
  SP  & Pre  & 5.16 (0.87) & 4.80 (0.60) & 4.95 (0.84) & 5.23 (0.81) & 4.92 (0.59) \\
      & Post & 5.35 (1.00) & 5.47 (1.00) & 4.93 (1.33) & 5.63 (0.98) & 4.77 (1.03) \\
  INV & Pre  & 4.40 (1.27) & 4.84 (1.09) & 4.80 (1.38) & 4.60 (1.04) & 4.29 (1.05) \\
      & Post & 5.09 (1.04) & 5.13 (1.30) & 4.84 (1.19) & 5.16 (1.19) & 5.18 (1.14) \\
  REAL& Pre  & 4.03 (1.04) & 4.43 (0.75) & 3.87 (1.32) & 4.47 (0.69) & 4.20 (0.96) \\
      & Post & 4.13 (1.30) & 4.43 (0.94) & 3.83 (0.94) & 4.53 (1.22) & 4.20 (1.01) \\
  \midrule
  \multicolumn{2}{l}{Q1} & 6.27 (1.03) & 6.40 (0.91) & 6.00 (0.85) & 5.80 (1.21) & 6.67 (0.49) \\
  \multicolumn{2}{l}{Q2} & 6.53 (1.06) & 6.33 (0.90) & 5.73 (1.58) & 6.00 (1.31) & 5.60 (1.99) \\
  \multicolumn{2}{l}{Q4} & 6.33 (0.72) & 5.47 (1.41) & 5.80 (1.61) & 5.93 (1.53) & 5.80 (1.52) \\
  \multicolumn{2}{l}{NASA-TLX} & 42.0 (13.5) & 36.1 (17.2) & 43.5 (17.4) & 44.0 (16.7) & 37.4 (23.0) \\
  \bottomrule
 \end{tabular}
\end{table}

\begin{table*}[p]
 \caption[Questionnaire tests, Groups A--D (ART)]{Questionnaire Tests for Groups A--D: Aligned Rank Transform ANOVA ($N = 60$; $F(1, 56)$). Values in parentheses are Holm-adjusted $p$ within the four IPQ component scores of the same score type, or within the four UES-SF factors}
 \label{S-table:tests_art}
 \scriptsize
 \centering
 \setlength{\tabcolsep}{3pt}
 \begin{tabular}{lccccccccc}
  \toprule
  & \multicolumn{3}{c}{Duration (LD/SD)} & \multicolumn{3}{c}{Transition (CT/DT)} & \multicolumn{3}{c}{Interaction} \\
  \cmidrule(lr){2-4}\cmidrule(lr){5-7}\cmidrule(lr){8-10}
  Measure & $F$ & $p$ & $\eta_p^2$ & $F$ & $p$ & $\eta_p^2$ & $F$ & $p$ & $\eta_p^2$ \\
  \midrule
IPQ change: ALL & 0.35 & .555 & 0.006 & 8.46 & .005 & 0.131 & 1.26 & .266 & 0.022 \\
IPQ change: G & 0.14 & .714 ($>$.99) & 0.002 & 12.16 & $<$.001 (.004) & 0.178 & 0.29 & .590 ($>$.99) & 0.005 \\
IPQ change: SP & 2.39 & .127 (.510) & 0.041 & 0.90 & .347 (.695) & 0.016 & 0.00 & .966 ($>$.99) & 0.000 \\
IPQ change: INV & 0.00 & .988 ($>$.99) & 0.000 & 2.16 & .147 (.441) & 0.037 & 3.26 & .076 (.305) & 0.055 \\
IPQ change: REAL & 0.00 & .965 ($>$.99) & 0.000 & 0.02 & .902 (.902) & 0.000 & 0.13 & .716 ($>$.99) & 0.002 \\
IPQ post-experience: ALL & 0.46 & .499 & 0.008 & 0.40 & .527 & 0.007 & 0.13 & .720 & 0.002 \\
IPQ post-experience: G & 0.55 & .460 (.921) & 0.010 & 1.22 & .275 ($>$.99) & 0.021 & 0.04 & .852 ($>$.99) & 0.001 \\
IPQ post-experience: SP & 2.07 & .155 (.466) & 0.036 & 0.19 & .666 ($>$.99) & 0.003 & 0.59 & .446 ($>$.99) & 0.010 \\
IPQ post-experience: INV & 0.33 & .565 (.921) & 0.006 & 0.12 & .730 ($>$.99) & 0.002 & 0.04 & .852 ($>$.99) & 0.001 \\
IPQ post-experience: REAL & 3.00 & .089 (.355) & 0.051 & 0.04 & .835 ($>$.99) & 0.001 & 0.51 & .478 ($>$.99) & 0.009 \\
IPQ post-practice: ALL & 0.30 & .585 & 0.005 & 1.58 & .214 & 0.027 & 0.00 & .954 & 0.000 \\
IPQ post-practice: G & 0.14 & .705 ($>$.99) & 0.003 & 7.17 & .010 (.039) & 0.114 & 0.10 & .749 ($>$.99) & 0.002 \\
IPQ post-practice: SP & 0.03 & .874 ($>$.99) & 0.000 & 0.32 & .575 ($>$.99) & 0.006 & 1.23 & .273 ($>$.99) & 0.021 \\
IPQ post-practice: INV & 0.08 & .774 ($>$.99) & 0.001 & 0.01 & .909 ($>$.99) & 0.000 & 1.23 & .273 ($>$.99) & 0.021 \\
IPQ post-practice: REAL & 3.88 & .054 (.215) & 0.065 & 0.05 & .816 ($>$.99) & 0.001 & 0.06 & .805 ($>$.99) & 0.001 \\
UES-SF: ALL & 0.71 & .404 & 0.012 & 0.29 & .595 & 0.005 & 0.35 & .556 & 0.006 \\
UES-SF: FA & 0.93 & .338 ($>$.99) & 0.016 & 0.32 & .573 ($>$.99) & 0.006 & 0.49 & .488 ($>$.99) & 0.009 \\
UES-SF: PU & 0.04 & .851 ($>$.99) & 0.001 & 0.14 & .707 ($>$.99) & 0.003 & 0.14 & .707 ($>$.99) & 0.003 \\
UES-SF: AE & 3.28 & .075 (.302) & 0.055 & 0.21 & .645 ($>$.99) & 0.004 & 1.31 & .258 ($>$.99) & 0.023 \\
UES-SF: RW & 0.24 & .624 ($>$.99) & 0.004 & 0.02 & .897 ($>$.99) & 0.000 & 0.75 & .391 ($>$.99) & 0.013 \\
Q2 & 0.12 & .728 & 0.002 & 3.67 & .061 & 0.061 & 2.08 & .155 & 0.036 \\
NASA-TLX overall & 0.46 & .499 & 0.008 & 1.16 & .286 & 0.020 & 0.35 & .556 & 0.006 \\
Q1 & 0.00 & .977 & 0.000 & 3.65 & .061 & 0.061 & 0.30 & .588 & 0.005 \\
Q4 & 0.84 & .363 & 0.015 & 0.88 & .352 & 0.015 & 1.73 & .194 & 0.030 \\
  \bottomrule
 \end{tabular}
\end{table*}

\begin{table*}[t]
 \caption[Questionnaire tests, Groups A--D (ANCOVA)]{ANCOVA of Post-Experience IPQ Scores with the Post-Practice Score as Covariate, Groups A--D ($F(1, 55)$), and Rank-Based ANCOVA (Quade). Parentheses give Holm-adjusted $p$ across the four components for each effect and method}
 \label{S-table:tests_ancova}
 \scriptsize
 \centering
 \setlength{\tabcolsep}{3pt}
 \begin{tabular}{lcccccccccccc}
  \toprule
  & \multicolumn{4}{c}{Duration (LD/SD)} & \multicolumn{4}{c}{Transition (CT/DT)} & \multicolumn{4}{c}{Interaction} \\
  \cmidrule(lr){2-5}\cmidrule(lr){6-9}\cmidrule(lr){10-13}
  IPQ & $F$ & $p$ & $\eta_p^2$ & Quade $p$ & $F$ & $p$ & $\eta_p^2$ & Quade $p$ & $F$ & $p$ & $\eta_p^2$ & Quade $p$ \\
  \midrule
ALL & 0.35 & .556 & 0.006 & .584 & 4.03 & .0496 & 0.068 & .140 & 0.54 & .467 & 0.010 & .690 \\
G & 0.59 & .445 ($>$.99) & 0.011 & .560 ($>$.99) & 3.11 & .083 (.333) & 0.054 & .060 (.239) & 0.00 & .945 ($>$.99) & 0.000 & .955 ($>$.99) \\
SP & 3.56 & .065 (.258) & 0.061 & .064 (.257) & 0.82 & .370 ($>$.99) & 0.015 & .371 ($>$.99) & 0.02 & .882 ($>$.99) & 0.000 & .834 ($>$.99) \\
INV & 0.19 & .662 ($>$.99) & 0.003 & .591 ($>$.99) & 0.32 & .575 ($>$.99) & 0.006 & .628 ($>$.99) & 1.20 & .278 ($>$.99) & 0.021 & .396 ($>$.99) \\
REAL & 0.69 & .409 ($>$.99) & 0.012 & .361 ($>$.99) & 0.06 & .805 ($>$.99) & 0.001 & .603 ($>$.99) & 0.33 & .570 ($>$.99) & 0.006 & .437 ($>$.99) \\
  \bottomrule
 \end{tabular}
\end{table*}

\begin{table*}[t]
 \caption[Questionnaire tests, Groups A and E]{Questionnaire Tests for Groups A and E: Wilcoxon Rank-Sum Tests (Tie-Corrected Normal Approximation), Effect Size $r = |Z|/\sqrt{N}$, and Mean Difference A$-$E with Welch 95\% CI. $Z > 0$: higher in Group A. Holm: adjusted within the four IPQ component scores of the same score type, or within the four UES-SF factors. The CI refers to the difference in means and the test to ranks, so the two can disagree near the significance threshold}
 \label{S-table:tests_ae}
 \scriptsize
 \centering
 \setlength{\tabcolsep}{4pt}
 \begin{tabular}{lccccccc}
  \toprule
  Measure & $n$ (A/E) & $U$ & $Z$ & $p$ & $p_\mathrm{Holm}$ & $r$ & Mean difference [95\% CI] \\
  \midrule
IPQ change: ALL & 15/15 & 128.5 & 0.66 & .507 & -- & 0.121 & +0.19 [-0.46, +0.84] \\
IPQ change: G & 15/15 & 137.5 & 1.08 & .280 & $>$.99 & 0.197 & +0.53 [-0.38, +1.45] \\
IPQ change: SP & 15/15 & 138.0 & 1.06 & .288 & $>$.99 & 0.194 & +0.33 [-0.43, +1.09] \\
IPQ change: INV & 15/15 & 104.5 & -0.33 & .738 & $>$.99 & 0.061 & -0.20 [-1.08, +0.68] \\
IPQ change: REAL & 15/15 & 108.0 & -0.19 & .850 & $>$.99 & 0.034 & +0.10 [-0.79, +0.99] \\
IPQ post-experience: ALL & 15/15 & 123.0 & 0.44 & .663 & -- & 0.080 & +0.22 [-0.46, +0.90] \\
IPQ post-experience: G & 15/15 & 131.0 & 0.79 & .428 & $>$.99 & 0.145 & +0.47 [-0.48, +1.41] \\
IPQ post-experience: SP & 15/15 & 147.5 & 1.46 & .146 & .582 & 0.266 & +0.57 [-0.19, +1.33] \\
IPQ post-experience: INV & 15/15 & 104.0 & -0.36 & .722 & $>$.99 & 0.065 & -0.09 [-0.91, +0.73] \\
IPQ post-experience: REAL & 15/15 & 110.5 & -0.09 & .932 & $>$.99 & 0.016 & -0.07 [-0.94, +0.81] \\
UES-SF: ALL & 15/15 & 152.5 & 1.67 & .095 & -- & 0.305 & +0.31 [-0.04, +0.66] \\
UES-SF: FA & 15/15 & 135.5 & 0.98 & .327 & .653 & 0.179 & +0.42 [-0.14, +0.99] \\
UES-SF: PU & 15/15 & 98.0 & -0.62 & .538 & .653 & 0.112 & -0.09 [-0.54, +0.36] \\
UES-SF: AE & 15/15 & 155.0 & 1.80 & .072 & .288 & 0.328 & +0.47 [-0.01, +0.95] \\
UES-SF: RW & 15/15 & 154.0 & 1.76 & .078 & .288 & 0.322 & +0.44 [+0.01, +0.88] \\
NASA-TLX overall & 15/15 & 125.5 & 0.54 & .590 & -- & 0.098 & +4.56 [-9.72, +18.84] \\
Q1 & 15/15 & 87.5 & -1.19 & .233 & -- & 0.218 & -0.40 [-1.02, +0.22] \\
Q2 & 15/15 & 147.0 & 1.63 & .103 & -- & 0.297 & +0.93 [-0.28, +2.14] \\
Q4 & 15/15 & 129.0 & 0.73 & .463 & -- & 0.134 & +0.53 [-0.37, +1.44] \\
  \bottomrule
 \end{tabular}
\end{table*}

Groups A and E did not differ in mean heart rate in any phase (VR phase: $t = -0.68$, $p = .505$; Welch's $t$-test) or in the event-locked response at VR onset ($p = .462$). Relative to the practice session, heart rate in the VR phase decreased in Group A ($-2.3$~bpm, $SD = 2.9$) and increased slightly in Group E ($+1.9$~bpm, $SD = 7.9$); this difference was at trend level in Welch's $t$-test ($p = .068$) and significant in a Wilcoxon rank-sum test ($p = .025$).

\section{Questionnaire Test Statistics}
\label{S-sec:supp_tests}
\Cref{S-table:tests_art,S-table:tests_ancova,S-table:tests_ae} give the ART and ANCOVA results for Groups A--D and the Wilcoxon rank-sum results for Groups A and E. The main paper reports the Steel--Dwass comparisons and the A--E ANCOVA (\cref{sec:results}). NASA-TLX subscales were not tested. For the overall IPQ score and G, we compared ANCOVA models with common and group-specific slopes across Groups A--D (partial $F$ test, 3 and 52 degrees of freedom). For Quade's rank-based ANCOVA, the post-experience and post-practice scores were each ranked across the 60 participants, the ranks of the former were regressed on those of the latter, and the residuals were analyzed with a $2 \times 2$ ANOVA (sum-to-zero coding, Type III sums of squares). The standardized difference $d$ in \cref{sec:results_robust} is the adjusted difference divided by the pooled within-cell SD of the post-experience score. The heart-rate tests are given in \cref{S-sec:supp_hr}.

\section{Additional Results}
\label{S-sec:supp_results}
\begin{figure}[t]
\centering
\subfloat[Overall score]{\includegraphics[width=0.65\columnwidth]{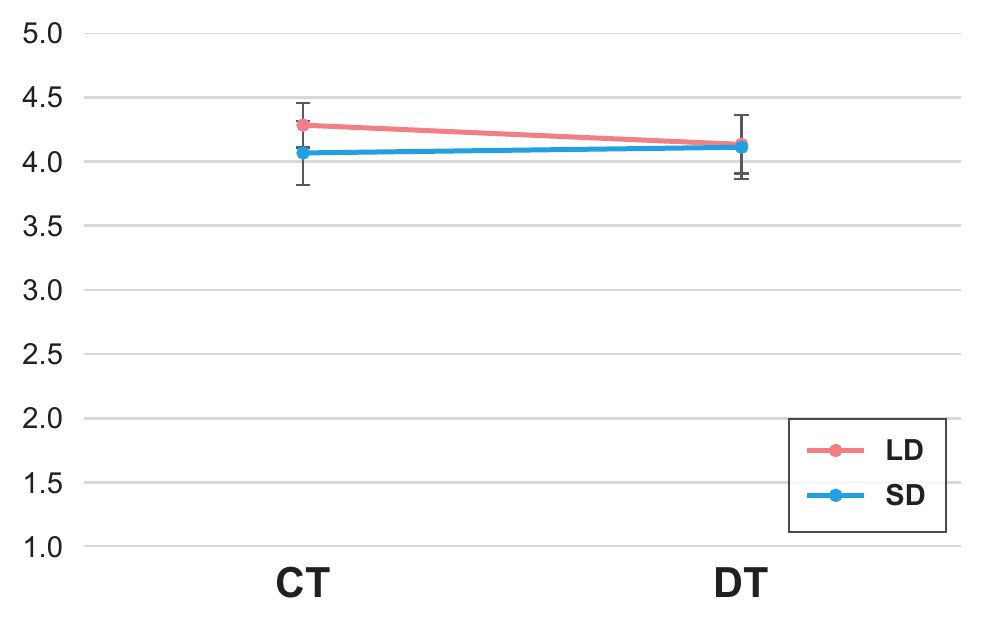}}

\subfloat[FA]{\includegraphics[width=0.48\columnwidth]{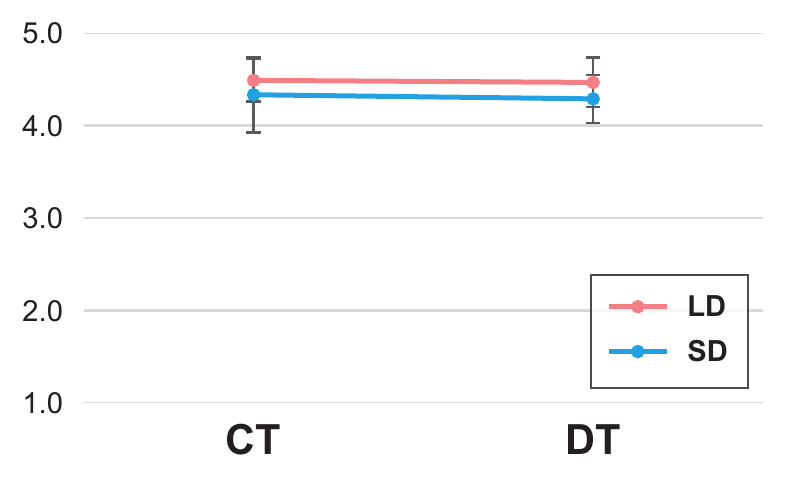}}
\hfil
\subfloat[PU]{\includegraphics[width=0.48\columnwidth]{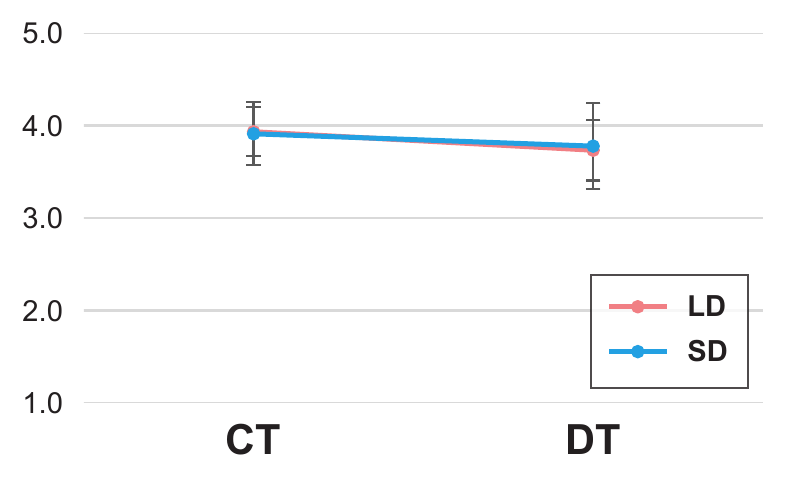}}

\subfloat[AE]{\includegraphics[width=0.48\columnwidth]{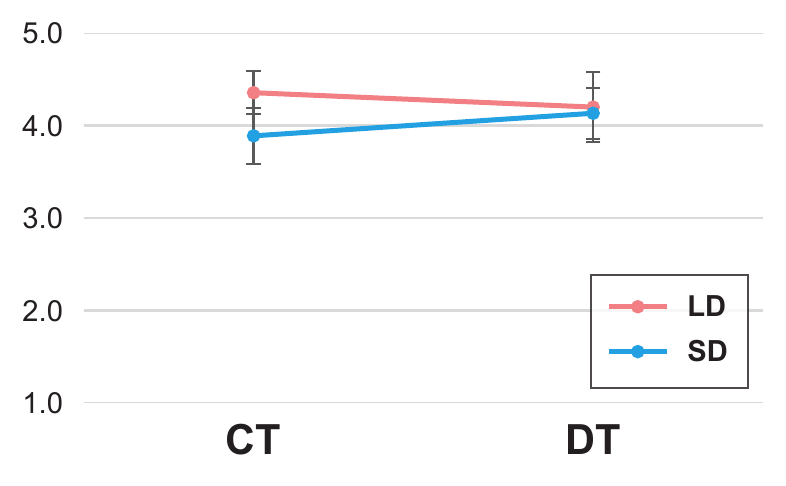}}
\hfil
\subfloat[RW]{\includegraphics[width=0.48\columnwidth]{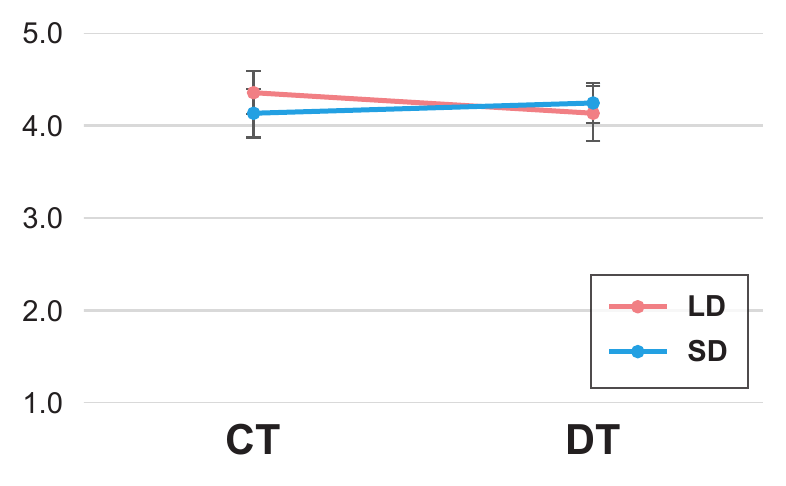}}
\caption[UES-SF results for Groups A--D]{UES-SF results for Groups A--D: (a) overall score and (b)--(e) factor scores (mean; error bars, 95\% confidence intervals). No significant effects were found for AR-phase duration, transition technique, or their interaction.}
\label{S-fig:result_ex1_UES}
\end{figure}

\begin{figure*}[t]
\centering
\subfloat[Overall]{\includegraphics[width=0.24\textwidth]{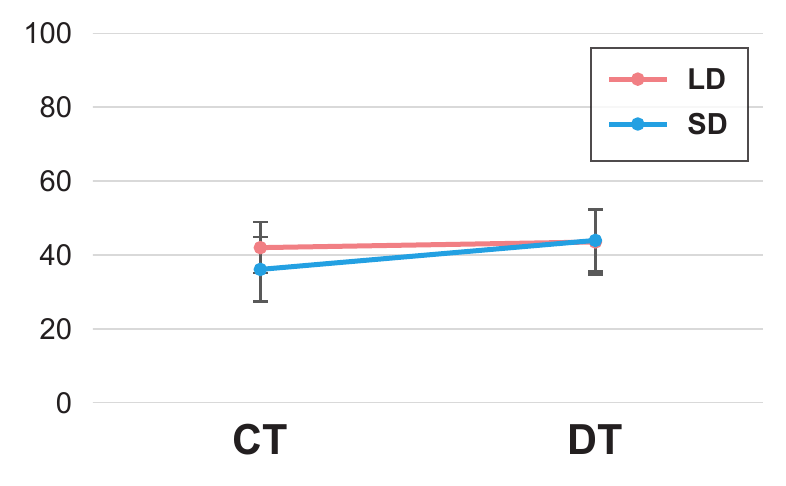}}
\hfil
\subfloat[MD]{\includegraphics[width=0.24\textwidth]{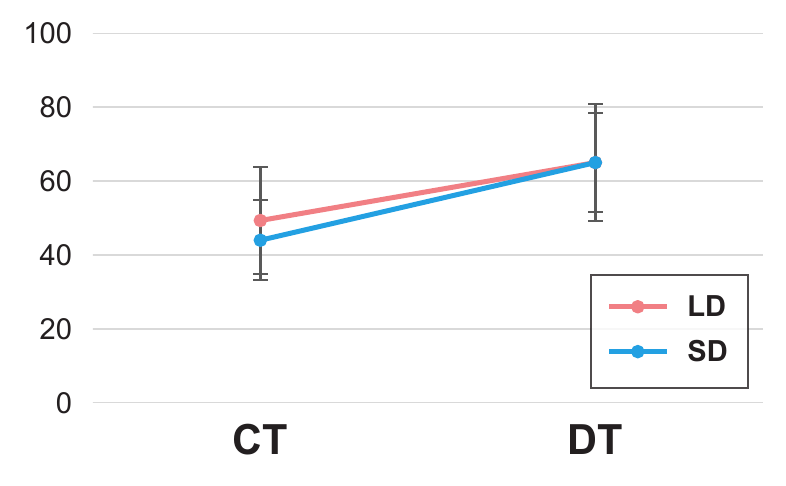}}
\hfil
\subfloat[PD]{\includegraphics[width=0.24\textwidth]{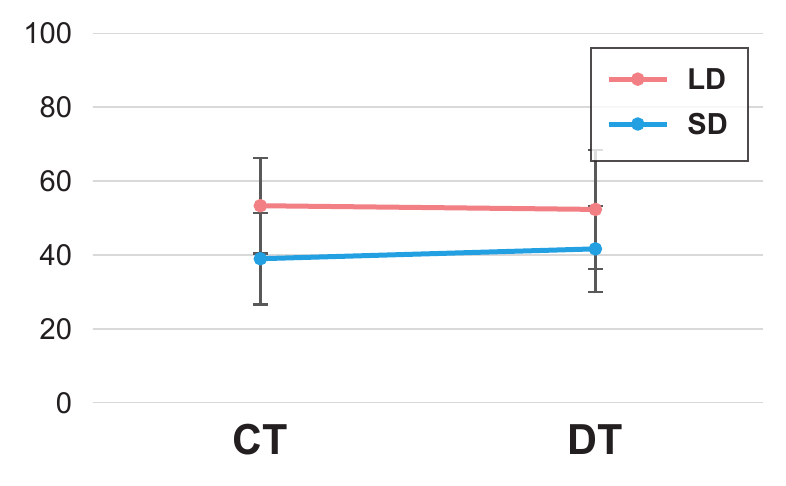}}
\hfil
\subfloat[TD]{\includegraphics[width=0.24\textwidth]{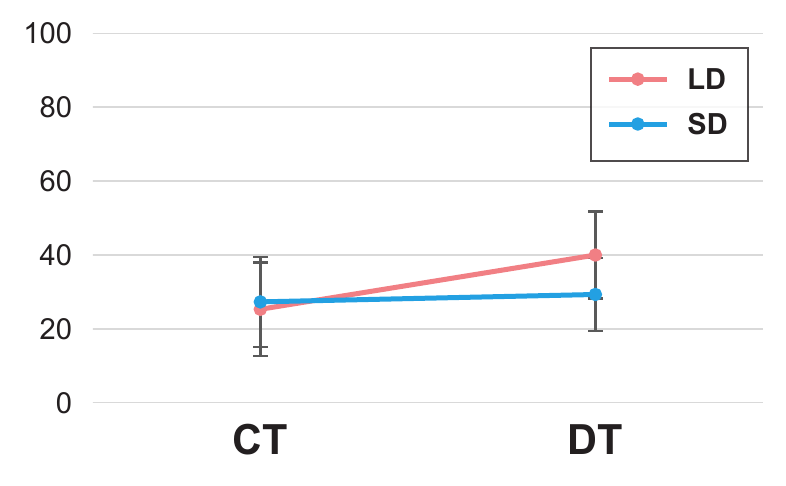}}

\subfloat[PE]{\includegraphics[width=0.24\textwidth]{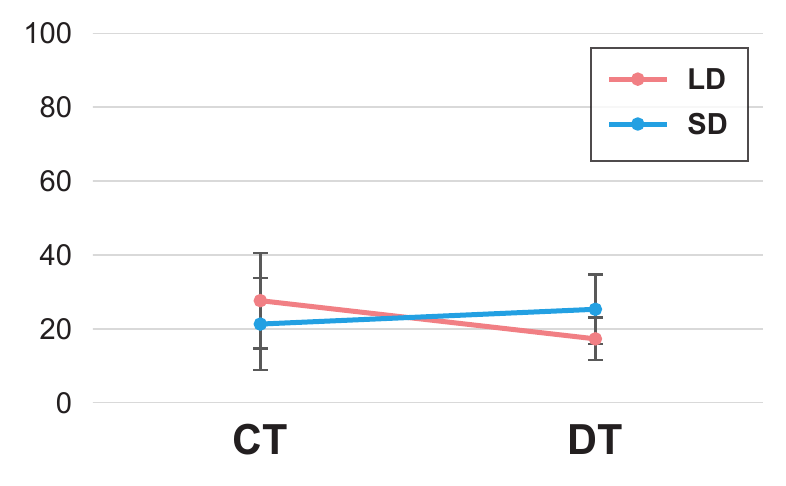}}
\hfil
\subfloat[EF]{\includegraphics[width=0.24\textwidth]{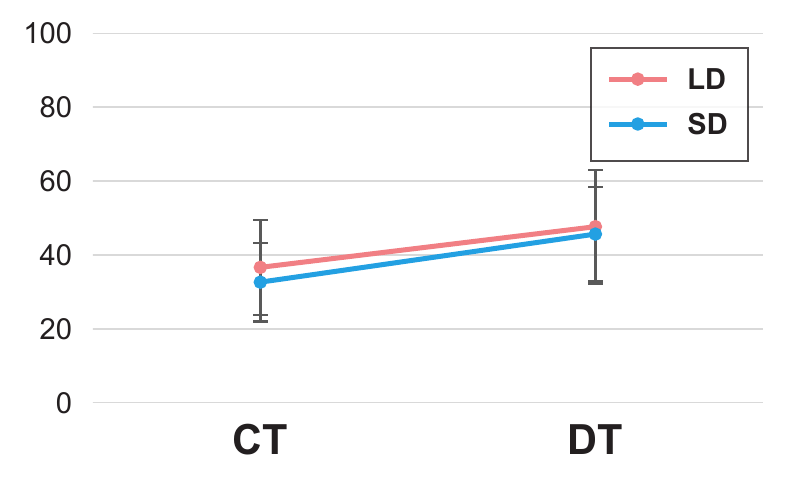}}
\hfil
\subfloat[FR]{\includegraphics[width=0.24\textwidth]{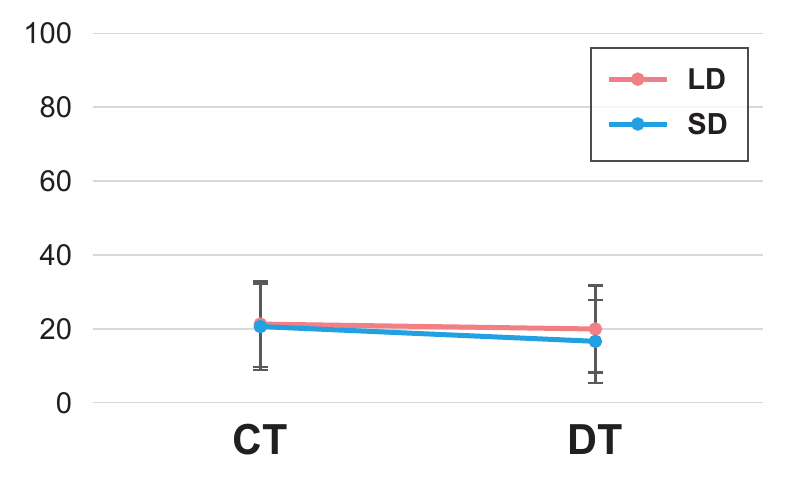}}
\caption[NASA-TLX results for Groups A--D]{NASA-TLX results for Groups A--D: (a) weighted overall score and (b)--(g) subscale ratings (mean; error bars, 95\% confidence intervals). No significant effects were found for the overall score; the subscales were not tested.}
\label{S-fig:result_ex1_NASATLX}
\end{figure*}

\begin{figure}[t]
\centering
\includegraphics[width=0.7\columnwidth]{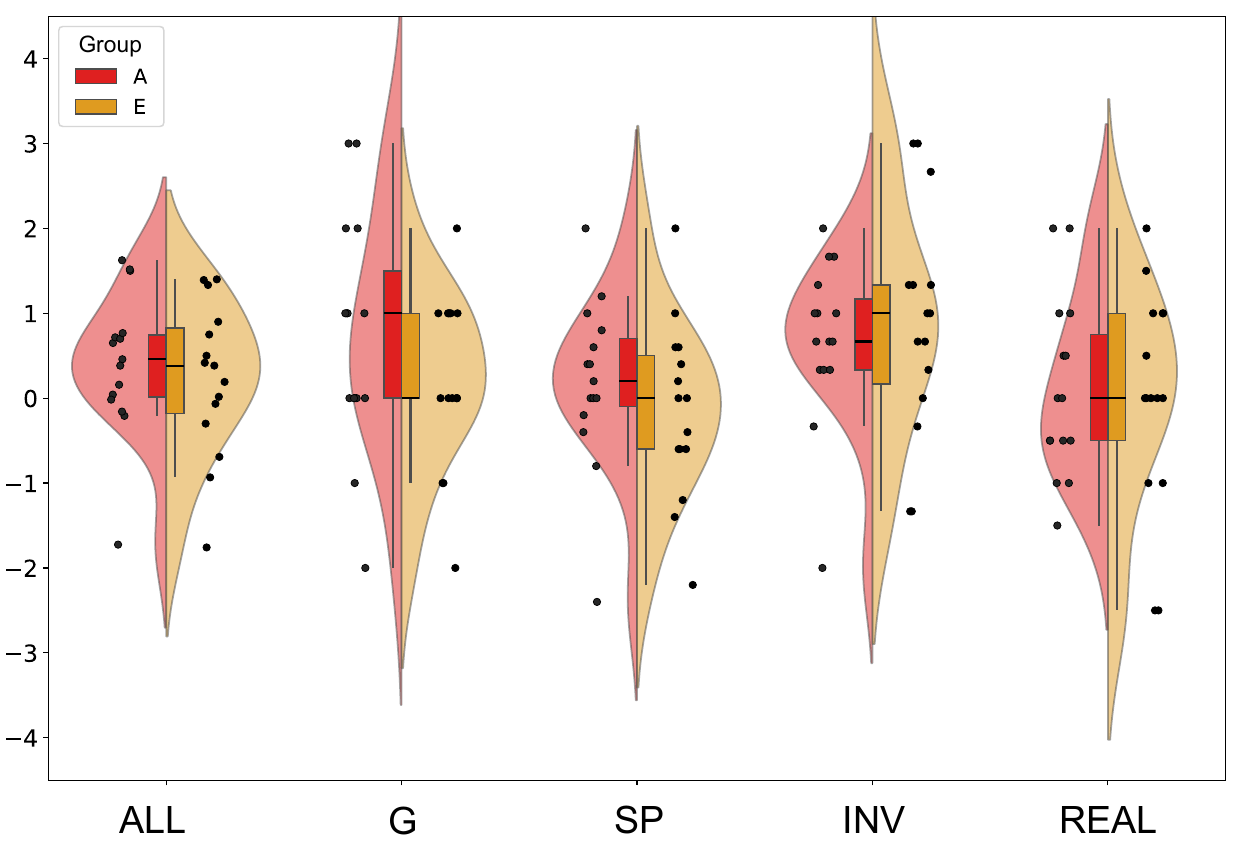}
\caption[IPQ results for Groups A and E]{IPQ change scores for Groups A and E. Half violins show the distribution of each group (A, left; E, right), boxes show the median and interquartile range, and dots show individual participants. No significant differences were found.}
\label{S-fig:result_ex2_IPQ}
\end{figure}

\begin{figure}[t]
\centering
\includegraphics[width=\columnwidth]{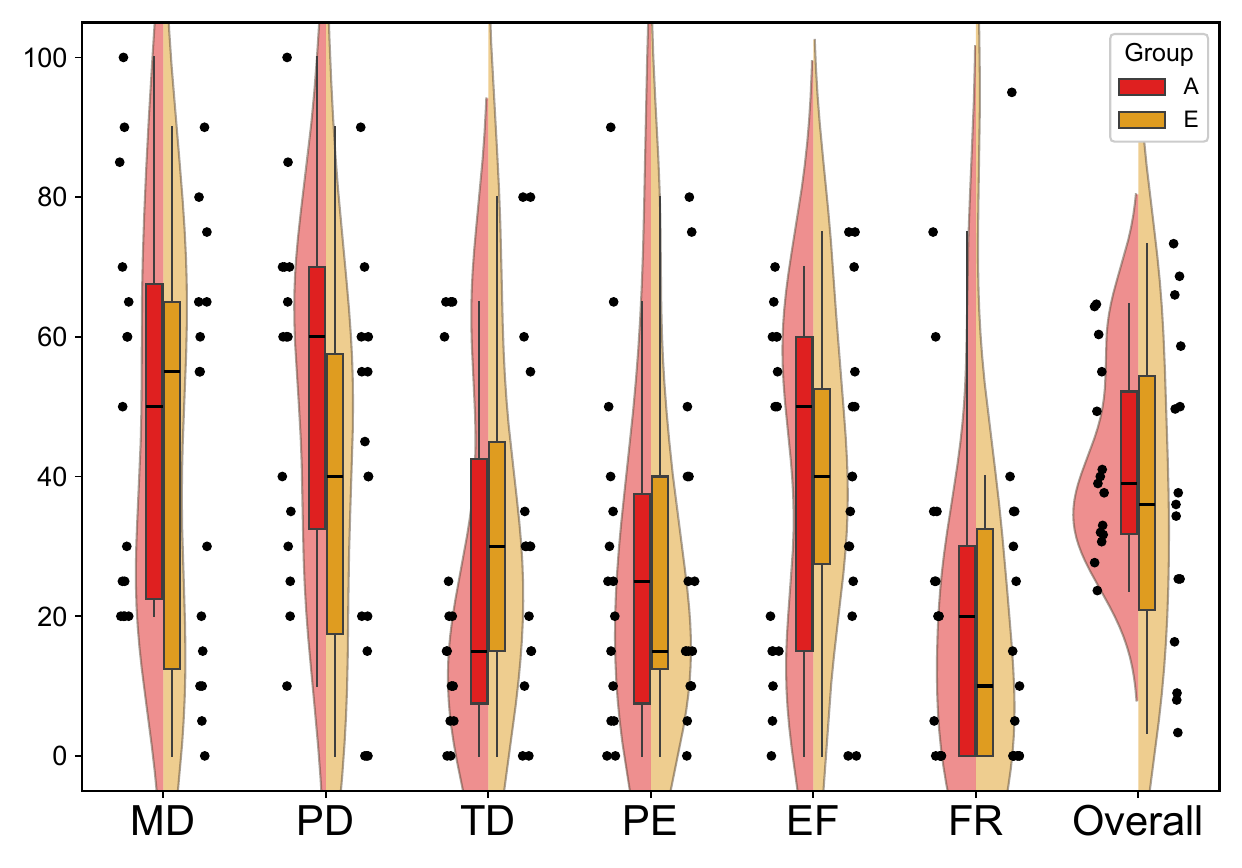}
\caption[NASA-TLX results for Groups A and E]{NASA-TLX results for Groups A and E (subscale ratings and weighted overall score). Half violins show the distribution of each group (A, left; E, right), boxes show the median and interquartile range, and dots show individual participants. The overall scores did not differ significantly; the subscales were not tested.}
\label{S-fig:result_ex2_NASATLX}
\end{figure}

\Cref{S-table:ipq_pre_post} lists IPQ scores at both administrations, Q1, Q2, Q4 (7-point scales, higher Q4 indicates less interference), and NASA-TLX (0--100). \Cref{S-fig:result_ex1_UES,S-fig:result_ex1_NASATLX} show the UES-SF and NASA-TLX results for Groups A--D, and \cref{S-fig:result_ex2_IPQ,S-fig:result_ex2_NASATLX} show the IPQ and NASA-TLX results for Groups A and E. The statistics are reported in the main paper (\cref{sec:results}) and in \cref{S-sec:supp_tests}.

\end{document}